\documentclass[10pt]{article}
\usepackage{blindtext}
\usepackage{epigraph}
\usepackage{subfigure}
\usepackage{lipsum}
\usepackage{natbib}
\usepackage{booktabs}
\usepackage[labelfont=bf]{caption}
\usepackage{subcaption}
\usepackage[dvipsnames]{xcolor}
\usepackage{graphicx}
\usepackage{authblk}
\usepackage{setspace}
\usepackage{amsmath}
\usepackage{amsfonts}
\usepackage{float}
\usepackage{booktabs}
\usepackage{threeparttable}

\usepackage[T1]{fontenc}
\graphicspath{ {./figures/} }

\usepackage{makecell}

\renewcommand{\thefootnote}{}
\usepackage[colorlinks=true,linkcolor=Blue, urlcolor  = Blue, citecolor=Blue]{hyperref}%
\usepackage[top=1in,right=1in,left=1in,bottom=1in]{geometry}

\begin{document}
\begin{singlespace}

\title{\vspace{-.3in}\LARGE{\textbf{Estimating the Cost of Informal Care with a Novel Two-Stage Approach to Individual Synthetic Control}}\footnotetext{\llap{\textsuperscript{}}{\textbf{\noindent For Correspondence:} Maria Petrillo (e: \href{mailto:m.petrillo@sheffield.ac.uk}{m.petrillo@sheffield.ac.uk}) and Daniel Valdenegro (e: \href{mailto:daniel.valdenegro@demography.ox.ac.uk}{daniel.valdenegro@demography.ox.ac.uk}). \textbf{Code Availability Statement}: A software library which accompanies this work can be found at \href{https://github.com/centre-for-care/costofcare}{https://github.com/centre-for-care/costofcare}. Please see the \texttt{readme.md} file within that repository for a Data Availability Statement. \textbf{Acknowledgements}: Funding is gratefully acknowledged from ESRC (Economic and Social Research Council) Centre for Care (Grant: ES/W002302/1), the Leverhulme Trust (Grant RC-2018-003) for the Leverhulme Centre for Demographic Science, and Nuffield College. Insightful comments were gratefully received from seminar participants at the Sheffield Economics Department, the British Society for Population Studies, the Oxford Institute of Population Aging, the Population Association of America conference (2024), and the Leverhulme Centre for Demographic Science.}} \vspace{.3in}}

\author[1,2$^*$]{Maria Petrillo}
\author[1,3$^*$]{Daniel Valdenegro}
\author[1,3,4]{Charles Rahal}
\author[1,5,6]{Yanan Zhang}
\author[1,7]{Gwilym Pryce}
\author[1,8]{Matthew R. Bennett}


\affil[1]{ESRC Centre for Care}
\affil[2]{CIRCLE,University of Sheffield}
\affil[3]{Leverhulme Centre for Demographic Science, University of Oxford}
\affil[4]{Nuffield College, University of Oxford}
\affil[5]{Oxford Institute of Population Aging, University of Oxford}
\affil[6]{University College, University of Oxford}
\affil[7]{Department of Economics, University of Sheffield}
\affil[8]{School of Social Policy, University of Birmingham} 
\affil[$^*$]{Denotes joint lead authorship}

\date{\today}
\maketitle

\begin{abstract}
Informal carers provide the majority of care for people living with challenges related to older age, long-term illness, or disability. However, the care they provide often results in a significant income penalty for carers, a factor largely overlooked in the economics literature and policy discourse. Leveraging data from the UK Household Longitudinal Study, this paper provides the first robust causal estimates of the caring income penalty using a novel individual synthetic control based method that accounts for unit-level heterogeneity in post-treatment trajectories over time. Our baseline estimates identify an average relative income gap of up to 45\%, with an average decrease of £162 in monthly income, peaking at £192 per month after 4 years, based on the difference between informal carers providing the highest-intensity of care and their synthetic counterparts. We find that the income penalty is more pronounced for women than for men, and varies by ethnicity and age. 
\end{abstract} \vspace{.2in}


\end{singlespace}
\newpage
\let\clearpage\relax

\renewcommand{\thefootnote}{\arabic{footnote}}
\setcounter{footnote}{0}

\section{Introduction} 

Informal (unpaid) carers provide the majority of care for family members, friends, and neighbours facing challenges due to older age, long-term illness, or disability \citep{humphries2022ending}. In the UK, over 6.5 million people are informal carers,  providing care valued at £162Bn in England and Wales, and £5.8Bn in Northern Ireland; the equivalent of a second National Health Service and five times the expenditure of publicly funded adult social care services \citep{petrillo2023valuing, zhang2023valuing}. The support carers provide often has significant implications for their financial well-being, health, and relationships \citep{keating2021sustainable, brimblecombe2022inequalities}. Balancing paid work with caring responsibilities often leads to reduced productivity, declining work performance, fewer working hours, and various opportunity costs, all of which negatively impact carers' income \citep{johnson2000trade, bolin2008your,martsolf2020work}. Many occupations require fixed work schedules, which are often incompatible with the unpredictable demands of caring, and flexible working arrangements can be challenging to secure. Strict eligibility criteria for state-funded formal care services further limit access to necessary support, leaving many carers with no option but to reduce their working hours or exit the labour market entirely \citep{lilly2007labor, keating2014taxonomy, glasby2021lost}. Wage discrimination against carers compounds these challenges, further undermining their professional engagement, motivation and financial stability \citep{heitmueller2007earnings}.\par

Several studies have attempted to estimate the income penalty of informal care -- referred to hereafter as the `caring income penalty' -- offering \textit{prima facie} evidence that this penalty may be substantial. For example, analysis of the 1990 General Household Survey by  \citet{carmichael2003opportunity} found that working-age female informal carers in the UK earned lower hourly wages than expected given their human capital, with a 9\% wage reduction linked to providing care for more than 10 hours per week. Using data from the British Household Panel Survey (BHPS), \citet{heitmueller2007earnings} observed a widening wage gap for informal carers since 1990. Estimation based on the Work, Family and Community Nexus (WFCN) Survey by \citet{earle2012cost} found a 29\% increase in the likelihood of wage loss for individuals combining informal care with paid employment, though this was mitigated by access to paid leave for family health needs or supportive line management. Research on the long-term effects of providing care has found cumulative disadvantages over time. \cite{schmitz2017informal} analyse the German Socio-economic Panel (SOEP) data, estimating the impact of caring responsibilities on labour market participation up to eight years after care provision among women. They found no short-term effects on hourly wage, but a considerable long-run wage penalty. Early-life disadvantages also have compounding effects over time \citep{carmichael2003opportunity, skira2015dynamic}, as do the number of caring episodes \citep{raiber2022wage}.\par

There are notable gaps in the literature. First, previous studies have failed to produce robust causal estimates of the caring income penalty due to inadequate control for the endogeneity associated with informal care provision. This failure may stem from data limitations. For instance, both \cite{earle2012cost} and \cite{carmichael2003opportunity} relied on cross-sectional data, which limits the ability to control for unobserved individual characteristics, such as personality traits, that could influence both employment and caring decisions \citep{zhang2024insights}. Additionally, reverse causality may occur, as economic circumstances -- particularly income differentials within households -- can shape caring decisions, with lower earners or the unemployed being more likely to assume caring roles. Simple least squares, as applied by \cite{heitmueller2007earnings}, cannot adequately address these issues. To mitigate this problem, more advanced techniques have been used, such as Propensity Score Matching (PSM) with inverse probability weighting employed by \cite{schmitz2017informal}. However, these methods have significant limitations, with assumptions of strong conditional independence. This assumption presumes that all factors influencing caring decisions are accounted for in the model, effectively treating caring as a randomized treatment based on the controls. Nonetheless, unobserved factors may still influence caring decisions, which these methods cannot fully capture. Second, these studies often focus exclusively on wages, thereby excluding informal carers whose employment was most disrupted by caring responsibilities. As a result, the analysis only applies to carers who remained in or re-entered the workforce. Third, existing research typically examines the impact on wages at a single point in time after caring responsibilities begin, neglecting the possibility that wage and income effects may accumulate over several years. Over time, these effects may diminish in magnitude as individuals and households adjust to the new circumstances \citep{raiber2022wage}. Therefore, there is a pressing need to better understand the dynamics of the caring penalty over time.\par

This paper aims to contribute new methodological advancements to the causal literature by advancing the Individual Synthetic Control approach (ISC) of 
\cite{vagni2021earnings}. Using data from the UK Household Longitudinal Study (UKHLS), we create individual-level synthetic counterfactuals to offer robust estimates of the causal impact of caring responsibilities on income. This methodology offers policymakers and practitioners a more complete understanding of the income penalty by generating a counterfactual scenario for each individual in the treatment group. These counterfactuals are based on weighted outcomes of carers who are otherwise almost identical in the covariate space considered, except for their lack of involvement in informal care responsibilities.\footnote{See Section \ref{methods} for a detailed discussion of the advancements and evolution within the causal inference literature} Our new two-stage approach modifies the conventional synthetic control method, achieving significant improvement in computational performance\footnote{Synthetic control methods are known to be notoriously computationally taxing due to a double convex optimization which increases exponentially in complexity as the donor pool sample increases; see, for example, \cite{becker2018fast}, \cite{malo2023computing} and Figure \ref{sc_baseline_properties} in Section \ref{methods} of this paper.} and treatment-control alignment. We reduce computational complexity whilst maintaining unique and local optimization solutions by algorithmically reducing the donor pool sample size. We achieve this by first calculating a distance metric between each treated case and its potential donor pool of control units in the space formed by the pre-treatment dependent variable and economically relevant covariates.\par

In addition to its methodological contribution, this study advances the literature by shedding light on the intersectional inequalities and heterogeneities in the caring income penalty, with a particular focus on sex, ethnicity and age. Previous literature has shown that the caring income penalty is highly stratified by demographic factors \citep{brimblecombe2022inequalities, Watkins_Overton_2024}. Women typically face a greater income penalty due to prevailing gender norms that often assign them primary caring responsibilities -- particularly in higher-intensity caring roles -- which disproportionately affects their income \citep{van2013effect, glauber2017gender}. Women are also more likely to self-select into more flexible/part-time occupations to balance caring responsibilities and work commitments, albeit at a cost to their income \citep{dunham2003if,ettner1996opportunity, smith2020male, carr2018association}. Ethnic group disparities further complicate the caring income penalty, with occupational segregation contributing to differences in income across ethnic groups. White people, who are more likely to hold higher-paying jobs, tend to experience greater income loss when taking on caring responsibilities \citep{semyonov2007segregated}. However, ethnic minorities may be more likely to take on caring roles \citep{pinquart2005ethnic, cohen2019differences}. Ethnic group differences in caring patterns are also shaped by cultural factors, which play a significant role in caring behaviours within ethnic minority communities \citep{clancy2020eldercare, pinquart2005ethnic, dilworth2002issues, aranda1997influence}. Age also plays a critical role in shaping the caring income penalty \citep{king2019young}. Young carers are particularly vulnerable, as caring responsibilities can disrupt early career development at a time when opportunities for education, training, and career progression are crucial for long-term financial stability \citep{becker2008young}. Early-stage career interruptions or reductions in work hours can have both immediate and lasting effects on earnings, making the opportunity costs especially high for younger individuals \citep{brimblecombe2020high, d2021intergenerational}.

Using nationally representative data from the UKHLS \footnote{\cite{UnderstandingSociety2023}}, and a new, novel, advanced econometric method, we find a notable income gap between informal carers and their synthetic counterparts, particularly among those providing high-intensity care. High-intensity informal carers experience an increasing income gap, with personal income decreasing by up to £192 per month after four years compared to their synthetic counterparts. This contributes to a substantial reduction in overall household income for these carers. Additionally, the relative caring income penalty is more pronounced for women than for men, and for white respondents compared to ethnic minorities. Young carers, aged 25 and below, experience the most severe caring penalty, with their income dropping by as much as £502 per month when compared to their synthetic counterfactuals. The paper is organised as follows. Section \ref{methods} introduces our two-stage ISC approach, situating it alongside other established and popular econometric techniques for estimating the Average Treatment Effect on the Treated (ATT). Section \ref{data} provides a detailed description of the data. Section \ref{results} presents our results, followed by robustness checks that include data contiguity, length of care episodes, placebo tests, and empirical comparisons with PSM, Difference-in-Differences (DID), Synthetic Control (SCM), and Synthetic Difference-in-Differences (SDID) approaches. Finally, Section \ref{conclusion} concludes.
\section{Empirical Strategy} \label{methods}

\subsection{Previous approaches}

Our empirical strategy to estimate the caring income penalty builds on established techniques and methodologies commonly used to estimate the ATT utilising dis-aggregated panel data. Traditionally, researchers have preferred matching techniques (e.g., PSM) and DID for their robustness and relative simplicity. More recently, the SCM and SDID have been developed to address some limitations of these traditional approaches. In this section, we briefly review the main advantages and shortcomings of each method to motivate the development of our ISC approach.

\subsubsection{A common structure}

It is useful to express the functionality of different estimation approaches in a common structure to appreciate their commonalities and differences. Let's start by setting the initial problem in which we have a panel dataset with treated and untreated units. More formally: assume that we observe $J + 1$ units over times $1, 2, \dots T$. Let unit $1$ be treated at times $T_{0} + 1, \dots, T$ with $T_{0}$ corresponding to the moment of treatment and $J$ be a set of untreated units. Let $Y^{I}_{1, t}$ be the outcome of variable $Y$ for unit $1$ at time $t \in T$ if unit $1$ is exposed to treatment (superscript $I$ denotes treatment), and $Y^{N}_{1, t}$ be the outcome of the same unit $1$ at time $t \in T$ in the absence of any treatment (superscript $N$ denotes non-treatment). Within this setting, the ideal estimator for the ATT is:

\begin{equation}
    \tau_{1t} = Y^{I}_{1, t} - Y^{N}_{1, t}
\end{equation}

\noindent $\forall t \geq T_{0}$. Note that unit $1$ cannot be treated and non-treated at the same time. The above expression operates in the ideal but impossible scenario of having the same unit $1$ treated and untreated. In reality only $Y^{I}_{1t} = Y_{1t}, \ \forall t \in T$ is observed, along with $Y_{jt}, \ \forall t \in T\ \&\ \forall j \in J$. The goal of the estimator is to find a weighted combination of units in $J$ that best approximates the unobserved $Y^{N}_{1t}$, so the ATT can be computed as follows:

\begin{equation}\label{gen_estimator}
    \hat{\tau}_{1t} = Y_{1t} - \sum^{J}_{j=1}\omega_{j}Y_{jt}, \ \forall t \geq T_{0}
\end{equation}

\noindent Now, let $\textbf{X}_{1}$ be a $(k \times 1)$ vector of linear combinations of pre-treatment characteristics inclusive of $Y$ for treated units. Similarly, let $\textbf{X}_{0}$ be a vector $(k \times J)$ of linear combinations of the same pre-intervention characteristics for the untreated units. Finally, let $\textbf{W}$ be a vector $(J \times 1)$ of weights ($\omega_{j}\in\textbf{W})$ found by solving the following minimization problem: 

\begin{equation}\label{gen_problem}
    \min_{W}||\textbf{X}_{1} - \textbf{X}_{0}\textbf{W}||
\end{equation}

\noindent More substantively, this suggests that we need to find a combination of the values of the control units $J$ that best resembles the values of the treated unit for the pre-intervention time. We can think of each of the methods below as attempting to solve Eq. \ref{gen_problem} with different restrictions.

\subsubsection{Matching}

Matching-based estimators -- including PSM -- approach Eq. \ref{gen_problem} by applying a kernel function $\mathcal{K()}$ which determines the weights $\omega$ of each control unit $j$ based on a distance metric applied over the hyperplane determined by the matrix of covariates $\textbf{X}$. There are many metrics for matching. PSM approaches -- usually calculated with a logit or probit estimator -- are most commonly used. Other common metrics include the Euclidean distance, Manhattan distance, and the Minkowski distance. What is relevant for the procedure of matching is that these metrics allow us to place each case in a hyperplane, so we can find the closest control(s) for each treated case.

\begin{equation}\label{matching_problem}
    \min_{W}||\textbf{X}_{1} - \textbf{X}_{0}\textbf{W}|| \quad \textrm{s.t.} 
    \quad \omega_{j}=\mathcal{K}(\textbf{X}_{j}); \quad \sum^{J}_{j=1}\omega_{j}=1; \quad \textrm{and} \quad \omega_{j} \ge 0 \quad \forall j.
\end{equation}

\noindent The shape and behavior of $\mathcal{K()}$ can vary. Common variations in the economic literature are 1-nearest neighbor (1-NN), caliper matching, and kernel matching. However, all of these are better understood as kernel variations. 1-NN, for example, can be understood as a uniform kernel that selects the closest match. Caliper matching adds a conditional limit to the range of distances for the match. Other kernels can select a fixed $K$ number of matches and weight them using a variety of functions (e.g., uniform, Gaussian, inverse distance, and so forth). The advantage of this approach is that it provides a local solution, meaning that greater weights are given to control units closer and more similar to the treated unit in the covariate space. The main disadvantages are that matching estimators are more susceptible to extrapolation bias \citep{kellogg2021combining} since the projected values are based on the raw or kernel weighted values of the donor units, and the fact that the computed weights are not optimized to minimize Eq. \ref{gen_problem}. The validity of the matching estimators relies on two critical aspects: the assumption that all factors influencing the likelihood of receiving the treatment are adequately accounted for in the list of measured characteristics, and the quality of the matching process. This assumption implies that there are no unobserved confounders affecting both the treatment assignment and the outcomes. If these conditions are violated, the matching estimator may yield biased and unreliable estimates. These are issues that synthetic control is designed to address (i.e., the constraint in the internal minimization problem effectively acts as a regularisation process, see \citealp{abadie2003economic, abadie2010synthetic}).

\subsubsection{Difference-in-Differences}

In its original formulation \citep{ashenfelter1984using, card1990impact}, DID can be thought of as solving the optimization problem proposed in Equation \ref{gen_problem} subject to the following restrictions:

\begin{equation}\label{dnd_problem}
    \min_{W}||\textbf{X}_{1} - \textbf{X}_{0}\textbf{W}|| \quad \textrm{s.t.} \quad 
\omega_{j}=\frac{1}{J}; \quad \sum^{J}_{j=1}\omega_{j}=1; \quad \textrm{and} \quad \omega_{j} \ge 0 \quad \forall j.
\end{equation}

\noindent DID estimation typically obtains an average of the values of the control group, which is then subtracted from the values of the treated unit for every time after $T_0$. In the cases of multiple treated units, the values of both groups are averaged. Another substantive feature is that DID allows for a non-zero intercept, reflecting permanent additive differences between the treatment and control groups. Hence, the credibility of this method is strained when the pre-treatment trends or characteristics of the untreated units differ significantly from those of the treated units. Finally, DID assumes that unobserved confounders have time-invariant effects on the outcome, which is more commonly known as the `parallel (pre-treatment) trends' assumption.
Even when statistical tests do not reject the parallel trends assumption, unobserved factors may still affect the outcome. Our next section shows (see Eq. \ref{synth_section}) how synthetic control-based approaches allows us to relax this assumption by allowing time-varying unobserved factors as long as the pre-treatment fit remains within acceptable statistical error.

\subsubsection{Synthetic Controls}

The SCM creates a temporally consistent counterfactual of a treated unit where counterfactuals cannot be directly observed. The original methodology was proposed in the context of natural experiments as an explicit alternative to matching estimators \citep{abadie2003economic}. \cite{abadie2010synthetic} generalized this by allowing it to be used in a wider set of contexts, such as policy evaluations and large-scale interventions, but always initially with the focus of estimating causal effects at an aggregated unit (such as regions, states or countries). The SCM approach to Eq. \ref{gen_problem} is to numerically find the optimal weights for each control unit ($j \in J$). More formally:

\begin{equation}\label{synth_section}
    \min_{W}||\textbf{X}_{1} - \textbf{X}_{0}\textbf{W}||, \quad \textrm{s.t.} \quad \sum^{J}_{j=1}\omega_{j}=1; \quad \textrm{and} \quad \omega_{j} \ge 0 \quad \forall j.
\end{equation}

\noindent We can assume that Eq. \ref{synth_section} holds if -- as proposed in \citet{abadie2010synthetic} -- we also assume that $Y^{N}_{j,t}$ follows the following factor model:

\begin{equation}\label{factor_model}
    Y^{N}_{j, t} = \delta_{t} + \boldsymbol\theta_{t}\textbf{Z}_{j} + \boldsymbol\lambda_{t}\boldsymbol\mu_{j} + \varepsilon_{j, t}.
\end{equation}

\noindent This assumption incorporates the influence of time-specific effects $\delta_{t}$, the interaction between time-varying factors $\boldsymbol\theta_{t}$ and covariates $\textbf{Z}_{j}$, unit-specific factors $\boldsymbol\lambda_{t}$ and their loadings $\boldsymbol\mu_{j}$, as well as an idiosyncratic error term $\varepsilon_{j, t}$ as specified in the following equation:

\begin{equation}\label{factor_model_2}
    \sum^{J}_{j=1}\omega_{j}Y_{jt} = \delta_{t} + \boldsymbol\theta_{t}\sum^{J}_{j=1}\omega_{j}\textbf{Z}_{j} + \boldsymbol\lambda_{t}\sum^{J}_{j=1}\omega_{j}\boldsymbol\mu_{j} + \sum^{J}_{j=1}\omega_{j}\varepsilon_{jt}
\end{equation}.

\noindent Note that the left-hand side term in Eq. \ref{factor_model_2} is identical to the right-hand side term in Eq. \ref{gen_estimator}; it represents the synthetic control estimator. An unbiased synthetic control estimator will satisfy:

\begin{equation}
\sum^{J}_{j=1}\omega_{j}\textbf{Z}_{j}=\textbf{Z}_{1}
\end{equation}

\noindent and

\begin{equation}
\sum^{J}_{j=1}\omega_{j}\boldsymbol\mu_{j}=\boldsymbol\mu_{1}.
\end{equation}

\noindent However, $\boldsymbol\mu_{1}$ is unobserved. \citet{abadie2010synthetic} provides evidence that the factor model in Eq. \ref{factor_model} can only fit $\textbf{Z}_{1}$ and a long set of outcomes $Y_{1t},\dots,Y_{1T_{0}}$ as long as it also fits its loadings $\boldsymbol\mu_{1}$. This implies that the synthetic control estimator is robust to the presence of unobserved time varying confounders, something which is not the case with DID estimators.

\subsubsection{Synthetic Difference-in-Differences}

Building on the strengths of both DID and SCM, the SDID methodology offers a hybrid approach that aims to combine the advantages of these two techniques while addressing some of their inherent limitations. Developed by \citet{arkhangelsky2021synthetic}, it introduces the computation of a set of weights for each pre-treatment time as well as unit weights as in traditional synthetic control. More formally, SDID modifies Eq. \ref{gen_problem} as follows:

\begin{equation}
      \min_{W}||\textbf{X}_{1} - (\textbf{X}_{0}\textbf{W}) \odot \boldsymbol\Lambda||, \quad \textrm{s.t.} \quad \sum^{J}_{j=1}\omega_{j}=1; \quad \textrm{and} \quad \omega_{j} \ge 0 \quad \forall j
\end{equation}

\noindent where $\boldsymbol{\Lambda}$ is a column vector containing each time weight $\lambda_{t}$, obtained by minimising the following expression:

\begin{equation}
  \min_{\lambda_0 \in \mathbb{R}, \lambda \in \Lambda} \sum_{i=1}^{N_0} \left( \lambda_0 + \sum_{t=1}^{T_{pre}} \lambda_t Y_{it} - \frac{1}{T_{post}} \sum_{t=T_{pre}+1}^{T} Y_{it} \right)^2
\end{equation}

\noindent such that:

\begin{equation}
  \boldsymbol\Lambda = \left\{ \lambda \in \mathbb{R}_+^{T_{pre}} : \sum_{t=1}^{T_{pre}} \lambda_t = 1, \, \lambda_t = T_{post}^{-1} \text{ for all } t = T_{pre} + 1, \ldots, T \right\}.
\end{equation}

SDID introduces this second optimization routine to obtain the set of weights that minimise the difference between outcomes of the pre-treatment time (T$_{pre}$) against the average of the outcomes of the post-treatment time (T$_{post}$). The benefits of this approach are discussed in detail by \citet{arkhangelsky2021synthetic}. It is important to note that this procedure does not alter the calculation of the individual weights, which are obtained in the same manner as in the original SCM. Finally, while this methodology does not have any of the shortcomings of matching or DID, it does propose a procedure with a significant increase in computational time, as it adds an extra minimisation problem.

\subsection{Individual Synthetic Control}

Although developed for cases in which treated and untreated units were large aggregations of individuals, little work has been done to make such methods amenable to situations with multiple treated units ($l \in \{1, 2, \dots L\}$). \citet{vagni2021earnings} estimate the ATT in a micro-level application to the motherhood penalty in the following way:

\begin{equation}\label{isc}
    ATT_{t} = L^{-1}\sum^{L}_{l=1}\hat\tau_{lt}, \quad \forall t \geq T_{0}
\end{equation}

\noindent where $\hat\tau_{lt}$ correspond to the outcome of Eq. \ref{gen_estimator} for all treated cases. Note, here, that the weights for each $j\in J$ are re-calculated for each possible donor for all treated units ($l\in L$). A very similar modification was proposed by \citet{abadie2021penalized} in which a penalisation factor ($\lambda$) is added in order to favour control units $j$ with the smallest pairwise difference between each treatment, calculated as follows:

\begin{equation}\label{penalized}
    \hat{\tau}_{lt} = Y_{lt} - \sum^{J}_{j=1}\omega_{j}(\lambda)Y_{jt}, \ \forall t \geq T_{0}
\end{equation}

\noindent \citet{abadie2021penalized} proposed Eq. \ref{penalized} with the specific intention of ensuring a unique solution for when there are multiple treated units. By penalizing pairwise discrepancies, the ISC approach favours control donors more similar to the treated one. However, with both methods, weights for all control units must be determined,  even if those weights are negligible. The premise of our computationally tractable approach set out below is to only compute the weights of control units that are economically meaningful.

\subsection{Outlining a Two-Stage Approach to Individual-level Synthetic Control}

\subsubsection{Our Approach to Individual-level Synthetic Control}\label{our_approach}

Having summarised the evolution of the causal literature up until now, we next describe our contribution which is essentially an enhancement of the ISC approach. Small donor pool sizes (i.e. $\|J\|$) are desirable when searching for local solutions. \cite{abadie2010synthetic}, \cite{abadie2021using}, and \cite{abadie2021penalized} all mention that restricting the donor pool to units most similar to the treated unit can help solve problems of uniqueness and interpolation bias. In addition to this, we also highlight the fact that reducing the donor pool size will reduce the computational complexity of the final computation, turning the ISC into an estimator that can be applied to high-dimensional scenarios. With this motivation in mind, we propose a modification to the idea of ISC that delivers more computationally tractable results. Substantively, we propose to find the $K$ nearest control units to each treated unit using some distance metric in the space of dependent variable and covariates (in our case, this would be household or personal income or income share plus age, sex, marital and employment status, ethnicity, educational level and household size) formed by $\textbf{X}$, effectively reducing the number of control units for which weights need to be calculated from $J$ (the total number control units) to $K$ (the number of nearest controls units), where $K<<J$. We can express this as a modification of the synthetic control problem in Eq. \ref{synth_section}:

\begin{equation}
    \min_{W}||\textbf{X}_{1} - \textbf{X}_{0}\widehat{\textbf{W}}||, \quad \textrm{s.t.} \quad \sum^{K}_{j=1}\widehat{\omega}_{j}=1; \quad \textrm{and} \quad \widehat{\omega}_{j} \ge 0 \quad \forall j
\end{equation}

\noindent where $\widehat{\textbf{W}}$ is a column vector $(K \times 1)$ of weights $\widehat{\omega}_{j}$, and $K$ is the length of the set if indices $S$ in $J$ that satisfy:

\begin{equation}\label{k_nearest}
    S = \{i|d_{i} \leq d_{j} \forall j \in J \land |S|=K\}
\end{equation}

\noindent where $d$ is some distance metric that outputs:

\begin{equation}\label{distance}
    d = ||\textbf{X}_{1} - \textbf{X}_{j}||, \quad\forall j \in J
\end{equation}

\noindent With this, our estimator, which we will call $\delta$ can be obtained as follows:

\begin{equation}\label{our_estimator}
    \delta_{lt} = Y_{lt} - \sum^{J}_{j=1}\widehat{\omega}_{j}Y_{jt}, \ \forall t \geq T_{0}, \ \forall l \in L
\end{equation}

\noindent while the ATT using our estimator can be obtained as follows:

\begin{equation}\label{isc_dv}
    ATT_{t} = L^{-1}\sum^{L}_{l=1}\delta_{lt}, \quad \forall t \geq T_{0}
\end{equation}

\noindent where $\delta_{lt}$ correspond to the outcome of Eq. \ref{our_estimator} for all treated cases $1, \dots L$.

With the ISC -- where there are potentially a large number of treated units -- the above procedure has significant savings in terms of computational complexity as only the $K$ `closest' best fitting controls to the treated in the covariate space are chosen to contribute to the synthetic control which acts as a counterfactual to the treated unit. In our procedure we explicitly favour reducing the pairwise distance between the treated and selected controls, to later find the optimal combination of weights to create synthetic counterfactuals. This favours the ecological validity of the synthetic control, but might affect the fit of it compared with a solution that uses all controls in the donor pool. Recently, similar approaches have been proposed using variable selection techniques (i.e., Lasso regressions, Singular Value Decomposition) to reduce the donor pool size \citep{hollingsworth2020tactics, amjad2018robust}. In principle, these alternate methods achieve the same reduction of donor pool size with one critical difference; the selected donors are not necessarily the closest in the covariate space. This is important, as it ensures maximum validity in the estimation of the synthetic counterfactual.

\subsubsection{The Choice of $K$}

Our approach proposes the use of ISC for disaggregated data, which may include hundreds of thousands of treated cases. We introduce this additional step where the donor pool sample size is reduced from the total number of non-treated cases to only the closest $K$ for each treated case. One step is still missing: determining the optimal size of $K$. Ideally, $K$ should be chosen to simultaneously minimize the difference between the treated and its synthetic control (i.e. RMSPE) for the pre-treatment time while balancing the need for computational tractability. This can be done individually for each treated unit (resulting in different values for $K$), or uniformly for all treated units (i.e., a general $K$ for all). In our approach, we use a general $K$=10 for all treated units, but also calculate the RMSPE in Section \ref{profiling} as shown in Eq. \ref{mrmspe}:

\begin{equation}\label{mrmspe}
    \text{RMSPE} = L^{-1}\sum_{l=1}^{L}\left(T_{0}^{-1} \sum_{t=1}^{T_{0}} (Y_{lt} - \hat{Y_{lt})^2}\right)^{\frac{1}{2}}
\end{equation}

\subsubsection{Algorithmic Profiling}\label{profiling}

We profiled the algorithm outlined in Section \ref{our_approach} in two ways. First, we analyzed the RMSPE and execution time -- both as a function of logarithmically gridded $K$ -- and also analyzed the frequency of selection of individual control units' selection into the donor pool using our two baseline models for (real) Individual and Household Income. We also simulated a population of 1,000 units composed of 25 subpopulations with 100 treated cases randomly assigned to any of the 25 subpopulations. This involves a measurement period of 100 steps using random walks \citep{pearson1905problem}. Each subpopulation had specific parameterisations for their random walks in order to simulate as closely as possible the variation in paths and across subpopulations. For the treated units, treatment occurred at the 50th step. At this point, the random walks were changed to increase the probability of having a downward movement, simulating a treatment and reducing the magnitude of the measured outcome. Then, for each of the 100 treated units, we computed their synthetic controls using our method 30 times. We repeat this simulation ten times, each with a different instantiation of our pseudo-random number generator. We again analysed the RMSPE and execution time -- both as a function of $K$ -- as well as the frequency of selection of into the donor pool.\par

\begin{figure}[!t]
\centering
\includegraphics[width=0.99\textwidth]{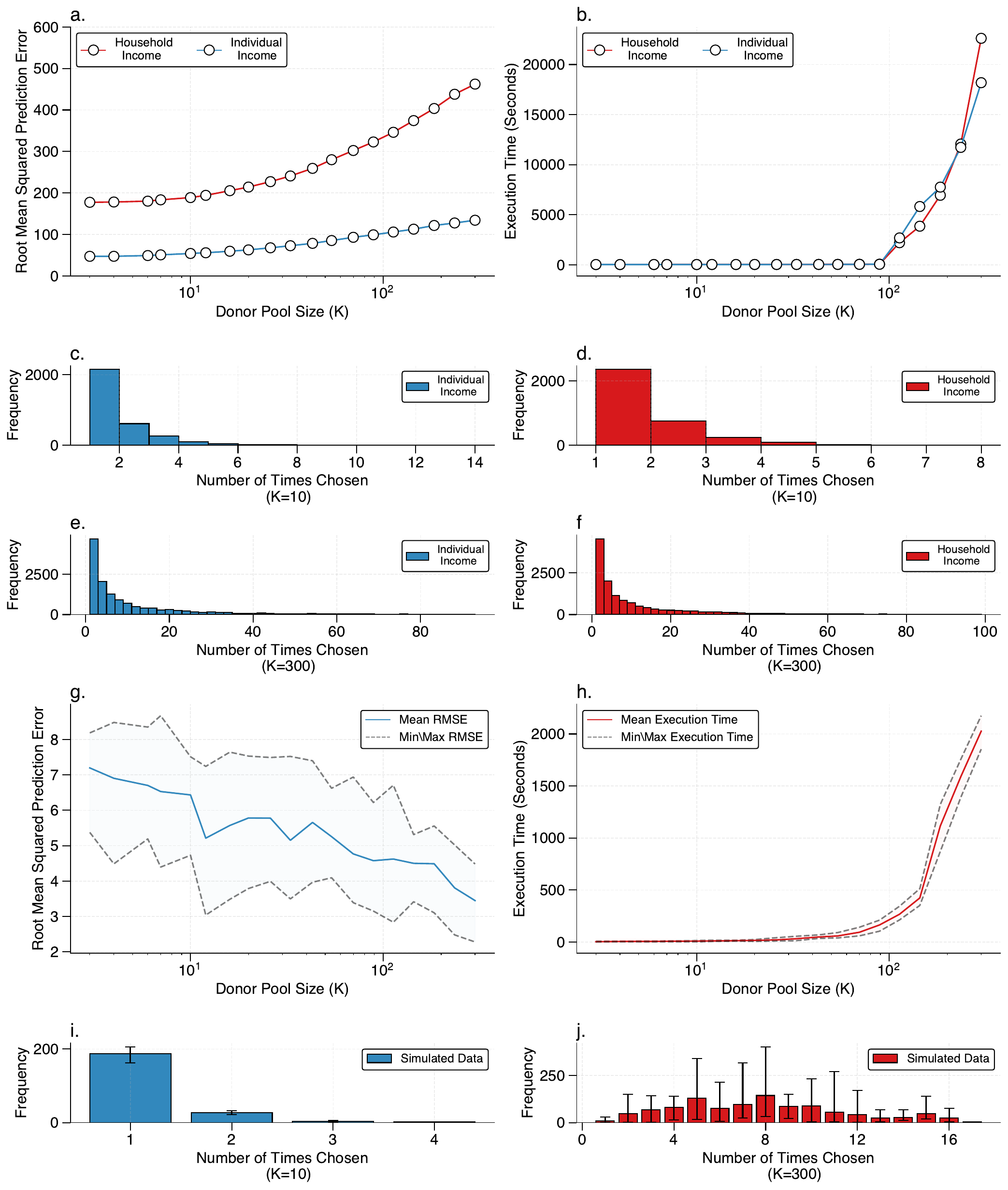}

\caption{\small{\textbf{Baseline and Simulated Performance of the Two-Stage Individual Synthetic Control Method:} Panels a. and g. represent the RMSPE against donor pool size for the empirical baseline and simulation respectively. Panels b. and h. represent execution time against donor pool size. Panels c.-e. and i-j. represent the frequency count by which the same individual forms part of a control group. Panels a., c., and e. are for individual income, while b., d. and f. are for household income. Panels g.-j. are from our simulation experiment, ran across ten seeds.}}
\label{sc_baseline_properties}
\end{figure}

Panels a. and b. in Figure \ref{sc_baseline_properties} show the results of our profiling for the baseline model, which suggest that the minimal RMSPE was obtained with donor pool sizes between 5 and 10, with very similar and efficient execution times. Given this, we set $K=10$ for all of our downstream analysis, as it provides a balance between diversity in the donor pool size and the minimization of the RMSPE. Panels g. and h. in Figure \ref{sc_baseline_properties} show the resulting mean RMSPE and mean execution time across the increasing donor pool sizes in our simulated scenario. Overall, we observe that execution time explodes after $K$>100, while the RMSPE has a much more gradual and highly variable reduction across the runs. Notably, the reduction in RMSPE is not monotonic as the the donor pool size increases. Comparing the execution time and RMSPE of our baseline and simulation, we can see that there are similarities in terms of execution time, but not in the RMSPE across runs. When using real data, the error increases monotonically as the donor pool size increases. We recognize this result is counter intuitive. Therefore, we ran this analysis using different optimization algorithms, with similar conclusions. A tentative explanation is that the complexity inherent in real-world data is far greater than in our generalised, overly simplistic simulation strategy. This complexity -- potentially due to geographical dis-continuities or other empirical sub-population phenomena -- might be driving the difference in the results.\footnote{Given this, we recommend researchers interested in using our approach run a similar grid-searched `fit check' over a set of increasing donor pools sizes in order to test the behaviour within their their specific data. Computing time measurements was done using a 11th Gen Intel® Core™ i7-1185G7 × 8 processor, a fairly common consumer level CPU found in many laptops, meant to represent the average computing power of an academic researcher.} In terms of selection into the donor pools (Figure \ref{sc_baseline_properties} Panels c.-f. and i.-j.), our results show a considerable mass in the distribution around people being selected only once into a donor pool size; with low values of K (i.e, 10), individuals are infrequently chosen more than once. This indicates that our algorithm is highly discerning in terms of the range of people that can potentially be selected as donors to each individual treated unit.

\subsubsection{Confidence Interval Estimation}

\cite{vagni2021earnings} propose a method for estimating confidence intervals for the cross-sectional estimates (each time point) in which within and between individual variances are estimated. Between individual variance follows the standard procedure. Within individual variance estimation is achieved by bootstrapping the synthetic control estimation for each treated case, resampling with replacement from the donor pool. Three scenarios are possible for the `fit' (implemented as the mean RMSPE) of the bootstrapped models:

\begin{enumerate}
    \item They have the same fit as the optimal solution with an overall different set of controls, but non-zero weights are assigned to the same set of controls, with replacements only in control units with null weights (i.e. the new chosen set of controls is different only in inconsequential units);
    \item They have the same fit as the optimal solution, but with a different set of weights, meaning that the solution is not unique;
    \item They have a worse fit than the optimal solution.
\end{enumerate}

\noindent The first scenario yields the optimal result, so it does not produce variation in the post-treatment outcomes. The second scenario is undesirable and recognized as a violation of the assumptions in \citet{abadie2010synthetic}. A solution was proposed to ensure local and unique solutions in \citet{abadie2021penalized}, similar to what we propose. Finally, the third scenario produces a synthetic control with worse fit, and hence, according to \citet{abadie2010synthetic}, is more biased. We, therefore, argue that this within variance estimation is unnecessary, and adhere to the fact that the synthetic control is not an estimation of a real population value, but a solution of best fit given the data. Following the reasoning presented in \citet{abadie2010synthetic}, the best fitting solution should yield less biased results, and hence computing solutions known to yield worse fit would introduce bias. Therefore, we conduct a bootstrap procedure to create confidence intervals at the between-level for each cross-sectional estimate as follows (simplified for one time period $t$ only):
\begin{itemize}
    \item Let $\Delta_{t} = \{\delta_{1t}, \delta_{2t}, ..., \delta_{Lt}\}$ be the collection of all the outputs of our estimator (Eq. \ref{our_estimator}) for each treated case up to $L$ in time $t$,
    \item Let the mean of $\Delta_{t}$ be $\bar{\Delta}_{t}= L^{-1}\sum^{L}_{l=1}\delta_{lt}$ as in Eq. \ref{isc_dv},
    \item Let $\Delta^{*}_{tb}$ be the $b_{th}$ bootstrap sample of size $n=L$ obtained by sampling with replacement from $\Delta_{t}$,
    \item Let $B$ represent the number of bootstrap samples to take.
\end{itemize}

\noindent This allows us to formalise our approach as:

\begin{equation}
\begin{split}
    \bar{\Delta}^{*}_{tb} &= n^{-1}\sum^{n}_{i=1}\delta^{*}_{tb} \quad \forall b \in B, \\
    \textbf{S}_t &= \{\bar{\Delta}^{*}_{t1}, \bar{\Delta}^{*}_{t2}, \dots, \bar{\Delta}^{*}_{tB} \}, \\
    [CI_{2.5\%}, CI_{97.5\%}]_{t} &= [Q_{2.5\%}(\textbf{S}_t), Q_{97.5\%}(\textbf{S}_t)] \quad \forall t \in T
\end{split}
\end{equation}

\noindent where $[CI_{2.5\%}, CI_{97.5\%}]_{t}$ is the 95\% confidence interval drawn from the 2.5th and 97.5th percentiles of the set $\textbf{S}_t$ ($[Q_{2.5\%}(\textbf{S}_t), Q_{97.5\%}(\textbf{S}_t)]$).
Each value from $t=1$ to $t=T$ in our standardized trajectories is an average of many individual synthetic controls. To obtain confidence intervals, we resampled with replacement 1,000 times, each yielding a new average. 
In Figures \ref{hh_and_ind_inc_baseline} and \ref{inc_share_fig} we show that our confidence interval estimation approach and the approach of \citet{vagni2021earnings} overlaps almost entirely. Finally, we apply our two-stage ISC approach to estimate the impact of informal caring on carers' income trajectories. The ISC method will effectively account for unobserved changes in income over time by creating a synthetic control group that closely mirrors the income pattern of informal carers. It is essential to acknowledge that the choice to provide informal care is not assumed to be random. There may indeed be unobserved variables affecting informal care decisions, yet we assume that these unobserved variables do not correlate with the income trajectory of informal carers after undertaking caring responsibilities. Essentially, the only bias unaddressed by this method arises from an unobserved variable that impacts both the decision to undertake caring responsibilities and the subsequent income trajectory, without affecting the income trajectory of those who assumed caring roles before the event \citep{vagni2021earnings}.
\section{Data}\label{data}

Our analysis draws upon twelve waves of panel data from the UKHLS, spanning 2009 to 2020 (see Supplementary Information \ref{ukhls} for more information). The UKHLS provides valuable insights into individuals’ `carer status', enabling us to examine the causal impact of caring responsibilities on income throughout one’s life. Individuals are treated if they provide informal care or special assistance to sick, disabled, or older adults, regardless of whether they reside within the same household or elsewhere. We do not consider the duration of the caring episode in our baseline model. Conversely, the control group comprises individuals who do not engage in informal caring activities throughout the longitudinal period of the panel. The UKHLS is unique as it allows us to quantify informal caring responsibilities per week, and in the process explore `threshold effects'; how increased intensities of caring impact upon the caring income penalty.\footnote{See Supplementary Information \ref{datapreparation} Section \ref{caredefinition} and Table \ref{table:variables} for detailed information on all independent and dependent variables question wordings, operationalisations and cleaning/coding.} We categorise informal carers into four groups based on the intensity of care they provide:\footnote{This categorization of informal carers by care intensity is directly shaped by the limitation of our data. For more information, see Supplementary Information Section \ref{caredefinition}.}

\begin{itemize}
    \item \textbf{High-intensity informal carers}: Individuals providing 50 hours or more per week;
    \item \textbf{Medium-high-intensity informal carers}: Individuals providing 20 to 49 hours per week;
    \item \textbf{Medium-low-intensity informal carers}: Individuals providing 5 to 19 hours per week;
    \item \textbf{Low-intensity informal carers}: Individuals providing less than 5 hours per week.
\end{itemize}

\noindent This categorization relies on the information provided during the first year of treatment to capitalise on the potential shock that providing informal care can cause in individuals' lives. Furthermore, we only include those individuals for whom we have a minimum of three data measurement points before the onset of the treatment. This criterion is essential to facilitate the reliable calculation of weights in Eq. \ref{synth_section}.\par

\begin{figure}[!t]
    \centering    \includegraphics[width=\linewidth]{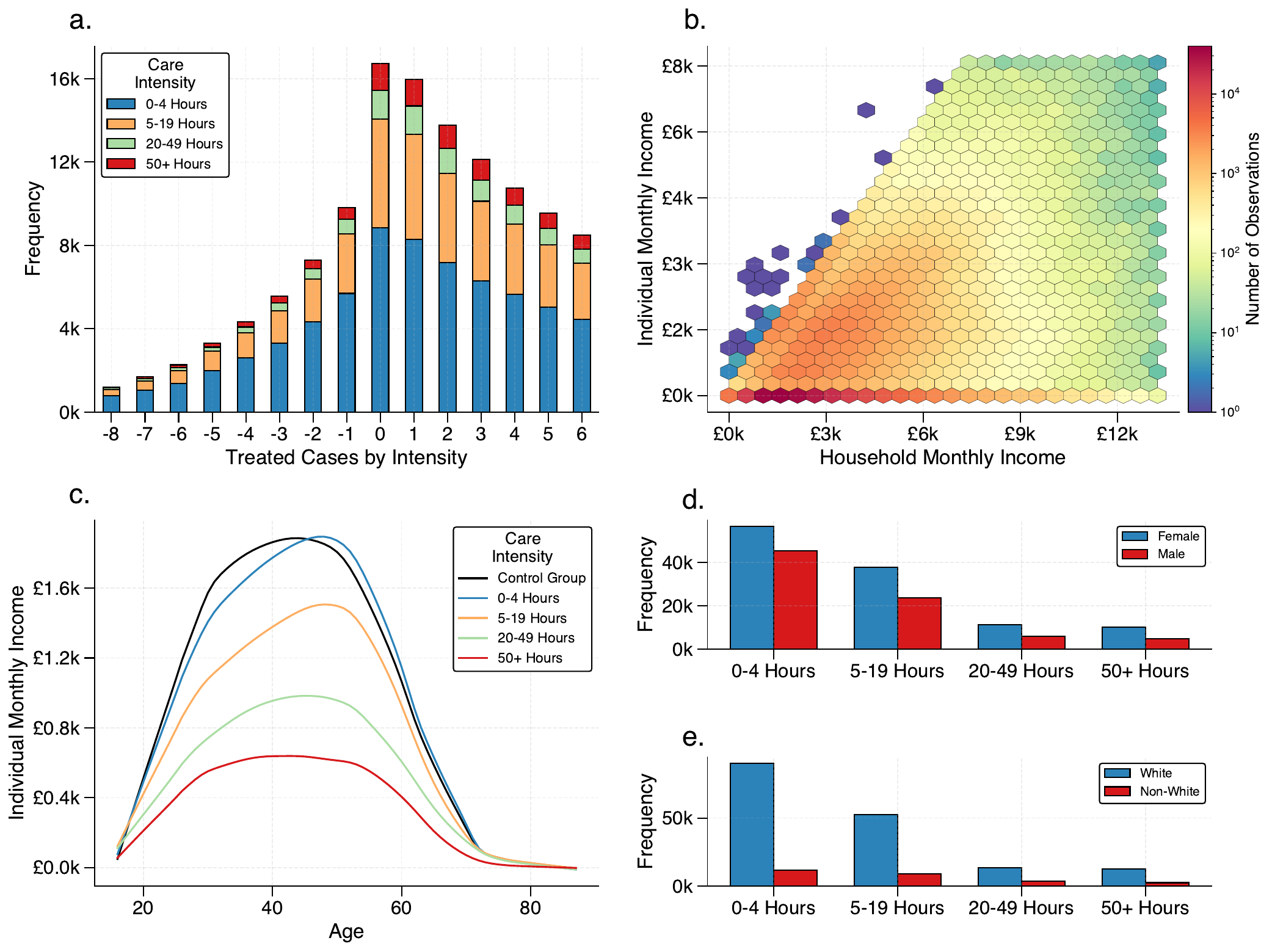}
    \caption{\small{\textbf{Care Intensity and Income Profiles.} Note: Panel a. plots the number of observations we have both before and after a treatment across our four different levels of treatment intensity. Panel b. plots monthly individual against household income. Panel c. plots the LOESS smoothed (frac=0.3) mean monthly individual income across ages by these same four different care intensities (as well as the control group). Panels d. and e. plot care intensity frequencies by both male and female (d.) and our `white' and `non-white' ethnicity groups (e.). Source: UKHLS data (years 2009-2020), author's calculations.}}
    \label{fig1}
\end{figure}

Our analysis focuses on dependent variables which reflect our threefold conceptualisation of the cost of providing care: i.) individual monthly income; ii.) household monthly; and iii.) income share.\footnote{All monetary amounts are adjusted for inflation (base year 2015) using a Consumer Price Index which includes owner-occupiers' housing costs (CPIH)} Table \ref{descr} provides an overview of the sample characteristics by presenting the mean values of controls used in our analysis for both the treatment groups (delineated by treatment intensity), and the control group.\footnote{For further detail on the variables used in the analysis, please refer to the Supplementary Information \ref{variablesdefinitions}.} Figure \ref{fig1} expands this further, primarily focusing on Care Intensity across treatment periods (Panel a.), income profiles (b.), care intensity by age (c.), and intersectional characteristics (d. and e.). High-intensity carers are less likely to be employed (23\%). The likelihood of being employed increases as the intensity of caring hours decreases; 56\% prevalence for low-intensity carers, while the control group who never provide care is 49\%. A similar trend is evident in income share where caring intensity has an impact on the share of household income. Low-intensity carers contribute nearly 25.97\% to household income, while high-intensity carers contribute only 8.86\% in contrast to the control group's 27.11\%. Informal carers are more likely to be married compared to our control group; 68\% and 64\% of low-intensity and high-intensity carers, respectively. High-intensity caring is predominantly provided by women rather than men; 64\% of the subsample are female carers. The curvature of the age-monthly income relationship is linked to the weekly commitment of informal care hours. Among individuals who provide 0-4 hours of informal care per week, average income is higher for those assuming caring responsibilities. However, a significant shift occurs when we focus on individuals undertaking more than 20 hours of caring. \par

\begin{table}[!t]
  \centering
    \begin{tabular}{cccccc}
    \toprule
    & (1) & (2) & (3) & (4) & (5)\\
    \midrule
    & L-Intensity & ML-Intensity & MH-Intensity & H-Intensity & Control \\
    \midrule
    \multicolumn{5}{l}{\textit{Target variables}} \\
    \midrule
    Ind Income & 1089.71 & 859.85 & 580.40 & 296.48 & 1115.15 \\
    Household Income & 3784.43 & 3359.74 & 2976.63 & 2606.10 & 3733.65 \\
    Income Share(\%) & 25.97 & 23.06 & 16.75 & 8.86 & 27.11 \\
    \midrule
    \multicolumn{5}{l}{\textit{Background characteristics}} \\
    \midrule
    Employed & 0.56 & 0.51 & 0.40 & 0.23 & 0.49 \\
    Age & 52.92 & 53.42 & 53.98 & 56.74 & 47.16 \\
    Male & 0.47 & 0.40 & 0.37 & 0.36 & 0.49 \\
    Married & 0.68 & 0.65 & 0.62 & 0.64 & 0.59 \\
    Household Size & 2.66 & 2.66 & 2.74 & 2.78 & 2.79 \\
    \midrule
    Ethnicity &  &  &  &  \\
    \midrule
    Asian & 0.03 & 0.04 & 0.06 & 0.0 &  0.05 \\
    Black & 0.01 & 0.01 & 0.02 & 0.02 & 0.02 \\
    White & 0.95 & 0.93 & 0.90 & 0.93 & 0.91 \\
    Mixed & 0.01 & 0.01 & 0.02 & 0.02 &0.02\\
    Other & & & & \\
    \midrule
    Education &  &  &  &  \\
    \midrule
    Lower education & 0.32 & 0.26 & 0.23 & 0.17 & 0.33 \\
    Intermediate education & 0.39 & 0.44 & 0.51 & 0.54 & 0.39 \\
    Advanced education & 0.29 & 0.30 & 0.26 & 0.29 & 0.29 \\
    \midrule
    N & 57226 & 36215 & 9991 & 8901 & 286975 \\
    \bottomrule
    \end{tabular}
    \caption{\textbf{Descriptive statistics.} The table shows the main set of controls considered in our analysis. The sample includes all women and men with non-missing information on individual controls. Source: UKHLS data (years 2009-2020).}
  \label{descr}
\end{table}
\section{Results}\label{results}
In Section \ref{baseline} we first present our main baseline results for the caring income penalty. This is followed by an overview of results from existing methodological approaches in Section \ref{existing}, and a more nuanced analysis of differentials by intersectional characteristics in Section \ref{intersectional}. Finally, we conclude with a series of robustness tests in Section \ref{robustness}.

\subsection{Two-Stage Individual Synthetic Control: Baseline Results}\label{baseline}

\subsubsection{Individual Income}

Figure \ref{hh_and_ind_inc_baseline} (Panels a.-d.) shows the estimated difference between the average individual income of treated individuals (informal carers) and their synthetic control over time, spanning eight years before and six years after treatment.\footnote{A longer time period would have significantly reduced the number of valid cases for extreme time points.} Blue bars and markers represent pre-treatment trends, and red represent post-treatment. The shaded overlay represents confidence intervals computed using the method of \cite{vagni2021earnings}. Each panel provides insight into the average caring income penalty for varying levels of caring intensity.
Before the treatment year, the 95\% confidence interval for the difference between the treatment groups and their synthetic counterparts consistently includes zero, irrespective of the intensity of the treatment. This implies that leading up to the treatment year, there was no statistically significant difference in personal income between the synthetic control and treatment groups for all intensity levels. The onset of treatment has a significant impact on income. High-intensity carers experience a gradually widening negative income gap post-treatment. For example, two years post-treatment, high-intensity carers report a decline in personal income of £166 per month compared to their synthetic counterparts, which further increases to nearly £192 per month four years post-treatment. For low- and medium-low-intensity informal carers the difference with their respective counterfactuals is approximately £33 and £139 per month, respectively, during the same time frame. High-intensity informal carers face a more pronounced relative average income penalty compared to low-intensity carers with a difference of £162 compared to £44, respectively.\footnote{For more details on the difference between treatment and control group in terms of individual income see Supplementary Information Table \ref{tab:ind_ATT}.} This difference can be attributed to their substantially lower average pre-treatment individual income of £362 as opposed to £1057. The results demonstrate a clear caring income penalty, and substantial 'threshold effects' where higher intensity carers experience a greater income penalty. High-intensity carers experience a 45\% income reduction compared to a 4\% income reduction for low-intensity carers.\footnote{For more details on the caring penalty see Supplementary Information Table \ref{caregiving_penalty}, and Figure \ref{ii_and_hh_main}.} Beyond the fourth year post-treatment, the difference between treatment and synthetic control groups begins to taper off (Figure \ref{hh_and_ind_inc_baseline}a-d). Previous work suggested this may be due to skill acquisition that transfers to the labor market and improves longer-term employment prospects and `employment resilience',  whereby carers adapt to the challenges of combining work and care and engage in more flexible employment opportunities \citep{raiber2022wage}. No substantial differences are observed for medium-high-intensity carers. This lack of distinction could be attributed to several factors. This group is characterized by the largest variability in hours per week, ranging from 20 to 49 hours. The wide range of hours might contribute to a diverse set of individual circumstances and experiences within the group, making it challenging to identify a consistent pattern or a significant difference in personal income.

\subsubsection{Household Income and Income Share}

\begin{figure}[htbp]
    \centering    \includegraphics[width=1\linewidth]{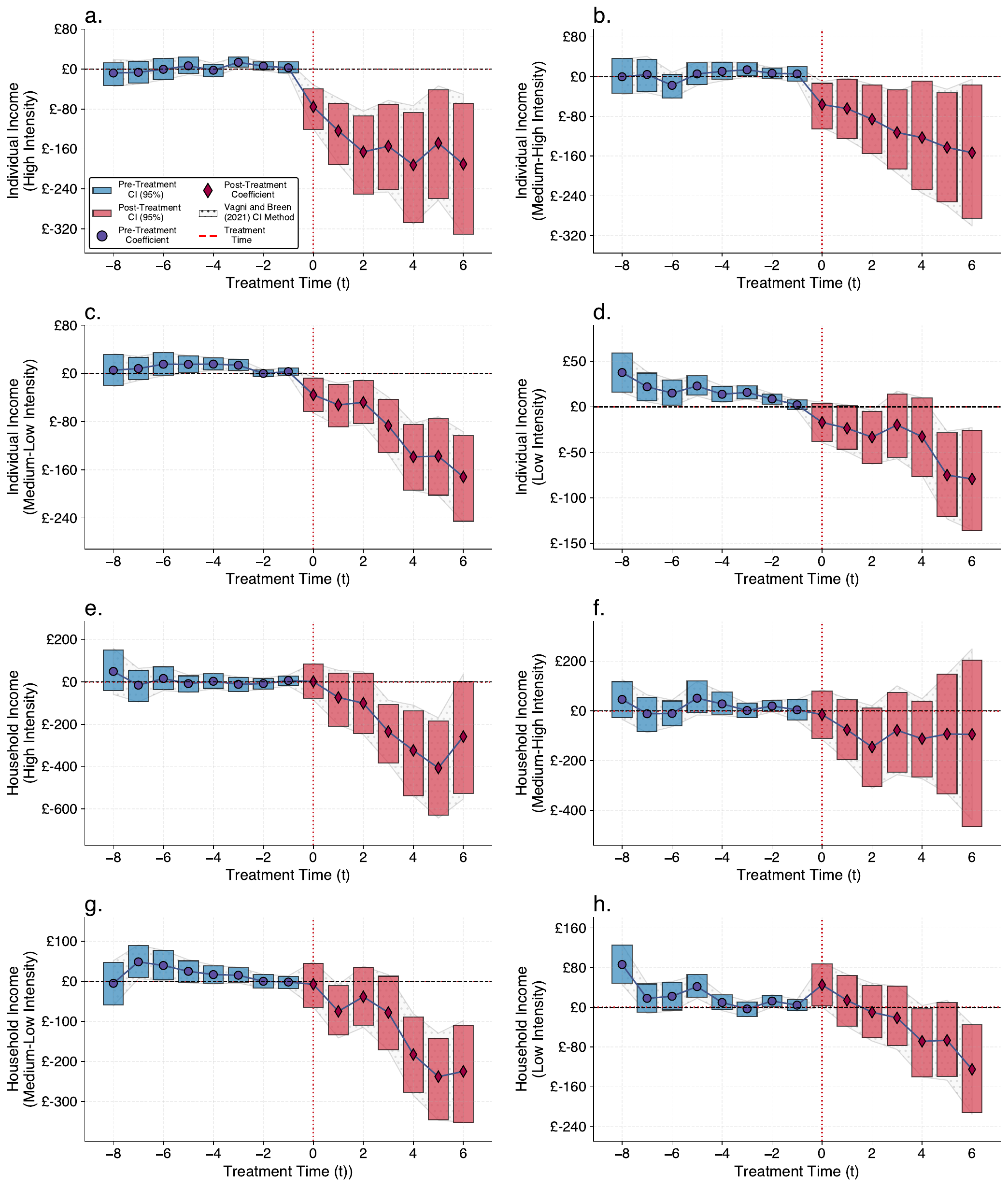}
    \caption{\small{\textbf{Inflation Adjusted Individual and Household Income.} 
    Average Treatment Effect on the Treated. The blue shaded areas and blue circles represent the pre-treatment confidence intervals at 95\% and the pre-treatment coefficients, respectively. The red shaded areas and red diamonds denote the post-treatment confidence intervals at 95\% and post-treatment coefficients, respectively. The shaded overlay represents confidence intervals computed using the method of \cite{vagni2021earnings}. For the full set of individual controls see Table \ref{descr}. Panels a. and e. represent the difference between high-intensity informal carers and their counterfactual; Panels b. and f. report medium-high-intensity informal carers; Panels c. and g. report medium-low-intensity informal carers; Panels d. and h. report low-intensity informal carers. Panels a.-d. represent individual income, while Panels e.-h. represent household income. Source: UKHLS data (2009-2020), authors' calculations.
    }}    \label{hh_and_ind_inc_baseline}
\end{figure}

\begin{figure}[!t]
    \centering    \includegraphics[width=1\linewidth]{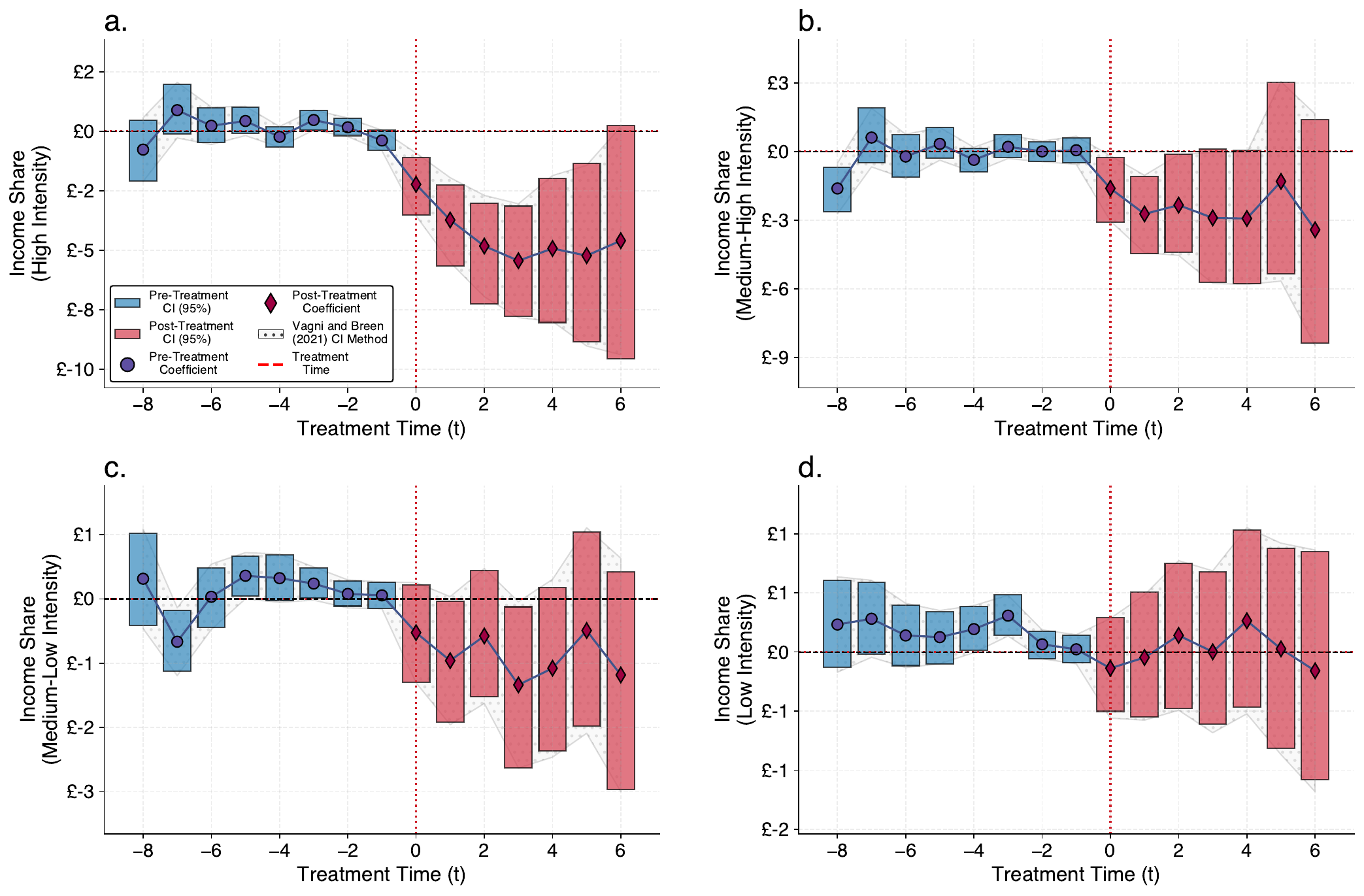}
    \caption{\small{\textbf{Income Share.}  Average Treatment Effect on the Treated. The blue shaded areas and blue circles represent the pre-treatment confidence intervals at 95\% and the pre-treatment coefficients, respectively. The red shaded areas and red diamonds denote the post-treatment confidence intervals at 95\% and post-treatment coefficients, respectively. The shaded overlay represents confidence intervals computed using the method of \cite{vagni2021earnings}. For the full set of individual controls see Table \ref{descr}.
    Panel a. reports the difference between high-intensity informal carers and their counterfactual; Panel b. reports medium-high-intensity informal carers; Panel c. reports medium-low-intensity informal carers; Panel d. reports low-intensity informal carers. Source: UKHLS data (2009-2020), authors' calculations.
     }}
    \label{inc_share_fig}
\end{figure}

The analysis of the caring income penalty for household income (Figure \ref{hh_and_ind_inc_baseline} Panels e.-h.) and income share (Figure \ref{inc_share_fig}) provides a comprehensive view of the broader economic impact of informal caring. Once more, a noteworthy contrast arises when considering high-intensity informal carers and their synthetic controls.
Building on the earlier discussed decline in personal income, high-intensity carers experience a substantial monthly reduction in their contribution to household income. This reduction stands at approximately 4.8\% in year 2 and 4.9\% in year 4, leading to a consequent decrease in overall household income by £100 and £324 in the second and fourth years, respectively.\footnote{For more details on the difference between treatment and control groups in terms of household income see Supplementary Information Table \ref{tab:hh_ATT}.} This translates into an overall household relative penalty of 12\%.\footnote{For more details on the household relative penalty see Supplementary Information Table \ref{caregiving_penalty} and Figure \ref{ii_and_hh_main}.} For carers providing support to a household member, particularly a spouse, the impact on household income is amplified as both the carer and the care recipient are often unable or partially unable to work, leading to a dual withdrawal from the labor market. This dual effect introduces potential confounding in assessing the income penalty, as the household may face a compounded economic strain. Over time, once a household member becomes a carer, compensating mechanisms often occur to adjust financial dynamics within the household. This could, for example, include redistributing financial responsibilities among family members, finding alternative sources of income, or adjusting spending patterns and financial priorities.

\subsection{Existing Methods}\label{existing}

In this section, we compare the findings from our novel ISC approach with the results from existing causal methodologies to highlight our unique contributions and the implications for inference and precision. 

\subsubsection{Matching}\label{existing_matching}

PSM was used to estimate the ATT of providing care at a certain intensity. 
We employed one-to-one nearest neighbour matching, pairing each treated individual with a control individual with the closest propensity score, following the method outlined by \cite{rosenbaum1983central}. To enhance match quality, we used the common support condition, which ensures better comparability between treated and control units \citep{becker2002estimation}. Additionally, we utilized the caliper matching method, setting a caliper width of 1\%, which limits the allowable difference in predicted probabilities between treated and control units for matching (see Eq. \ref{matching_problem}). The results are displayed in Tables \ref{tab:psm_ii}-\ref{tab:psm_hh2}. The PSM estimators reveal a clear negative impact of caring on both individual and household income, varying across levels of care intensity.  
Compared to the ISC estimators, the PSM estimator tends to show a larger magnitude of income loss. Unlike the ISC estimator, PSM does not indicate a significant trend in the influence of care provision over time and intensity (e.g., the income penalty does not consistently increase with time and care intensity). To evaluate the robustness of our PSM estimates, we conducted both a balance test and a Rosenbaum bounds sensitivity analysis to assess the quality of the matching process.\footnote{Results are available upon request.} Our balance tests reveal that -- despite employing nearest neighbour matching with a caliper width of 1\% -- there are significant differences in some covariates between the treated and control groups. This indicates that the matching process did not fully achieve balance, and some covariates remain imbalanced, potentially biasing the treatment effect estimates and violating the common support condition. The Rosenbaum bounds sensitivity analysis demonstrates that the estimated treatment effects are significantly influenced by unobserved factors.\footnote{For instance, with a Gamma value of 1.1 -- indicating a 10\% increase in the likelihood of receiving the treatment due to unobserved confounders -- the treatment effect loses its significance. This suggests that even a slight degree of hidden bias can substantially affect the estimated treatment effects.}

\subsubsection{Difference-in-Differences and Parallel Trends Violation}\label{existing_did}

\begin{figure}[hbtp]
    \centering
    \includegraphics[width=1\linewidth]{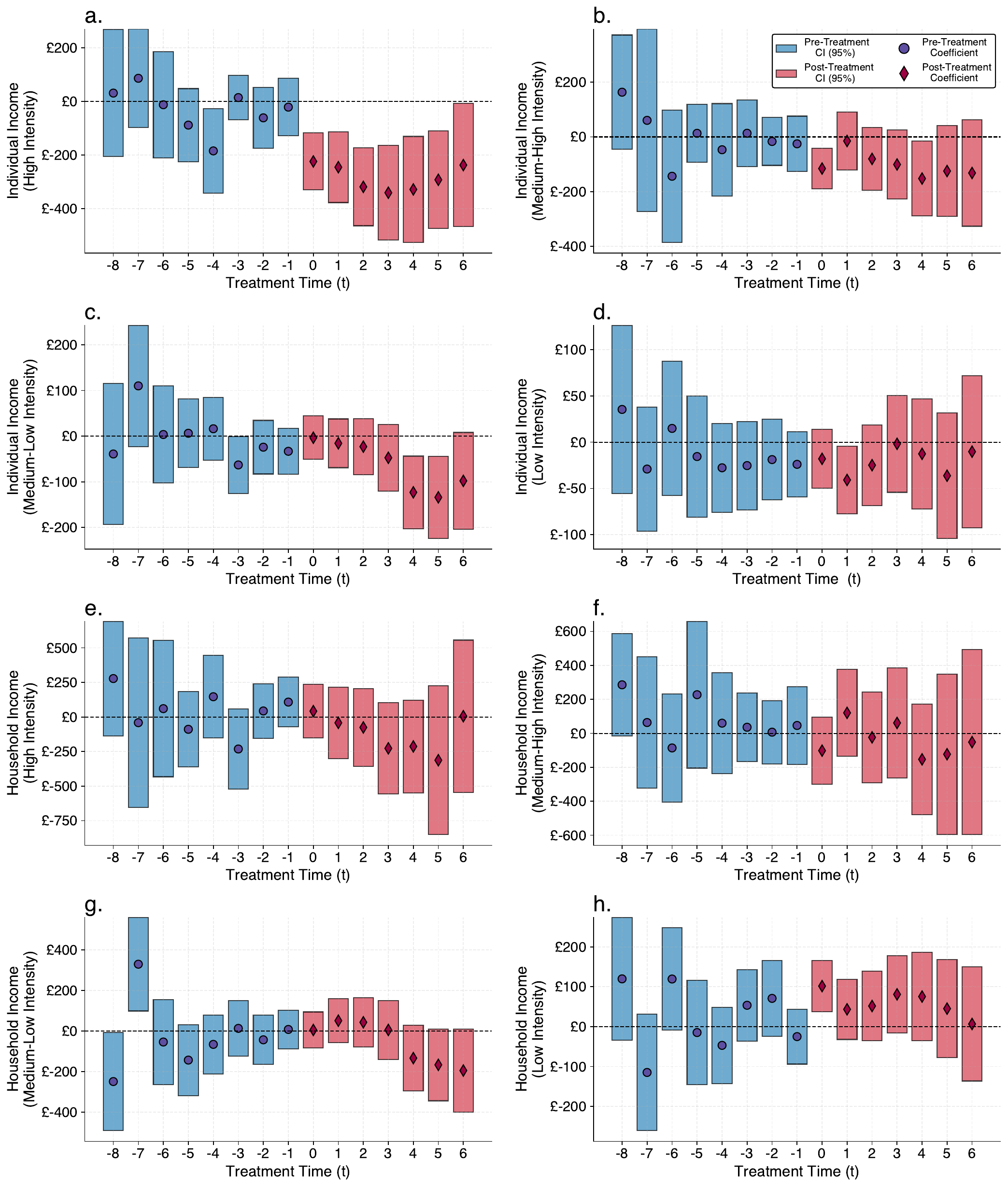}
    \caption{\small{\textbf{Doubly Roboust Difference-in-Differences.} Average treatment effect on the treated. The blue shaded areas and blue circles represent the pre-treatment confidence intervals at 95\% and the pre-treatment coefficients, respectively. The red shaded areas and red diamonds denote the post-treatment confidence intervals at 95\% and post-treatment coefficients, respectively. For the full set of individual controls see Table \ref{descr}.
    Panels a. and e. represent the difference between high-intensity informal carers and their counterfactual; Panels b. and f. report medium-high-intensity informal carers; Panel c. and g. report medium-low-intensity informal carers; Panels d. and h. report low-intensity informal carers. Panels a.-d. represent individual income, while Panels e.-h. represent household income. Source: UKHLS data (2009-2020), authors' calculations.
    }} \label{did_figure}
\end{figure}

We estimate a doubly robust DID estimator for the ATT based on inverse probability tilting and weighted least squares \citep{sant2020doubly}. The effectiveness of the DID framework hinges upon the validity of the common trend assumption, which posits that the individual or household income trajectories of informal carers and non-carers would have moved in tandem in the absence of the treatment.  Figure \ref{did_figure} shows the difference in individual and household income between treated individuals and those yet to receive treatment. In general, the income trajectories observed using the DID approach exhibit a similar trend to those derived from the ISC. However, there are clear violations of the common trend assumption at several points in the pre-treatment period (see Tables \ref{tab:diff-in-diff}-\ref{tab:diff-in-diff-household}). To demonstrate this, we estimate the \({\chi}^2\) statistic under the null hypothesis that all pre-treatment average effects on the treated are equal to zero (see Table \ref{table:paralleltrend}). The limitations of the DID approach are reflected in its RMSPE for the pre-treatment period. This suggests that the ISC delivers estimations with lower bias, as explained in Section \ref{synth_section}.

\subsubsection{Synthetic Difference-in-Differences}\label{existing_sdid}
\begin{figure}[hbtp]
    \centering
    \includegraphics[width=1\linewidth]{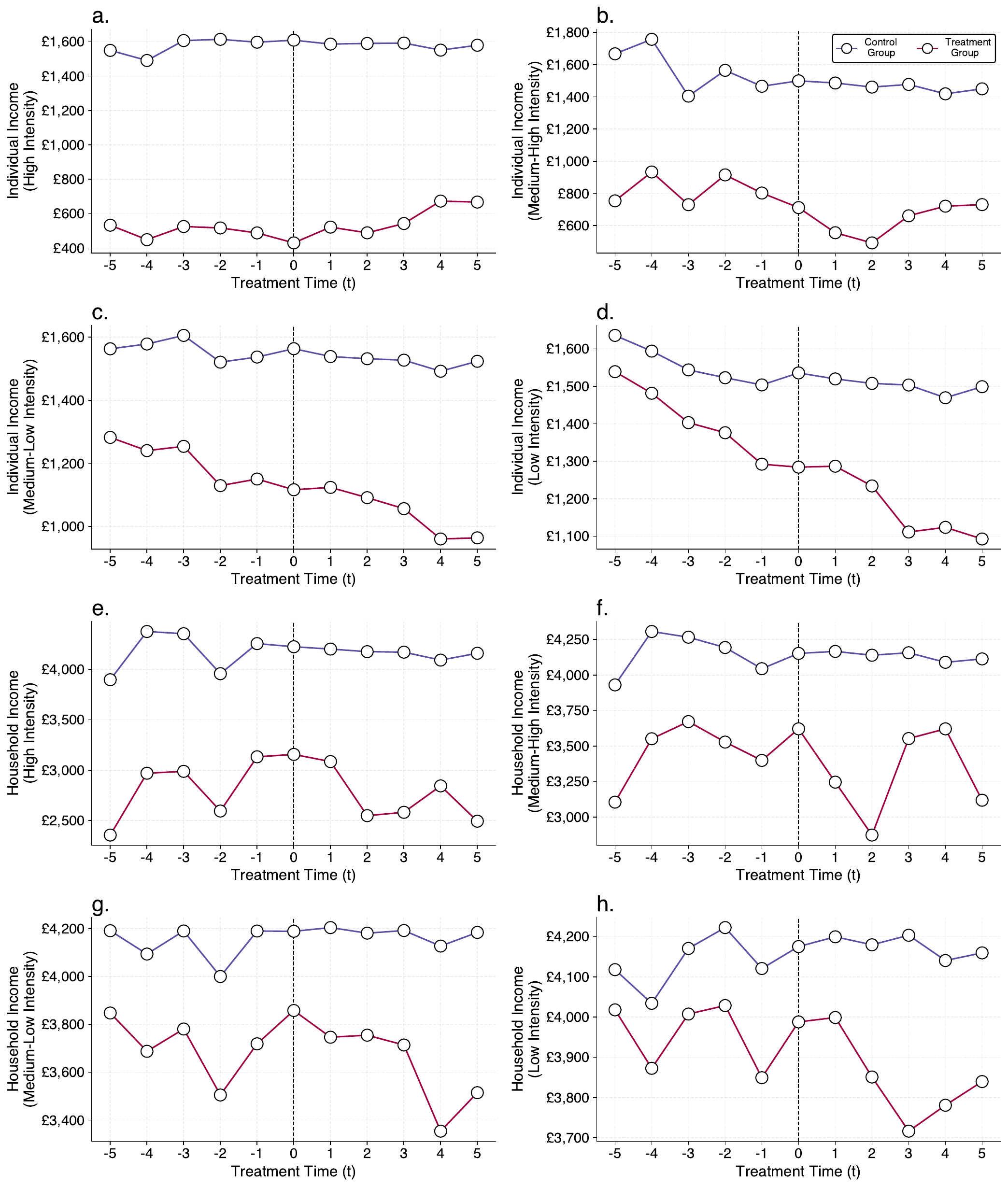}
    \caption{\small{\textbf{Synthetic Differences-in-Differences.} Average treatment effect on the treated. The blue line represents non-carers' income trajectories; the red line represents the income trajectory of unpaid carers. For the full set of individual controls see Table \ref{descr}.
    Panels a. and e. represent the difference between high-intensity informal carers and their counterfactual; Panels b. and f. report medium-high-intensity informal carers; Panels c. and g. report medium-low-intensity carers; Panels d. and h. report low-intensity informal carers. Panels a.-d. represent individual income, while Panels e.-h. represent household income. Source: UKHLS data (2009-2020), authors' calculations.
    }} \label{sdid_figure}
\end{figure}

The DID approach assumes that, without the treatment, outcomes of units in the treatment and control groups would have moved in tandem. However, if pre-event trends are not parallel, the DID estimate may be unreliable, as demonstrated in Section \ref{existing_did}. In contrast, SCM re-weights the control units so that their combined weighted outcomes closely match those of the treated units before the event, attributing any post-event differences to the event itself. The SDID further refines this estimate by adjusting the weights of the control units to ensure their time trends are parallel to those of the treated units before the event, and then applies a DID approach to the re-weighted data \citep{arkhangelsky2021synthetic}. Figure \ref{sdid_figure} shows that after weighting, the model achieves a parallel trend, effectively addressing issues related to differing pre-treatment trends between treated and control groups by constructing a synthetic control that mimics the pre-treatment characteristics of the informal carers, thereby reducing bias from pre-existing trends.
However, this methodology requires strongly balanced datasets, \footnote{To achieve this, we considered a subsample of individuals for which we had 10 years' worth of data, five years before and five years after the Treatment time (t).} which explains the differences in the magnitude of the results obtained when compared with those achieved by implementing the ISC results. Additionally, the SDID method is computationally intensive, particularly with large datasets, a staggered treatment, or complex donor pools, as is the case in potential applications to micro-level longitudinal data.

\subsection{Two-Stage Synthetic Control: Intersectional Differences}\label{intersectional}
We next explore intersectional inequalities and variations in the caring income penalty with a particular focus on sex (Section \ref{sec:gender}), ethnicity (Section \ref{sec:ethnicity}), and age (Section \ref{sec:age}). Finally, Table \ref{ATT_year3} provides an overview of the ATT for all different specifications considered in the following sections. 

\subsubsection{Sex Differences}\label{sec:gender}
Our analysis focuses exclusively on two levels of caring intensity due to sample size constraints. We designate carers who spend more than 20 hours per week on caring duties as `high-intensity carers', and those who contribute less than 20 hours per week are categorised as `low-intensity carers'. Figure \ref{ii_gender} displays the main results of our analysis. Men (Panels b. and d.) generally have higher pre-treatment individual incomes compared to women (Panels a. and c.) across both high- and low-intensity caring roles. Both men and women experience income loss after assuming caring responsibilities. However, the relative individual caring penalties -- calculated as the percentage decrease in individual income post-treatment -- reveal significant disparities between men and women and intensity levels. Women face a higher individual income penalty for high-intensity caring compared to men (30\% versus 25\%). Conversely, in low-intensity caring roles, men experience a slightly higher penalty compared to women (6\% versus 5\%).\footnote{For additional insights on the average treatment effect for individual and household income and on the relative caring penalty by sex, refer to Figures \ref{ci_ii_sex} and Tables \ref{tab:caregiving_penalty_gender}-\ref{tab:caregiving_penalty_gender_hh}.} \par

\subsubsection{Ethnic Group Differences}\label{sec:ethnicity}
Due to limitations in sample size, our analysis focuses on comparing `White' versus `non-White' ethnic groups, with the latter encompassing Asian, Black, Mixed, and other ethnic backgrounds -- acknowledging that this aggregated grouping obscures heterogeneities  between the constituent social groups \citep{alcoff2003}. Once again, we categorise caring intensity into high- and low-intensity levels. We find that both sets of ethnic groups experience income losses, but to varying degrees (Figure \ref{ii_eth}). The `White' ethnic category tends to face higher penalties, particularly in high-intensity caring roles; the relative individual caring gap for high-intensity carers stands at 32\% for `Whites' and 20\% for `non-Whites'. Among low-intensity carers, it is 5\% for `Whites' and only 4\% for `non-Whites'.\footnote{For additional insights on the ATT for individual and household income by ethnicity, see Figures \ref{ci_ii_eth} and Tables \ref{tab:caregiving_penalty_ethnicity}-\ref{tab:caregiving_penalty_ethnicity_hh}.}

\subsubsection{Age}\label{sec:age}
We distinguish three age groups: below 25 years of age, 25 to 65 years of age, and ages 65 and above as well as again between low-intensity and high-intensity caring roles(Figure \ref{ci_ii_age}). While low-intensity caring responsibilities appear to have a negligible impact on individual and household income, the situation changes substantially for high-intensity carers. Young carers face a significant caring penalty; after just two years of becoming a carer, they experience a reduction of £502 per month in their individual income compared to their counterfactual, registering an 181\% relative caring penalty.\footnote{For additional information on the relative individual caring penalty by age groups, please see Table \ref{tab:caregiving_penalty_age} and Figure \ref{age_ind} Panels a.-f.} The individual income penalty translates into a reduction in household income of £484 in the third year.\footnote{For additional information on the relative household caring penalty see Table \ref{tab:caregiving_penalty_age_hh} and Figure \ref{age_ind} Panels g.-l.)} 
We also observe a decrease in individual income for high-intensity carers aged 25-64. By the fifth year of caring they experience a reduction of nearly £170 per month in their individual income; an average relative caring penalty of 17\%, with a corresponding decrease of £297 in household income. This decrease -- although less severe than that experienced by younger carers -- is still significant and highlights the broader economic impact of high-intensity carers across different age groups. 
In contrast, we observe no significant caring penalty for individuals aged 65 and older. This outcome is expected, as the primary source of income for this age group is less likely to be from employment and more likely to come from pensions or retirement savings.

\subsection{Robustness Checks}\label{robustness}

We perform a series of robustness checks to ensure the reliability of our findings and test the sensitivity of our results to various assumptions. These checks include data contiguity (Section \ref{sec:contiguity}), placebo tests (Section \ref{sec:placebo}), employment status subsample analysis (Section \ref{sec:empstat}), and the examination of caring duration (Section \ref{sec:lengthepisode}).

\begin{table}[!t]
\begin{small}
\begin{center}
\begin{tabular}{cccccccc}\toprule
Dependent        & Care Intensity       & Sex & Ethnicity & Age & $ATE_{t+3}$  & Lower CI & Upper CI \\ \midrule
HH Income & High                   & All    & All  & All     & -£235 & -£375     & -£96      \\
HH Income & Medium High          & All    & All   & All      & -£78 & -£256     & £88      \\
HH Income & Medium Low           & All    & All   & All      & -£78 & -£172     & £12      \\
HH Income & Low                  & All    & All  & All       & -£21 & £84     & £38      \\
HH Income & High and Medium High & Male   & All   & All      & -£306 & -£536     & -£79      \\
HH Income & Low and Medium Low   & Male   & All  & All       & -£61 & -£144     & £22      \\
HH Income & High and Medium High & Female & All  & All       & -£95 & -£220     & £24      \\
HH Income & Low and Medium Low   & Female & All   & All      & -£41 & -£99     & £26     \\ 
HH Income & High and Medium High  &  All & White  & All     & -£262 & -£412     & -£120    \\
HH Income & Low and Medium Low  & All & White  & All   & -£37 &   -£93   &   £18  \\
HH Income &  High and Medium High  & All & non-White  & All      & -£53 & -£239     & £134    \\ 
HH Income &  Low and Medium Low  & All & non-White  & All     & -£62 & -£171     & £45    \\ 
HH Income & High and Medium High  &  All & All  & Below 25     & -£484 & -£1295     & £101    \\
HH Income & Low and Medium Low  & All & All & Below 25    & -£250 &   -£720   &   £165  \\
HH Income &  High and Medium High  & All & All  & 25-65      & -£107 & -£249     & £29    \\ 
HH Income &  Low and Medium Low  & All & All &  25-65     & -£89 & -£155     & -£24    \\ 
HH Income &  High and Medium High  & All & All  & 65 up      & -£145 & -£241     & -£43    \\ 
HH Income &  Low and Medium Low  & All & All &  65 up    & £12 & -£44     & £68    \\ 
Ind. Income & High                   & All    & All  & All     & -£154 & -£251     & -£62      \\
Ind. Income & Medium High          & All    & All   & All      & -£112 & -£186     & -£30      \\
Ind. Income & Medium Low           & All    & All   & All      & -£87 & -£128     & -£39      \\
Ind. Income & Low                  & All    & All  & All       & -£20 & -£57     & £14      \\
Ind. Income & High and Medium High & Male   & All   & All      & -£146 & -£284     & -£8      \\
Ind. Income & Low and Medium Low   & Male   & All  & All       & -£48 & -£103     & £12      \\
Ind. Income & High and Medium High & Female & All  & All       & -£105 & -£160     & -£54      \\
Ind. Income & Low and Medium Low   & Female & All   & All      & -£31 & -£60     & -£3     \\ 
Ind. Income & High and Medium High  &  All & White  & All     & -£132 & -£199     & -£71    \\
Ind. Income & Low and Medium Low  & All & White  & All   & -£31 &   -£62   &   £3  \\
Ind. Income &  High and Medium High  & All & non-White  & All      & -£77 & -£208     & £14    \\ 
Ind. Income &  Low and Medium Low  & All & non-White  & All     & -£38 & -£88     & £6    \\ 
Ind. Income & High and Medium High  &  All & All  & Below 25     & -£355 & -£813     & £51    \\
Ind. Income & Low and Medium Low  & All & All & Below 25    & -£54 &  -£205  &   £101  \\
Ind. Income &  High and Medium High  & All & All  & 25-65      & -£171 & -£254     & -£80    \\ 
Ind. Income &  Low and Medium Low  & All & All &  25-65     & -£77 & -£115     & -£37    \\ 
Ind. Income &  High and Medium High  & All & All  & 65 up      & -£8 & -£30     & £24    \\ 
Ind. Income &  Low and Medium Low  & All & All &  65 up     & -£7 & -£20     & £7    \\ 
\bottomrule
\end{tabular}
\caption{\textbf{Aggregated Results for Inflation Adjusted Individual and Household Income.} This table shows the Average Treatment effect on treated at time t+3 for all the different specifications considered in the analysis. Source: UKHLS data (years 2009-2020), authors' calculations.}
\label{ATT_year3}
\end{center}
\end{small}
\end{table}

\subsubsection{Data Contiguity}\label{sec:contiguity}

Our analysis thus far has included individuals with a minimum of three pre-treatment data points (as discussed in Section \ref{data}). We set this threshold based on previous studies which suggest a minimum number of time points pre-intervention to correctly estimate the ISC \citep{vagni2021earnings, abadie2021using}. However, in addition, we conduct sensitivity analyses incorporating various pre-treatment observation period lengths. We examine scenarios where treatment data spanned at least three consecutive waves ($T=\{-3 -2 -1\}$) in Figure \ref{hh_and_ii_rc_3}, and five consecutive waves ($T=\{-5 -4 -3 -2 -1\}$) in Figure \ref{ii_rc_5}. We observe no significant deviations in the magnitude of the ATT estimated in any of these scenarios. However, carrying out this analysis with longer pre-treatment periods significantly reduces the sample size and, consequently, the statistical power of the estimation.

\subsubsection{Placebo Tests}\label{sec:placebo}

We conduct placebo tests to evaluate the robustness of the ISC estimations by simulating fake treatments for individuals in the donor pool \citep{abadie2010synthetic}. Specifically, in our baseline estimations, there are \textit{n} units in the donor pool for each treated individual. We consider these \textit{n} control units as if they received the intervention at the same time and with the same intensity as the treated unit they act as a counterfactual for, including the actual treated unit within the donor pool. This results in \textit{n} placebo estimations for each treated individual. We then average the placebo estimations for each treatment unit. Finally, we aggregate these averages across all treatments to derive the final placebo test results.
Figure \ref{placebo_ind} shows that the placebo treatment has no effect and the ATT remains small in magnitude and not statistically significant in all the specifications considered. 

\subsubsection{Employment Status}\label{sec:empstat}

In our main specification, we consider both unemployed and employed individuals to ensure a comprehensive understanding of financial dynamics and to accurately capture income inequality. Focusing exclusively on employed individuals to make inferences about the entire population would lead to inconsistent estimations. This bias arises because any variable influencing the `income-earner' status could potentially correlate with the error term, skewing the results. By including the unemployed -- who often have systematically different characteristics -- we avoid the selection bias that would result from excluding this sub-group. We conduct separate analyses on the two sub-samples -- employed and unemployed -- enabling us to identify specific factors and trends within each sub-sample, providing more nuanced and detailed insight into income-related dynamics (see Figure \ref{employed_ind}).
As expected, our analysis reveals that while there is no significant difference for unemployed individuals, employed carers experience notable financial impacts. This is particularly pronounced for high-intensity carers who devote more time and energy to caring responsibilities, thereby further compromising their employment situation. For high-intensity employed carers, there is a reduction in individual income of £154 per month by the fourth year of caring compared to their synthetic counterfactuals (Figure \ref{employed_ind}a).
In contrast, low-intensity caring while still impactful, may require fewer work schedule adjustments and may allow carers to better manage their dual roles. However, even this level of caring results in a measurable decrease in income, with employed carers facing a reduction of £99 per month by the fourth year (Figure \ref{employed_ind}d). 
This reduction in individual income translates to a more substantial impact on household income. For high-intensity employed carers, household income decreases by £425 per month by the fourth year (Figure \ref{employed_ind}e), while for low-intensity employed carers, the household income reduction is £154 per month (Figure \ref{employed_ind}h). The lack of impact on unemployed carers is expected, as our analysis focuses on income derived from employment. 

\subsubsection{Length of care episode}\label{sec:lengthepisode}

In our main specification, we consider individuals as treated if they report any episode of caring without considering the length of the caring episode (measured in consecutive years of caring). In this section, we explore two additional specifications by computing the ATT for individuals who provide care for three consecutive years (Figure \ref{ci_hh_and_ind_3consecutive}) and for those who provide care for five consecutive years (Figure \ref{ci_ind_5consecutive}). We then compare the results from these two specifications with our baseline results.
While our baseline models report a decrease in individual income of £124 two years post-treatment for those undertaking high-intensity care responsibilities, individuals providing care for three consecutive years report a £224 loss in income compared to their counterfactual (Figure \ref{ci_hh_and_ind_3consecutive}a). The gap goes up to £372 for those individuals who provide care for five consecutive years (Figure \ref{ci_ind_5consecutive}a). For low-intensity carers, the income penalty is £122 and £179 for individuals who provide care for three and five consecutive years (Figure \ref{ci_hh_and_ind_3consecutive}d and \ref{ci_ind_5consecutive}d), respectively (compared to £26 reported in our baseline model). 
Even if the ATT in terms of household income is not statistically significant at 95\% confidence interval, the patterns suggest that carers who provide care for five consecutive years report a lower average income penalty compared to those providing care for three consecutive years and our baseline model. 
\section{Conclusion} \label{conclusion} 

Our study provides the first robust estimates of the causal impact of informal caring on income through innovative methodological advancements in causal inference; a novel two-stage approach to individual synthetic control. Our findings reveal a negative and statistically significant income gap between informal carers and their synthetic counterparts, which is particularly pronounced among high-intensity carers. We also provide the first robust estimates of how the dynamics of the carer penalty evolve over time. We find that income disparities persist for several years following the onset of caring, indicating enduring economic challenges faced by carers. Moreover, our analysis sheds light on the broader economic consequences of caring, including its effect on household income and income share. There is some evidence of income share recovering, but the effect is modest and not statistically significant. Additionally, the analysis explores differentials in the ATT by intersectional characteristics. We show that the financial impact of caring is significantly higher for women compared to men, and for White carers relative to those from non-white backgrounds. Young carers face the most substantial income reduction, with the penalty reaching as much as £502 per month when compared to their counterfactual.

The substantial decline in income as a result of high-intensity informal care observed in our study underscores the pressing need for policy interventions aimed at alleviating the financial burdens faced by carers. Whilst the decision to become an unpaid carer is partly driven by a sense of duty, personal responsibility and compassion, the economic disincentives to providing unpaid care implied by our causal estimates are not trivial. The challenges faced by informal carers are also being compounded by demographic shifts that place further pressures on a social care system already experiencing rising unmet needs, extensive reliance on self-funded services, substandard care quality, financially strained care providers, and rising pressures on both carers and care sector organisations, and in urgent need or reform \citep{glasby2021lost}. As the UK population ages, it faces an under-supply of labour due to ill health, retirement, and people leaving the labour market to informally care for relatives and friends with long-term illness, or disability. Policies that help unpaid carers remain in the labour market could therefore have potentially far reaching economic benefits that are likely to become increasingly important as these shifts continue to unfold. The implementation of flexible work arrangements (e.g. working from home and paid care leave), robust support systems (e.g. respite care and formal services), and targeted financial assistance (e.g improving the eligibility criteria and benefits as part of carers allowance) could mitigate the adverse economic consequences of caring, enabling carers to remain in the labour market.

\newpage
\bibliography{refs}

\begin{thebibliography}{}

\bibitem[\protect\citeauthoryear{Abadie}{Abadie}{2021}]{abadie2021using}
Abadie, A. (2021).
\newblock Using synthetic controls: Feasibility, data requirements, and methodological aspects.
\newblock {\em Journal of Economic Literature\/}~{\em 59\/}(2), 391--425.

\bibitem[\protect\citeauthoryear{Abadie, Diamond, and Hainmueller}{Abadie et~al.}{2010}]{abadie2010synthetic}
Abadie, A., A.~Diamond, and J.~Hainmueller (2010).
\newblock Synthetic control methods for comparative case studies: Estimating the effect of california’s tobacco control program.
\newblock {\em Journal of the American statistical Association\/}~{\em 105\/}(490), 493--505.

\bibitem[\protect\citeauthoryear{Abadie and Gardeazabal}{Abadie and Gardeazabal}{2003}]{abadie2003economic}
Abadie, A. and J.~Gardeazabal (2003).
\newblock The economic costs of conflict: A case study of the basque country.
\newblock {\em American economic review\/}~{\em 93\/}(1), 113--132.

\bibitem[\protect\citeauthoryear{Abadie and L’hour}{Abadie and L’hour}{2021}]{abadie2021penalized}
Abadie, A. and J.~L’hour (2021).
\newblock A penalized synthetic control estimator for disaggregated data.
\newblock {\em Journal of the American Statistical Association\/}~{\em 116\/}(536), 1817--1834.

\bibitem[\protect\citeauthoryear{Alcoff}{Alcoff}{2003}]{alcoff2003}
Alcoff, L.~M. (2003).
\newblock Latino/as, asian americans, and the black-white binary.
\newblock {\em The Journal of Ethics\/}~{\em 7\/}(1), 5--27.

\bibitem[\protect\citeauthoryear{Amjad, Shah, and Shen}{Amjad et~al.}{2018}]{amjad2018robust}
Amjad, M., D.~Shah, and D.~Shen (2018).
\newblock Robust synthetic control.
\newblock {\em Journal of Machine Learning Research\/}~{\em 19\/}(22), 1--51.

\bibitem[\protect\citeauthoryear{Aranda and Knight}{Aranda and Knight}{1997}]{aranda1997influence}
Aranda, M.~P. and B.~G. Knight (1997).
\newblock The influence of ethnicity and culture on the caregiver stress and coping process: A sociocultural review and analysis.
\newblock {\em The Gerontologist\/}~{\em 37\/}(3), 342--354.

\bibitem[\protect\citeauthoryear{Arkhangelsky, Athey, Hirshberg, Imbens, and Wager}{Arkhangelsky et~al.}{2021}]{arkhangelsky2021synthetic}
Arkhangelsky, D., S.~Athey, D.~A. Hirshberg, G.~W. Imbens, and S.~Wager (2021).
\newblock Synthetic difference-in-differences.
\newblock {\em American Economic Review\/}~{\em 111\/}(12), 4088--4118.

\bibitem[\protect\citeauthoryear{Ashenfelter and Card}{Ashenfelter and Card}{1984}]{ashenfelter1984using}
Ashenfelter, O.~C. and D.~Card (1984).
\newblock Using the longitudinal structure of earnings to estimate the effect of training programs.

\bibitem[\protect\citeauthoryear{Becker and Becker}{Becker and Becker}{2008}]{becker2008young}
Becker, F. and S.~Becker (2008).
\newblock Young adult carers in the uk.
\newblock {\em Experiences, needs and services for carers aged\/}, 16--24.

\bibitem[\protect\citeauthoryear{Becker and Kl{\"o}{\ss}ner}{Becker and Kl{\"o}{\ss}ner}{2018}]{becker2018fast}
Becker, M. and S.~Kl{\"o}{\ss}ner (2018).
\newblock Fast and reliable computation of generalized synthetic controls.
\newblock {\em Econometrics and Statistics\/}~{\em 5}, 1--19.

\bibitem[\protect\citeauthoryear{Becker and Ichino}{Becker and Ichino}{2002}]{becker2002estimation}
Becker, S.~O. and A.~Ichino (2002).
\newblock Estimation of average treatment effects based on propensity scores.
\newblock {\em The stata journal\/}~{\em 2\/}(4), 358--377.

\bibitem[\protect\citeauthoryear{Bolin, Lindgren, and Lundborg}{Bolin et~al.}{2008}]{bolin2008your}
Bolin, K., B.~Lindgren, and P.~Lundborg (2008).
\newblock Your next of kin or your own career?: Caring and working among the 50+ of europe.
\newblock {\em Journal of health economics\/}~{\em 27\/}(3), 718--738.

\bibitem[\protect\citeauthoryear{Brimblecombe and Cartagena~Farias}{Brimblecombe and Cartagena~Farias}{2022}]{brimblecombe2022inequalities}
Brimblecombe, N. and J.~Cartagena~Farias (2022).
\newblock Inequalities in unpaid carer's health, employment status and social isolation.
\newblock {\em Health \& Social Care in the Community\/}~{\em 30\/}(6), e6564--e6576.

\bibitem[\protect\citeauthoryear{Brimblecombe, Knapp, King, Stevens, and Cartagena~Farias}{Brimblecombe et~al.}{2020}]{brimblecombe2020high}
Brimblecombe, N., M.~Knapp, D.~King, M.~Stevens, and J.~Cartagena~Farias (2020).
\newblock The high cost of unpaid care by young people: health and economic impacts of providing unpaid care.
\newblock {\em BMC Public Health\/}~{\em 20}, 1--11.

\bibitem[\protect\citeauthoryear{Card}{Card}{1990}]{card1990impact}
Card, D. (1990).
\newblock The impact of the mariel boatlift on the miami labor market.
\newblock {\em Ilr Review\/}~{\em 43\/}(2), 245--257.

\bibitem[\protect\citeauthoryear{Carmichael and Charles}{Carmichael and Charles}{2003}]{carmichael2003opportunity}
Carmichael, F. and S.~Charles (2003).
\newblock The opportunity costs of informal care: does gender matter?
\newblock {\em Journal of health economics\/}~{\em 22\/}(5), 781--803.

\bibitem[\protect\citeauthoryear{Carr, Murray, Zaninotto, Cadar, Head, Stansfeld, and Stafford}{Carr et~al.}{2018}]{carr2018association}
Carr, E., E.~T. Murray, P.~Zaninotto, D.~Cadar, J.~Head, S.~Stansfeld, and M.~Stafford (2018).
\newblock The association between informal caregiving and exit from employment among older workers: prospective findings from the uk household longitudinal study.
\newblock {\em The Journals of Gerontology: Series B\/}~{\em 73\/}(7), 1253--1262.

\bibitem[\protect\citeauthoryear{Clancy, Fisher, Daigle, Henle, McCarthy, and Fruhauf}{Clancy et~al.}{2020}]{clancy2020eldercare}
Clancy, R.~L., G.~G. Fisher, K.~L. Daigle, C.~A. Henle, J.~McCarthy, and C.~A. Fruhauf (2020).
\newblock Eldercare and work among informal caregivers: A multidzhangiplinary review and recommendations for future research.
\newblock {\em Journal of Business and Psychology\/}~{\em 35}, 9--27.

\bibitem[\protect\citeauthoryear{Cohen, Sabik, Cook, Azzoli, and Mendez-Luck}{Cohen et~al.}{2019}]{cohen2019differences}
Cohen, S.~A., N.~J. Sabik, S.~K. Cook, A.~B. Azzoli, and C.~A. Mendez-Luck (2019).
\newblock Differences within differences: Gender inequalities in caregiving intensity vary by race and ethnicity in informal caregivers.
\newblock {\em Journal of Cross-Cultural Gerontology\/}~{\em 34}, 245--263.

\bibitem[\protect\citeauthoryear{Dilworth-Anderson, Williams, and Gibson}{Dilworth-Anderson et~al.}{2002}]{dilworth2002issues}
Dilworth-Anderson, P., I.~C. Williams, and B.~E. Gibson (2002).
\newblock Issues of race, ethnicity, and culture in caregiving research: A 20-year review (1980--2000).
\newblock {\em The Gerontologist\/}~{\em 42\/}(2), 237--272.

\bibitem[\protect\citeauthoryear{Dunham and Dietz}{Dunham and Dietz}{2003}]{dunham2003if}
Dunham, C.~C. and B.~E. Dietz (2003).
\newblock "{If} i'm not allowed to put my family first”: Challenges experienced by women who are caregiving for family members with dementia.
\newblock {\em Journal of women \& aging\/}~{\em 15\/}(1), 55--69.

\bibitem[\protect\citeauthoryear{D’Amen, Socci, and Santini}{D’Amen et~al.}{2021}]{d2021intergenerational}
D’Amen, B., M.~Socci, and S.~Santini (2021).
\newblock Intergenerational caring: A systematic literature review on young and young adult caregivers of older people.
\newblock {\em BMC geriatrics\/}~{\em 21\/}(1), 105.

\bibitem[\protect\citeauthoryear{Earle and Heymann}{Earle and Heymann}{2012}]{earle2012cost}
Earle, A. and J.~Heymann (2012).
\newblock The cost of caregiving: Wage loss among caregivers of elderly and disabled adults and children with special needs.
\newblock {\em Community, Work \& Family\/}~{\em 15\/}(3), 357--375.

\bibitem[\protect\citeauthoryear{Ettner}{Ettner}{1996}]{ettner1996opportunity}
Ettner, S.~L. (1996).
\newblock The opportunity costs of elder care.
\newblock {\em Journal of Human Resources\/}, 189--205.

\bibitem[\protect\citeauthoryear{Glasby, Zhang, Bennett, and Hall}{Glasby et~al.}{2021}]{glasby2021lost}
Glasby, J., Y.~Zhang, M.~R. Bennett, and P.~Hall (2021).
\newblock A lost decade? a renewed case for adult social care reform in england.
\newblock {\em Journal of Social Policy\/}~{\em 50\/}(2), 406--437.

\bibitem[\protect\citeauthoryear{Glauber}{Glauber}{2017}]{glauber2017gender}
Glauber, R. (2017).
\newblock Gender differences in spousal care across the later life course.
\newblock {\em Research on aging\/}~{\em 39\/}(8), 934--959.

\bibitem[\protect\citeauthoryear{Heitmueller and Inglis}{Heitmueller and Inglis}{2007}]{heitmueller2007earnings}
Heitmueller, A. and K.~Inglis (2007).
\newblock The earnings of informal carers: Wage differentials and opportunity costs.
\newblock {\em Journal of health economics\/}~{\em 26\/}(4), 821--841.

\bibitem[\protect\citeauthoryear{Hollingsworth and Wing}{Hollingsworth and Wing}{2020}]{hollingsworth2020tactics}
Hollingsworth, A. and C.~Wing (2020).
\newblock Tactics for design and inference in synthetic control studies: An applied example using high-dimensional data.
\newblock {\em Available at SSRN 3592088\/}.

\bibitem[\protect\citeauthoryear{Humphries}{Humphries}{2022}]{humphries2022ending}
Humphries, R. (2022).
\newblock {\em Ending the social care crisis: A new road to reform}.
\newblock Policy Press.

\bibitem[\protect\citeauthoryear{Johnson and Sasso}{Johnson and Sasso}{2000}]{johnson2000trade}
Johnson, R.~W. and A.~T.~L. Sasso (2000).
\newblock The trade-off between hours of paid employment and time assistance to elderly parents at midlife.

\bibitem[\protect\citeauthoryear{Keating, McGregor, and Yeandle}{Keating et~al.}{2021}]{keating2021sustainable}
Keating, N., J.~A. McGregor, and S.~Yeandle (2021).
\newblock Sustainable care: theorising the wellbeing of caregivers to older persons.
\newblock {\em International Journal of Care and Caring\/}~{\em 5\/}(4), 611--630.

\bibitem[\protect\citeauthoryear{Keating, Fast, Lero, Lucas, and Eales}{Keating et~al.}{2014}]{keating2014taxonomy}
Keating, N.~C., J.~E. Fast, D.~S. Lero, S.~J. Lucas, and J.~Eales (2014).
\newblock A taxonomy of the economic costs of family care to adults.
\newblock {\em The Journal of the Economics of Ageing\/}~{\em 3}, 11--20.

\bibitem[\protect\citeauthoryear{Kellogg, Mogstad, Pouliot, and Torgovitsky}{Kellogg et~al.}{2021}]{kellogg2021combining}
Kellogg, M., M.~Mogstad, G.~A. Pouliot, and A.~Torgovitsky (2021).
\newblock Combining matching and synthetic control to tradeoff biases from extrapolation and interpolation.
\newblock {\em Journal of the American statistical association\/}~{\em 116\/}(536), 1804--1816.

\bibitem[\protect\citeauthoryear{King~McLaughlin, Greenfield, Hasche, and De~Fries}{King~McLaughlin et~al.}{2019}]{king2019young}
King~McLaughlin, J., J.~C. Greenfield, L.~Hasche, and C.~De~Fries (2019).
\newblock Young adult caregiver strain and benefits.
\newblock {\em Social Work Research\/}~{\em 43\/}(4), 269--278.

\bibitem[\protect\citeauthoryear{Lilly, Laporte, and Coyte}{Lilly et~al.}{2007}]{lilly2007labor}
Lilly, M.~B., A.~Laporte, and P.~C. Coyte (2007).
\newblock Labor market work and home care's unpaid caregivers: a systematic review of labor force participation rates, predictors of labor market withdrawal, and hours of work.
\newblock {\em The Milbank Quarterly\/}~{\em 85\/}(4), 641--690.

\bibitem[\protect\citeauthoryear{Malo, Eskelinen, Zhou, and Kuosmanen}{Malo et~al.}{2023}]{malo2023computing}
Malo, P., J.~Eskelinen, X.~Zhou, and T.~Kuosmanen (2023).
\newblock Computing synthetic controls using bilevel optimization.
\newblock {\em Computational economics\/}, 1--24.

\bibitem[\protect\citeauthoryear{Martsolf, Kandrack, Rodakowski, Friedman, Beach, Folb, and James~III}{Martsolf et~al.}{2020}]{martsolf2020work}
Martsolf, G.~R., R.~Kandrack, J.~Rodakowski, E.~M. Friedman, S.~Beach, B.~Folb, and A.~E. James~III (2020).
\newblock Work performance among informal caregivers: a review of the literature.
\newblock {\em Journal of aging and health\/}~{\em 32\/}(9), 1017--1028.

\bibitem[\protect\citeauthoryear{Pearson}{Pearson}{1905}]{pearson1905problem}
Pearson, K. (1905).
\newblock The problem of the random walk.
\newblock {\em Nature\/}~{\em 72\/}(1867), 342--342.

\bibitem[\protect\citeauthoryear{Petrillo and Bennett}{Petrillo and Bennett}{2023}]{petrillo2023valuing}
Petrillo, M. and M.~R. Bennett (2023).
\newblock Valuing carers 2021: England and {W}ales.
\newblock {\em London: Carers UK\/}.

\bibitem[\protect\citeauthoryear{Pinquart and S{\"o}rensen}{Pinquart and S{\"o}rensen}{2005}]{pinquart2005ethnic}
Pinquart, M. and S.~S{\"o}rensen (2005).
\newblock Ethnic differences in stressors, resources, and psychological outcomes of family caregiving: A meta-analysis.
\newblock {\em The Gerontologist\/}~{\em 45\/}(1), 90--106.

\bibitem[\protect\citeauthoryear{Raiber, Visser, and Verbakel}{Raiber et~al.}{2022}]{raiber2022wage}
Raiber, K., M.~Visser, and E.~Verbakel (2022).
\newblock The wage penalty for informal caregivers from a life course perspective.
\newblock {\em Advances in Life Course Research\/}~{\em 53}, 100490.

\bibitem[\protect\citeauthoryear{Rosenbaum and Rubin}{Rosenbaum and Rubin}{1983}]{rosenbaum1983central}
Rosenbaum, P.~R. and D.~B. Rubin (1983).
\newblock The central role of the propensity score in observational studies for causal effects.
\newblock {\em Biometrika\/}~{\em 70\/}(1), 41--55.

\bibitem[\protect\citeauthoryear{Sant’Anna and Zhao}{Sant’Anna and Zhao}{2020}]{sant2020doubly}
Sant’Anna, P.~H. and J.~Zhao (2020).
\newblock Doubly robust difference-in-differences estimators.
\newblock {\em Journal of econometrics\/}~{\em 219\/}(1), 101--122.

\bibitem[\protect\citeauthoryear{Schmitz and Westphal}{Schmitz and Westphal}{2017}]{schmitz2017informal}
Schmitz, H. and M.~Westphal (2017).
\newblock Informal care and long-term labor market outcomes.
\newblock {\em Journal of health economics\/}~{\em 56}, 1--18.

\bibitem[\protect\citeauthoryear{Semyonov and Herring}{Semyonov and Herring}{2007}]{semyonov2007segregated}
Semyonov, M. and C.~Herring (2007).
\newblock Segregated jobs or ethnic niches?: The impact of racialized employment on earnings inequality.
\newblock {\em Research in Social Stratification and Mobility\/}~{\em 25\/}(4), 245--257.

\bibitem[\protect\citeauthoryear{Skira}{Skira}{2015}]{skira2015dynamic}
Skira, M.~M. (2015).
\newblock Dynamic wage and employment effects of elder parent care.
\newblock {\em International Economic Review\/}~{\em 56\/}(1), 63--93.

\bibitem[\protect\citeauthoryear{Smith, Cawley, Williams, and Mustard}{Smith et~al.}{2020}]{smith2020male}
Smith, P.~M., C.~Cawley, A.~Williams, and C.~Mustard (2020).
\newblock Male/female differences in the impact of caring for elderly relatives on labor market attachment and hours of work: 1997--2015.
\newblock {\em The Journals of Gerontology: Series B\/}~{\em 75\/}(3), 694--704.

\bibitem[\protect\citeauthoryear{{University of Essex, Institute for Social and Economic Research}}{{University of Essex, Institute for Social and Economic Research}}{2023}]{UnderstandingSociety2023}
{University of Essex, Institute for Social and Economic Research} (2023).
\newblock Understanding society: Waves 1-13, 2009-2022 and harmonised bhps: Waves 1-18, 1991-2009. [data collection].
\newblock SN: 6614.

\bibitem[\protect\citeauthoryear{Vagni and Breen}{Vagni and Breen}{2021}]{vagni2021earnings}
Vagni, G. and R.~Breen (2021).
\newblock Earnings and income penalties for motherhood: estimates for british women using the individual synthetic control method.
\newblock {\em European Sociological Review\/}~{\em 37\/}(5), 834--848.

\bibitem[\protect\citeauthoryear{Van~Houtven, Coe, and Skira}{Van~Houtven et~al.}{2013}]{van2013effect}
Van~Houtven, C.~H., N.~B. Coe, and M.~M. Skira (2013).
\newblock The effect of informal care on work and wages.
\newblock {\em Journal of health economics\/}~{\em 32\/}(1), 240--252.

\bibitem[\protect\citeauthoryear{Watkins and Overton}{Watkins and Overton}{2024}]{Watkins_Overton_2024}
Watkins, M. and L.~Overton (2024).
\newblock The cost of caring: a scoping review of qualitative evidence on the financial wellbeing implications of unpaid care to older adults.
\newblock {\em Ageing and Society\/}, 1–28.

\bibitem[\protect\citeauthoryear{Zhang, Petrillo, and Bennett}{Zhang et~al.}{2023}]{zhang2023valuing}
Zhang, J., M.~Petrillo, and M.~Bennett (2023).
\newblock Valuing {Ca}rers 2021: Northern {I}reland.
\newblock {\em Belfast: Carers Northern Ireland\/}.

\bibitem[\protect\citeauthoryear{Zhang and Bennett}{Zhang and Bennett}{2024}]{zhang2024insights}
Zhang, Y. and M.~R. Bennett (2024).
\newblock Insights into informal caregivers’ well-being: A longitudinal analysis of care intensity, care location, and care relationship.
\newblock {\em The Journals of Gerontology: Series B\/}~{\em 79\/}(2), gbad166.

\end{thebibliography}
\newpage
\renewcommand{\thesection}{S\Roman{section}}
\renewcommand{\thetable}{S\arabic{table}}
\renewcommand{\thefigure}{S\arabic{figure}}
\renewcommand{\theequation}{S\arabic{equation}}

\section*{Supplementary Information}\label{data_prep_appendix}
\setcounter{section}{0}
\setcounter{figure}{0} 
\setcounter{table}{0}

\subsection{Data Preparation}\label{datapreparation}

\subsubsection{The UKHLS} \label{ukhls}

The UK Household Longitudinal Study (UKHLS), initiated in 2009, is a comprehensive household panel survey designed to follow the same individuals and households over time. Building upon the British Household Panel Survey (BHPS), the UKHLS aims to represent the population residing in UK households. With an initial sample size of approximately 40,000 households, it stands as the largest household panel survey of its kind.
The UKHLS employs a multi-stage stratified random sampling method. This involves dividing the population into distinct groups (or strata) and then randomly selecting samples from each group. This approach ensures the sample is representative of the population across various dimensions, including region, urban or rural location, and household composition.
A common issue in longitudinal studies like the UKHLS is panel attrition, which refers to the proportion of participants who discontinue their involvement in the study over time. Reasons for attrition include relocation, loss of interest, or death. Attrition rates have varied across different waves of the survey, with some waves experiencing higher rates than others. Detailed information on attrition rates for each wave is available in the technical reports accessible on the official UKHLS website (https://www.understandingsociety.ac.uk/).

\subsubsection{Definition of informal carers and care intensity} \label{caredefinition}

Respondents are defined as informal carers if they answer `yes' to any of the following two questions:

\begin{quote}
``\textit{Is there anyone living with you who is sick, disabled or elderly whom you look after or give special help to (for example, a sick, disabled or elderly relative, husband, wife or friend etc)?}''
\end{quote}

or

\begin{quote}``
\textit{Do you provide some regular service or help for any sick, disabled or elderly person not living with you?}''
\end{quote}

The intensity of care provided has been identified with the following question:

\begin{quote}``\textit{Now thinking about everyone who you look after or provide help for, both those living with you and not living with you - in total, how many hours do you spend each week looking after or helping them? i.) 0-4 hours per week, ii.) 5-9 hours per week, iii.) 10-19 hours per week, iv.) 20-34 hours per week, v.) 35-49 hours per week, vi.) 50-99 hours per week, vii.) 100 or more hours per week/continuous care, viii.) Varies under 20 hours, ix.) Varies 20 hours or more, x.) Other.}``\end{quote}

We excluded participants who fell into the categories 8, 9 and 10 from the analysis.

\subsubsection{Variable Definitions} 

See Table \ref{table:variables} for more details on the variables used in the analysis.
\label{variablesdefinitions}
\begin{table}[h!]
\caption{Variable Descriptions}
\label{table:variables}
\centering
\small
\begin{tabular}{|>{\raggedright\arraybackslash}p{3cm}|>{\raggedright\arraybackslash}p{12cm}|}
\hline
\textbf{Variable} & \textbf{Description} \\
\hline
Individual income & Total personal monthly income gross. To limit the influence of outliners, this analysis trims the bottom and the top one per cent of the wage distribution. The variable is adjusted for inflation (base year 2015) using a Consumer Price Index which includes owner-occupiers’ housing costs (CPIH). \\
\hline
Household Income & Total gross household labour income in the month before the interview. This is described as the sum of total personal monthly income from labour income received by all household members. To limit the influence of outliners, this analysis trims the bottom and the top one per cent of the wage distribution. The variable is adjusted for inflation (base year 2015) using a Consumer Price Index which includes owner-occupiers’ housing costs (CPIH). \\
\hline
Income share & It is derived as the ratio between individual income and household income. \\
\hline
Low-Intensity Care & A dummy variable equal to one if the respondent spends less than 5 hours per week on caring, and zero otherwise. \\
\hline
Medium-Low-Intensity Care & A dummy variable equal to one if the respondent spends 5-19 hours per week on caring, and zero otherwise. \\
\hline
Medium-High-Intensity Care & A dummy variable equal to one if the respondent spends 20-49 hours per week on caring, and zero otherwise. \\
\hline
High-Intensity Care & Dummy variable, equal to one if the respondent spends 50+ hours per week on caring, and zero otherwise. \\
\hline
Age & Age of the respondent. \\
\hline
Male & A dummy variable equal to one if the respondent is male, zero if female. \\
\hline
Married & A dummy variable equal to one if the respondent is married or cohabits with his/her partner, and zero otherwise. \\
\hline
Asian & A dummy variable equal to one if the respondent has one of the following ethnicities: Indian, Pakistani, Bangladeshi or any other Asian background. It takes the value zero otherwise. \\
\hline
Black & A dummy variable equal to one if the respondent has one of the following ethnicities: African, Caribbean or any other black background. It takes the value zero otherwise. \\
\hline
White & A dummy variable equal to one if the respondent has one of the following ethnicities British, English, Scottish, Welsh, Northern Irish, Irish, Gypsy or Irish traveller or any other white background. It takes the value zero otherwise. \\
\hline
Mixed & A dummy variable equal to one if the respondent has one of the following ethnicities: White and black Caribbean, White and Asian, White and Black African, any other mixed background. It takes the value zero otherwise. \\
\hline
Others & A dummy variable equal to one if the respondent has one of the following ethnicities: Arabs or any other ethnic group. It takes the value zero otherwise. \\
\hline
Household Size & The number of people in the household. \\
\hline
Employed & A dummy variable equal to one if the respondent is self-employed or employed, on maternity leave, on apprenticeship, or on a government training scheme. The dummy variable takes the value of zero if the individual is unemployed, full-time student, sick or disabled, on furlough, in unpaid family business or temporarily laid off. \\
\hline
Lower Education & A dummy variable equal to one if the respondent has as the highest qualification achieved one of the following qualifications: cse, other school certification, gcse/o level, standard/o/level. It takes a value of zero otherwise. \\
\hline
Intermediate Education & A dummy variable equal to one if the respondent has as the highest qualification achieved one of the following qualifications: a level, as level, Highers(scot), certificate 6th-year studies, I ‘national baccalaureate, Welsh baccalaureate, diploma in higher education, nursing/other med qualification, a teaching qualification (not pgce). It takes a value of zero otherwise. \\
\hline
Advanced Education & A dummy variable equal to one if the respondent has as the highest qualification achieved one of the following qualifications: 1st degree or equivalent, higher degree, other higher degree. It takes a value of zero otherwise. \\
\hline
\end{tabular}
\end{table}

\newpage

\subsubsection{Temporal Alignment}

In order to be able to apply the Individual Synthetic Control (ISC) approach, several data preparation steps regarding timing were performed. Consider the example of one treated unit and how the controls for that treated unit were prepared. Assume a treated unit measured in the time span between 2010 to 2020, where this treated unit declared treatment $T0$ in 2015. Let’s also assume also that this unit is female and that we are only interested in comparing this unit with other females. Finally, let’s assume that this treated unit did not participate in all the ten annual waves, with -- for example -- Wave 2011 and 2017 missing. 
The first step is to select all the female control units (units that never declared unit caring responsibilities) that have measurements in the same years as our treated unit. Control units with measurement in years 2010 to 2020 will be used, but we will not consider their measurement in years 2011 and 2017. Simultaneously, this means that units that have missing values in any of the years that the treated unit does have will be ignored. The second step is to transform each year to a relative year with the origin point at $T0$. In the case of our example, 2015 will be now relative to year 0 for the treated unit and for all its selected set of controls. Years before $T0$ will be negative, and years after $T0$ will be positive. Additionally, these relative years keep their relative original position in the sequence. For example: 2015=$T0$, 2016=$T1$, and 2018=$T3$. Notice that 2018 is equal to relative year 3, because even though 2017 is a missing time point for this particular case, its relative position is respected and kept. These relative years allow us to centre the results around the point of treatment, while keeping the length of measurement point one year apart. Finally, this treated unit and its set of controls are sent to be used in the synthetic control. This procedure is done separately for each treated unit, each with its own $T0$. Since the set of control units -- although large -- is limited, all control units are possible candidates to be used for all treated units. For example, a control unit with flawless participation record between 2009 and 2021 could be used as a control unit for all treated units as it has observations in all periods within the sequence. However, each time a synthetic control is performed for each treated unit, the weights of the selected control units are recalculated, giving each synthetic control its unique set of weights. 

\newpage
\subsection{Supplementary Tables}
\begin{table}[htbp]
  \centering
  \caption{Individual Synthetic Control - Inflation Adjusted Individual Income.}
  \label{tab:ind_ATT}
  \begin{tabular}{l@{\hskip 1em}>{\centering\arraybackslash}p{3.5cm}@{\hskip 1em}>{\centering\arraybackslash}p{3.5cm}@{\hskip 1em}>{\centering\arraybackslash}p{3.5cm}@{\hskip 1em}>{\centering\arraybackslash}p{3.5cm}}
    \toprule
    & \multicolumn{4}{c}{Intensity} \\
    \cmidrule(lr){2-5}
    & L-Intensity & ML-Intensity & MH-Intensity & H-Intensity \\
    \midrule
    Tm8 & 37.67*** & 5.35 & -0.28 & -7.83 \\
    & (3.57) & (0.41) & (-0.02) & (-0.66) \\
    Tm7 & 22.02** & 8.15 & 4.33 & -6.39 \\
    & (3.57) & (0.85) & (0.25) & (-0.58) \\    
    Tm6 & 15.13* & 15.36 & -17.55 & -0.36 \\
    & (2.15) & (1.62) & (-1.41) & (-0.04) \\
    Tm5 & 22.90*** & 15.14* & 5.48 & 6.62 \\
    & (4.26) & (2.18) & (0.49) & (0.78) \\
    Tm4 & 13.83** & 15.55** & 10.59 & -2.20 \\
    & (3.07) & (3.08) & (1.14) & (-0.34) \\
    Tm3 & 15.84*** & 13.72** & 13.74* & 13.64* \\
    & (4.40) & (2.82) & (1.98) & (2.55) \\
    Tm2 & 8.65** & 0.00 & 6.74 & 5.98 \\
    & (2.96) & (0.00) & (1.29) & (1.43) \\
    Tm1 & 2.19 & 3.11 & 5.45 & 2.71 \\
    & (0.82) & (1.06) & (0.74) & (0.45) \\
    Tp0 & -16.99 & -35.52* & -56.25* & -75.63*** \\
    & (-1.54) & (-2.40) & (-2.49) & (-3.54) \\
    Tp1 & -23.54 & -52.20** & -64.19* & -123.70*** \\
    & (-1.86) & (-2.94) & (-2.11) & (-3.81) \\
    Tp2 & -33.29* & -48.02** & -85.87* & -165.92*** \\
    & (-2.30) & (-2.61) & (-2.50) & (-4.11) \\
    Tp3 & -19.86 & -86.88*** & -112.38** & -154.41*** \\
    & (-1.09) & (-3.71) & (-2.79) & (-3.40) \\
    Tp4 & -32.76 & -138.55*** & -122.73* & -192.01** \\
    & (-1.55) & (-5.07) & (-2.21) & (-3.27) \\
    Tp5 & -74.95** & -137.35*** & -142.60* & -148.24** \\
    & (-3.17) & (-4.26) & (-2.48) & (-2.67) \\
    Tp6 & -79.08 & -171.83*** & -153.05* & -190.23** \\
    & (-2.86) & (-4.67) & (-2.11) & (-2.81) \\
    \midrule
    \multicolumn{5}{p{16cm}}{\footnotesize Note: The table shows results for the Individual Synthetic Control estimator, Average treatment effect on the treated. For the full set of individual controls, see Table \ref{descr}. t-statistics in parentheses. * p < 0.05, ** p < 0.01, *** p < 0.001. RMSPE Low-Intensity=59.5. RMSPE Medium-Low-Intensity=36.3. RMSPE Medium-High-Intensity=22.4. RMSPE High-Intensity= 15.1. Source: UKHLS data (years 2009-2020), authors' calculations.
    }
    \\
  \end{tabular}
\end{table}

\newpage

\begin{table}[htbp]
  \centering
  \caption{Individual Synthetic Control - Inflation Adjusted Household Income.}
  \label{tab:hh_ATT}
  \begin{tabular}{l@{\hskip 1em}>{\centering\arraybackslash}p{3.5cm}@{\hskip 1em}>{\centering\arraybackslash}p{3.5cm}@{\hskip 1em}>{\centering\arraybackslash}p{3.5cm}@{\hskip 1em}>{\centering\arraybackslash}p{3.5cm}}
    \toprule
    & \multicolumn{4}{c}{Intensity} \\
    \cmidrule(lr){2-5}
    & L-Intensity & ML-Intensity & MH-Intensity & H-Intensity \\
    \midrule
    Tm8 & 86.39*** & -4.70 & 46.49 & 49.59 \\
    & (4.44) & (-0.17) & (1.22) & (0.94) \\
    Tm7 & 18.15 & 48.88* & -11.36 & -15.20 \\
    & (1.25) & (2.40) & (-0.31) & (-0.39) \\
    Tm6 & 22.34 & 39.76* & -10.02 & 16.37 \\
    & (1.54) & (2.16) & (-0.40) & (0.59) \\
    Tm5 & 41.90*** & 25.20 & 51.31 & -7.90 \\
    & (3.62) & (1.82) & (1.54) & (-0.41) \\
    Tm4 & 9.85 & 17.05 & 28.36 & 3.21 \\
    & (1.25) & (1.51) & (1.24) & (0.18) \\
    Tm3 & -3.11 & 15.46 & 1.93 & -11.17 \\
    & (-0.40) & (1.63) & (0.14) & (-0.66) \\
    Tm2 & 12.73* & 0.28 & 19.90 & -7.79 \\
    & (2.10) & (0.03) & (1.77) & (-0.61) \\
    Tm1 & 4.49 & -1.57 & 4.43 & 6.75 \\
    & (0.80) & (-0.21) & (0.21) & (0.56) \\
    Tp0 & 45.33* & -7.04 & -14.78 & 1.69 \\
    & (2.07) & (-0.25) & (-0.30) & (0.04) \\
    Tp1 & 14.14 & -74.70* & -75.97 & -73.49 \\
    & (0.54) & (-2.30) & (-1.23) & (-1.17) \\
    Tp2 & -9.63 & -38.27 & -145.57 & -100.35 \\
    & (-0.34) & (-1.02) & (-1.77) & (-1.38) \\
    Tp3 & -21.20 & -77.72 & -77.94 & -234.50** \\
    & (-0.68) & (-1.67) & (-0.88) & (-3.28) \\
    Tp4 & -68.42 & -182.41*** & -111.97 & -323.87** \\
    & (-1.93) & (-3.64) & (-1.45) & (-3.07) \\
    Tp5 & -66.37 & -237.77*** & -92.79 & -406.64*** \\
    & (-1.68) & (-4.49) & (-0.78) & (-3.51) \\
    Tp6 & -125.04** & -224.68*** & -94.31 & -258.40 \\
    & (-2.83) & (-3.63) & (-0.55) & (-1.79) \\
    \midrule
    \multicolumn{5}{p{16cm}}{\footnotesize Note: The table shows results for the Individual Synthetic Control estimator. Average treatment effect on the treated. For the full set of individual controls, see Table \ref{descr}. t-statistics in parentheses. * p < 0.05, ** p < 0.01, *** p < 0.001. RMSPE Low-Intensity=34.6. RMSPE Mediumg-Low-Intensity=24.9. RMSPE Medium-High-Intensity=27.7. RMSPE High-Intensity= 20.8. Source: UKHLS data (years 2009-2020), authors' calculations.
    }
    \\
  \end{tabular}
\end{table}

\newpage

\begin{table}[htbp]
\centering
\caption{Pre- and Post-Treatment Average Income and Penalty}
\begin{tabular}{llcc}
\toprule
 &  & Individual Income & Household Income \\
\midrule
{High-Intensity} & Pre-treatment average income & £362 & £1,959 \\
 & Post-treatment average loss & £162 & £232 \\
 & Penalty & 45\% & 12\% \\
\midrule
{Medium-High-Intensity} & Pre-treatment average income & £518 & £2,208 \\
 & Post-treatment average loss & £113 & £100 \\
 & Penalty & 22\% & 5\% \\
\midrule
{Medium-Low-Intensity} & Pre-treatment average income & £848 & £2,630 \\
 & Post-treatment average loss & £106 & £139 \\
 & Penalty & 13\% & 5\% \\
\midrule
{Low-Intensity} & Pre-treatment average income & £1,057 & £3,042 \\
 & Post-treatment average loss & £44 & £46 \\
 & Penalty & 4\% & 2\% \\
\bottomrule
\multicolumn{4}{p{0.9\linewidth}}{\footnotesize Source: UKHLS data (years 2009-2020), authors' calculations.}
\label{caregiving_penalty}
\end{tabular}
\end{table}

\newpage

\begin{table}[ht]
\centering
\caption{Propensity Score Matching - Inflation Adjusted Individual Income.}
\label{tab:psm_ii}
\begin{threeparttable}
\begin{tabular}{lcccc}
\toprule
 & (1) & (2) & (3) & (4) \\
 & L-Intensity & ML-Intensity & MH-Intensity & H-Intensity \\
\midrule
Tp0 & -57.88* & -98.68*** & -220.04*** & -124.85** \\
 & (25.92) & (29.83) & (45.34) & (47.48) \\
N & 255,321 & 252,033 & 247,673 & 245,304 \\
Tp1 & -53.50 & -60.01 & -128.07* & -88.20 \\
 & (41.31) & (41.70) & (60.48) & (58.51) \\
N & 235,052 & 233,607 & 224,946 & 228,696 \\
Tp2 & -40.40 & -99.09* & -246.73*** & -198.04** \\
 & (44.08) & (45.88) & (65.86) & (67.63) \\
N & 208,687 & 207,436 & 199,625 & 197,580 \\
Tp3 & -22.95 & -94.80 & -188.75* & -235.23** \\
 & (47.64) & (51.68) & (79.90) & (78.40) \\
N & 185,104 & 183,854 & 179,448 & 173,445 \\
Tp4 & -57.83 & -100.99 & -183.47* & -226.07** \\
 & (53.15) & (54.30) & (93.63) & (84.75) \\
N & 162,747 & 162,276 & 158,349 & 153,497 \\
Tp5 & -42.64 & -56.33 & -296.23** & -86.96 \\
 & (58.11) & (58.69) & (93.55) & (86.88) \\
N & 140,192 & 139,834 & 136,148 & 127,829 \\
Tp6 & -83.27 & -162.21* & -158.75 & -164.40 \\
 & (65.44) & (66.00) & (113.24) & (108.95) \\
N & 118,524 & 119,326 & 112,584 & 113,598 \\
\bottomrule
\end{tabular}
\begin{tablenotes}
\item Note: The table shows results for the Propensity Score Matching estimator. Average treatment effect on the treated. Probit regressions were initially estimated to assess the likelihood of caring across different intensities. For the full set of individual controls, see Table \ref{descr}. The resulting propensity scores were then applied to match non-carers with carers who shared similar characteristics t-statistics in parentheses. * p < 0.05, ** p < 0.01, *** p < 0.001. Source: UKHLS data (years 2009-2020), authors' calculations.
\end{tablenotes}
\end{threeparttable}
\end{table}

\begin{table}[!htbp] \centering
\caption{Propensity Score Matching - Inflation Adjusted Household Income.}
\label{tab:psm_hh2}
\begin{threeparttable}
\begin{tabular}{lcccc}
\toprule
 & \textbf{L-Intensity} & \textbf{ML-Intensity} & \textbf{MH-Intensity} & \textbf{H-Intensity} \\
\midrule
Tp0 & 2.20 & -281.10*** & -532.75*** & -483.25*** \\
 & (44.24) & (53.85) & (87.25) & (94.05) \\
 N & 255,573 & 252,245 & 247,920 & 245,435 \\
\midrule
Tp1 & 53.24 & -157.36** & -472.30*** & -661.62*** \\
 & (68.94) & (76.60) & (115.85) & (124.7) \\
 N & 235,325 & 233,461 & 228,311 & 226,887 \\
\midrule
Tp2 & -46.71 & -149.66* & -345.15** & -602.38*** \\
 & (75.61) & (83.28) & (130.59) & (135.35) \\
N & 208,745 & 204,990 & 199,579 & 198,442 \\
\midrule
Tp3 & 76.02 & -78.24 & -226.57 & -744.88*** \\
 & (82.16) & (89.80) & (150.59) & (155.64) \\
 N & 185,218 & 184,002 & 179,906 & 173,430 \\
\midrule
Tp4 & -82.32 & -234.65* & -390.98*** & -648.67*** \\
 & (94.24) & (104.19) & (176.88) & (179.77) \\
 N & 162,834 & 162,349 & 158,348 & 153,476 \\
\midrule
Tp5 & -97.38 & 14.46 & -204.57 & -322.84 \\
 & (99.37) & (109.99) & (190.17) & (197.88) \\
N & 142,106 & 140,033 & 135,398 & 118,796 \\
\midrule
Tp6 & -119.89 & -53.13 & -91.45 & -208.88 \\
 & (100.99) & (121.24) & (213.06) & (200.81) \\
N & 118,753 & 119,434 & 112,197 & 113,818 \\
\bottomrule
\end{tabular}
\begin{tablenotes}
\item \textit{Note:} The table shows results for the Propensity Score Matching estimator. Average treatment effect on the treated. Probit regressions were initially estimated to assess the likelihood of caring across different intensities. For the full set of individual controls, see Table \ref{descr}. The resulting propensity scores were then applied to match non-carers with carers who shared similar characteristics. t-statistics in parentheses. * p < 0.05, ** p < 0.01, ***p < 0.001. Source: UKHLS data (years 2009-2020), authors' calculations.
\end{tablenotes}
\end{threeparttable}
\end{table}

\newpage

\begin{table}[htbp]
  \centering
  \caption{Difference-in-differences - Inflation Adjusted Individual Income.}
  \label{tab:diff-in-diff}
\begin{tabular}{l@{\hskip 1em}>{\centering\arraybackslash}p{3.5cm}@{\hskip 1em}>{\centering\arraybackslash}p{3.5cm}@{\hskip 1em}>{\centering\arraybackslash}p{3.5cm}@{\hskip 1em}>{\centering\arraybackslash}p{3.5cm}}
\toprule
    & (1) & (2) & (3) & (4) \\
    \midrule
    & L-Intensity & ML-Intensity & MH-Intensity & H-Intensity \\
    \midrule
    Tm8 & 35.32 & -39.07 & 163.36* & -31.64 \\
    & (0.76) & (-0.50) & (1.53) & (-0.26) \\
    Tm7 & -29.09 & 110.06 & 60.46 & 86.81 \\
    & (-0.85) & (1.63) & (0.36) & (0.93) \\
    Tm6 & 14.83 & 3.60 & -143.88 & -11.82 \\
    & (-0.40) & (0.07) & (-1.17) & (-0.12) \\
    Tm5 & -15.55 & 6.29 & 13.06 & -87.84 \\
    & (-0.47) & (0.16) & (0.24) & (-1.26) \\
    Tm4 & -27.69 & 16.16 & -46.96 & -184.11* \\
    & (-1.13) & (0.46) & (-0.55) & (-2.28) \\
    Tm3 & -25.33 & -63.01 & 13.33 & 14.86 \\
    & (-1.04) & (-1.99) & (0.21) & (0.35) \\
    Tm2 & -18.84 & -24.17 & -16.78 & -60.73 \\
    & (-0.85) & (-0.81) & (-0.38) & (-1.05) \\
    Tm1 & -23.97 & -33.17 & -25.28 & -20.81 \\
    & (-1.34) & (-1.30) & (-0.49) & (-0.38) \\
    Tp0 & -18.02 & -3.46 & -155.67** & -223.51*** \\
    & (-1.11) & (-0.14) & (-3.03) & (-4.13) \\
    Tp1 & -40.86* & -15.68 & -14.98 & -244.98*** \\
    & (-2.19) & (-0.58) & (-0.28) & (-3.64) \\
    Tp2 & -24.88 & -23.01 & -80.42 & -318.32*** \\
    & (-1.12) & (-0.74) & (-1.38) & (-4.30) \\
    Tp3 & -1.82 & -47.35 & -100.88 & -340.23*** \\
    & (-0.07) & (-1.28) & (-1.56) & (-3.78) \\
    Tp4 & -12.64 & -123.05** & -152.26* & -327.93** \\
    & (-0.42) & (-3.04) & (-2.18) & (-3.25) \\
    Tp5 & -36.24 & -133.98** & -124.79 & -291.91** \\
    & (-1.05) & (-2.91) & (-1.48) & (-3.15) \\
    Tp6 & -10.29 & -97.90 & 131.97 & -237.04* \\
    & (-0.25) & (-1.81) & (-1.33) & (-2.02) \\
    \midrule
    N & 149,931 & 131,027 & 118,334 & 117,277 \\
    \midrule
    \multicolumn{5}{p{1\linewidth}}{\footnotesize Note: The table shows results for the Doubly Robust Difference-in-Difference estimator. Average treatment effect on the treated. For the full set of individual controls, see Table \ref{descr}. t-statistics in parentheses. * p < 0.05, ** p < 0.01, *** p < 0.001. RMSPE Low-Intensity=27.2. RMSPE Medium-Low-Intensity=58.8. RMSPE Medium-High-Intensity=78.1. RMSPE High-Intensity= 82.6. Source: UKHLS data (years 2009-2020), authors' calculations.
    }
    \\
  \end{tabular}
\end{table}


\newpage

\begin{table}[htbp]
  \centering
  \caption{Difference-in-differences - Inflation Adjusted Household Income.}
  \label{tab:diff-in-diff-household}
  \begin{tabular}{l@{\hskip 1em}>{\centering\arraybackslash}p{3.5cm}@{\hskip 1em}>{\centering\arraybackslash}p{3.5cm}@{\hskip 1em}>{\centering\arraybackslash}p{3.5cm}@{\hskip 1em}>{\centering\arraybackslash}p{3.5cm}}
    \toprule
    & (1) & (2) & (3) & (4) \\
    \midrule
    & L-Intensity & ML-Intensity & MH-Intensity & H-Intensity \\
    \midrule
    Tm8 & 119.78 & -249.51* & 285.62 & 227.90 \\
    & (1.52) & (-2.02) & (1.86) & (1.32) \\
    Tm7 & -114.84 & 329.35** & 63.62 & -41.24 \\
    & (-1.54) & (2.80) & (0.32) & (-0.13) \\
    Tm6 & 119.40 & -54.43 & -86.00 & 60.23 \\
    & (1.83) & (-0.51) & (-0.53) & (0.24) \\
    Tm5 & -14.79 & -143.72 & 227.36 & -89.18 \\
    & (-0.22) & (-1.61) & (1.03) & (-0.64) \\
    Tm4 & -46.95 & -66.16 & 59.82 & 146.47 \\
    & (-0.96) & (-0.89) & (0.39) & (0.96) \\
    Tm3 & 53.57 & 12.74 & 35.84 & -231.68 \\
    & (1.17) & (0.18) & (0.35) & (-1.57) \\
    Tm2 & 70.85 & -43.67 & 6.81 & 43.52 \\
    & (1.46) & (-0.70) & (0.07) & (0.43) \\
    Tm1 & -24.96 & 7.50 & 46.24 & 107.77 \\
    & (-0.71) & (0.15) & (0.40) & (1.18) \\
    Tp0 & 101.61** & 4.82 & -101.99 & 42.07 \\
    & (3.09) & (0.11) & (-1.01) & (0.42) \\
    Tp1 & 43.39 & 50.68 & 120.51 & -42.47 \\
    & (1.13) & (0.92) & (0.92) & (-0.32) \\
    Tp2 & 51.69 & 42.69 & -23.64 & -75.75 \\
    & (1.17) & (0.69) & (-0.17) & (-0.53) \\
    Tp3 & 80.91 & 5.07 & 61.06 & -227.01 \\
    & (1.16) & (0.07) & (0.37) & (-1.35) \\
    Tp4 & 75.38 & -133.90 & -153.25 & -213.32 \\
    & (1.33) & (-1.62) & (-0.92) & (-1.25) \\
    Tp5 & 45.15 & -167.17 & -123.20 & -312.94 \\
    & (0.72) & (-1.85) & (-0.51) & (-1.14) \\
    Tp6 & 6.65 & -195.27 & -51.82 & 5.46 \\
    & (0.09) & (-1.87) & (-0.19) & (0.02) \\
    \midrule
    N & 150,139 & 131,164 & 118,524 & 117,435 \\
    \midrule
    \multicolumn{5}{p{1\linewidth}}{\footnotesize Note: The table shows results using the Doubly Robust Difference-in-difference estimator. Average treatment effect on the treated. For the full set of individual controls, see Table \ref{descr}. t-statistics in parentheses. * p < 0.05, ** p < 0.01, *** p < 0.001. RMSPE Low-Intensity=86.0. RMSPE Medium-Low-Intensity=159.8. RMSPE Medium-High-Intensity=137.7. RMSPE High-Intensity=149.6.
    Source: UKHLS data (years 2009-2020), authors' calculations.
    }
    \\
  \end{tabular}
\end{table}

\newpage

\begin{table}[htbp]
\caption{Difference in differences - Parallel trend Assumption.}
\label{table:paralleltrend}
\centering
\begin{tabular}{lcc}
\toprule
\textbf{Individual Income} & \textbf{\({\chi}^2\)}& \textbf{P value} \\
  \midrule
 High-Intensity & \({\chi}^2\)(65)= 138.66 & 0.000 \\
Medium-high-Intensity & \({\chi}^2\)(65)= 186.49 & 0.000  \\
Medium-low-Intensity &	\({\chi}^2\)(65)= 75.02	& 0.186 \\
Low-Intensity &	\({\chi}^2\)(65)= 93.73 &	0.011 \\
  \midrule
\textbf{Household Income} &	\textbf{\({\chi}^2\)}	& \textbf{P value} \\
  \midrule
High-Intensity & 	\({\chi}^2\)(63)= 341.82 &	0.000 \\
Medium-high-Intensity &	\({\chi}^2\)(64)= 140.43& 	0.000 \\
Medium-low-Intensity &	\({\chi}^2\)(65)= 91.795 &	0.016 \\
Low-Intensity	& \({\chi}^2\)(65)= 84.95 &	0.049 \\
\bottomrule
\multicolumn{3}{p{0.7\linewidth}}{\footnotesize Note: \({\chi}^2\) statistic of the null hypothesis that all pre-treatment ATTGT's are statistically equal to zero. Source: UKHLS data (years 2009-2020), authors' calculations.}
\end{tabular}
\end{table}

\newpage

\begin{table}[htbp]
  \centering
  \caption{Relative caring penalty by sex - Individual Income.}
  \label{tab:caregiving_penalty_gender}
  \begin{tabular}{lcccccc}
    \toprule
    & \multicolumn{2}{c}{Pre-Treatment Av. Income} & \multicolumn{2}{c}{Post-Treatment Av. Loss} & \multicolumn{2}{c}{Penalty}\\
    \cmidrule(lr){2-3} \cmidrule(lr){4-5} \cmidrule(lr){6-7}
    & HI & LI & HI & LI & HI & LI \\
    \midrule
    Women & £399 & £753 & £121 & 38 & 30\% & 5\% \\
    Men & £545 & 1310 & £137 & £84 & 25\% & 6\% \\
    \midrule
    \multicolumn{6}{p{0.7\linewidth}}{\footnotesize Note: HI = High-Intensity, LI = Low-Intensity.\par Source: UKHLS data (years 2009-2020), authors' calculations.}

  \end{tabular}
\end{table}

\newpage

\begin{table}[htbp]
  \centering
  \caption{Relative caring penalty by sex - Household Income.}
  \label{tab:caregiving_penalty_gender_hh}
  \begin{tabular}{lcccccc}
    \toprule
    & \multicolumn{2}{c}{Pre-Treatment Av. Income} & \multicolumn{2}{c}{Post-Treatment Av. Loss} & \multicolumn{2}{c}{Penalty}\\
    \cmidrule(lr){2-3} \cmidrule(lr){4-5} \cmidrule(lr){6-7}
    & HI & LI & HI & LI & HI & LI \\
    \midrule
    Women & £2,005 & £2,665 & £122 & 86 & 6\% & 3\% \\
    Men & £2,273 & 3,220 & £193 & £69 & 8\% & 2\% \\
    \midrule
    \multicolumn{6}{p{0.9\linewidth}}{\footnotesize Note: HI = High-Intensity, LI = Low-Intensity. Source: UKHLS data (years 2009-2020), authors' calculations.}
  \end{tabular}
\end{table}

\newpage

\begin{table}[htbp]
  \centering
  \caption{Relative caring penalty by ethnicity - Individual Income.}
  \label{tab:caregiving_penalty_ethnicity}
  \begin{tabular}{lcccccc}
    \toprule
    & \multicolumn{2}{c}{Pre-Treatment Av. Income} & \multicolumn{2}{c}{Post-Treatment Av. Loss} & \multicolumn{2}{c}{Penalty}\\
    \cmidrule(lr){2-3} \cmidrule(lr){4-5} \cmidrule(lr){6-7}
    & HI & LI & HI & LI & HI & LI \\
    \midrule
    White & £480 & £1,060 & £153 & £57 & 32\% & 5\% \\
    Non-white & £288 & £484 & £57 & £20 & 20\% & 4\% \\
    \midrule
    \multicolumn{6}{p{0.9\linewidth}}{\footnotesize Note: HI = High-Intensity, LI = Low-Intensity. Source: UKHLS data (years 2009-2020), authors' calculations.}
  \end{tabular}
\end{table}

\newpage

\begin{table}[htbp]
  \centering
  \caption{Relative caring penalty by ethnicity - Household Income.}
  \label{tab:caregiving_penalty_ethnicity_hh}
  \begin{tabular}{lcccccc}
    \toprule
    & \multicolumn{2}{c}{Pre-Treatment Av. Income} & \multicolumn{2}{c}{Post-Treatment Av. Loss} & \multicolumn{2}{c}{Penalty}\\
    \cmidrule(lr){2-3} \cmidrule(lr){4-5} \cmidrule(lr){6-7}
    & HI & LI & HI & LI & HI & LI \\
    \midrule
    White & 2,300 & £3,100 & £176 & £69 & 8\% & 2\% \\
    Non-white & £1,137 & £1,555 & £56 & £7 & 5\% & 0\% \\
    \midrule
    \multicolumn{6}{p{0.9\linewidth}}{\footnotesize Note: HI = High-Intensity, LI = Low-Intensity. Source: UKHLS data (years 2009-2020), authors' calculations.}
  \end{tabular}
\end{table}

\newpage

\begin{table}[htbp]
  \centering
  \caption{Relative caring penalty by age - Individual Income.}
  \label{tab:caregiving_penalty_age}
  \begin{tabular}{lcccccc}
    \toprule
    & \multicolumn{2}{c}{Pre-Treatment Av. Income} & \multicolumn{2}{c}{Post-Treatment Av. Loss} & \multicolumn{2}{c}{Penalty}\\
    \cmidrule(lr){2-3} \cmidrule(lr){4-5} \cmidrule(lr){6-7}
    & HI & LI & HI & LI & HI & LI \\
    \midrule
    Aged 25 and below & £247 & £316 & £447 & 40 & 181\% & 13\% \\
    Aged 26-64 & £596 & 1,257 & £152 & £104 & 17\% & 8\% \\
    Aged 65 and above & £65 & 86 & £(-6) & £(-3) & (-9)\% & (-3)\% \\
    \midrule
    \multicolumn{6}{p{0.9\linewidth}}{\footnotesize Note: HI = High-Intensity, LI = Low-Intensity. Source: UKHLS data (years 2009-2020), authors' calculations.}
  \end{tabular}
\end{table}

\newpage

\begin{table}[htbp]
  \centering
  \caption{Relative caring penalty by age - Household Income.}
  \label{tab:caregiving_penalty_age_hh}
  \begin{tabular}{lcccccc}
    \toprule
    & \multicolumn{2}{c}{Pre-Treatment Av. Income} & \multicolumn{2}{c}{Post-Treatment Av. Loss} & \multicolumn{2}{c}{Penalty}\\
    \cmidrule(lr){2-3} \cmidrule(lr){4-5} \cmidrule(lr){6-7}
    & HI & LI & HI & LI & HI & LI \\
    \midrule
    Aged 25 and below & £1,843 & £2,893 & £351 & 35 & 19\% & 1\% \\
    Aged 26-64 & £2,208 & 3,143 & £213 & £154 & 7\% & 5\% \\
    Aged 65 and above & £1,793 & 1,902 & £39 & £(-14) & 2\% & (-1)\% \\
    \midrule
    \multicolumn{6}{p{0.9\linewidth}}{\footnotesize Note: HI = High-Intensity, LI = Low-Intensity. Source: UKHLS data (years 2009-2020), authors' calculations.}
  \end{tabular}
\end{table}

\clearpage
\subsection{Supplementary Figures}

\begin{figure}[H]\label{additional_results}
    \centering
    \includegraphics[width=0.955\linewidth]{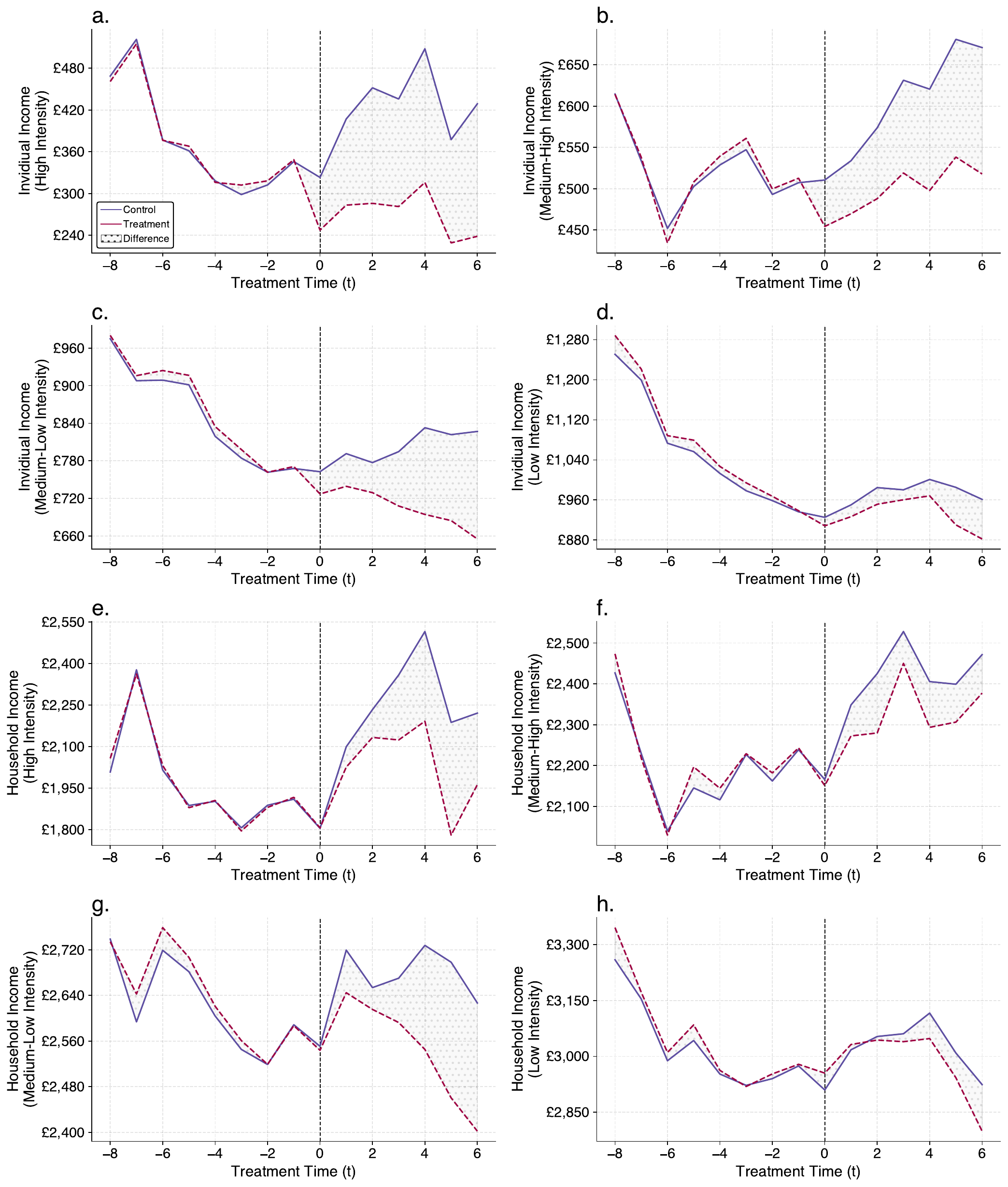}
     \caption{\small{\textbf{Inflation Adjusted Individual and Household Income - Difference between treatment and control groups.} Individual Synthetic Control. The solid violet line depicts non-carers' income trajectories; the burgundy dashed line represents informal carers'. The grey area represents the difference. For the full set of individual controls see Table \ref{descr}. 
    Panels a. and e. report the difference between high-intensity informal carers and their counterfactual; Panels b. and f. report medium-high-intensity informal carers; Panels c. and g. report medium-low-intensity informal carers; Panels d. and h. report low-intensity informal carers. Panels a.-d. report individual income, while e.-h. report household income. Source: UKHLS data (years 2009-2020), authors' calculations.
    }}
        \label{ii_and_hh_main}
\end{figure}

\begin{figure}[H]
 \centering
    \includegraphics[width=0.955\linewidth]{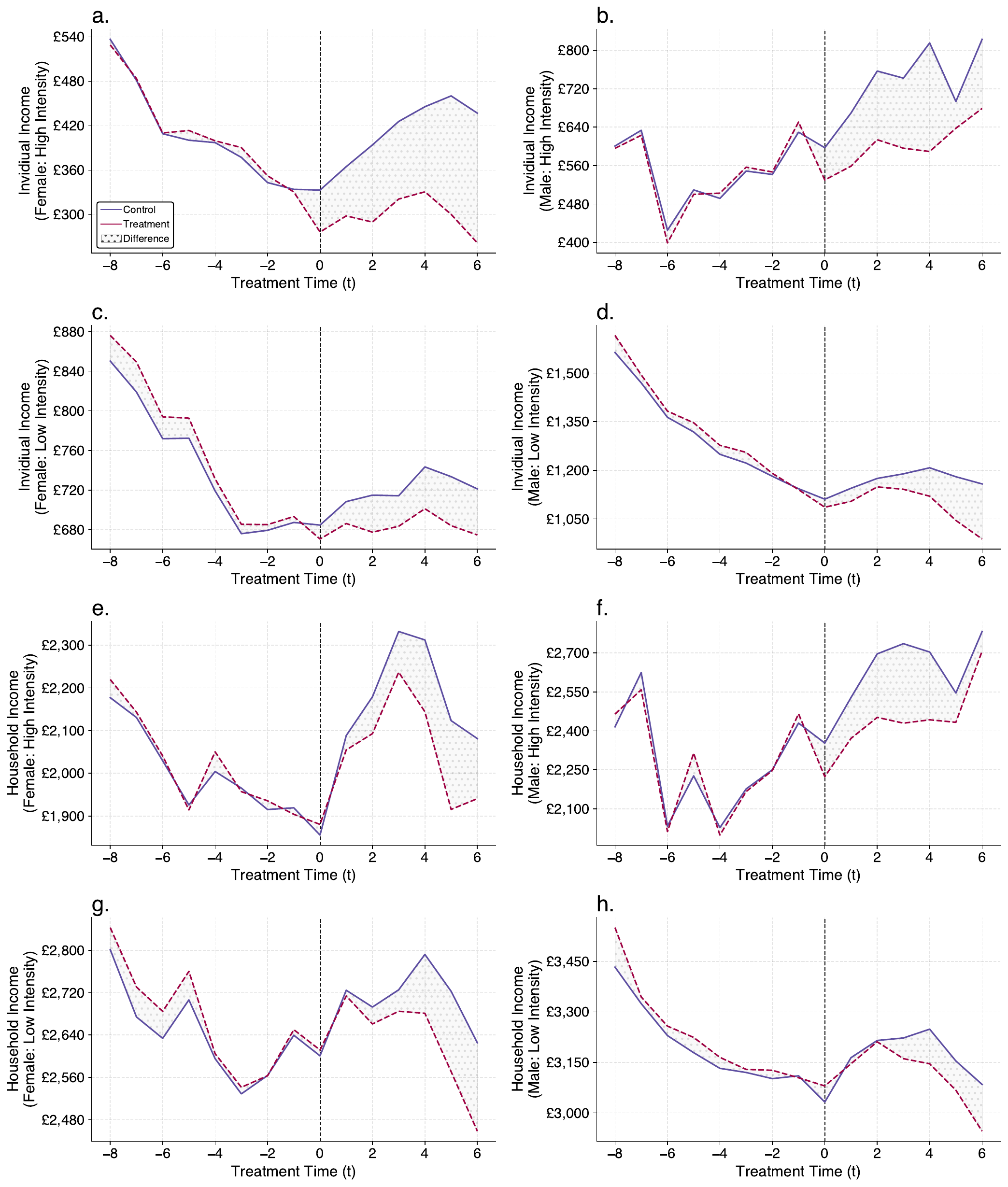}
    \caption{\small{\textbf{Inflation Adjusted Individual and Household Income - Difference between treatment and control groups by sex.}
    Individual Synthetic Control. The solid violet line depicts non-carers' income trajectories; the burgundy dashed line represents informal carers'. The grey area represents the difference. For the full set of individual controls see Table \ref{descr}. 
    Panels a. and e. report high-intensity informal female carers and their counterfactual; Panels b. and f. report high-intensity informal male carers; Panels c. and g. report low-intensity informal female carers; Panels d. and h. report low-intensity informal male carers. Panels a.-d. report individual income, while e.-h. report household income. Source: UKHLS data (years 2009-2020), authors' calculations.
    }}\label{ii_gender}
\end{figure}

\newpage 
\begin{figure}[H]
    \centering
    \includegraphics[width=0.955\linewidth]{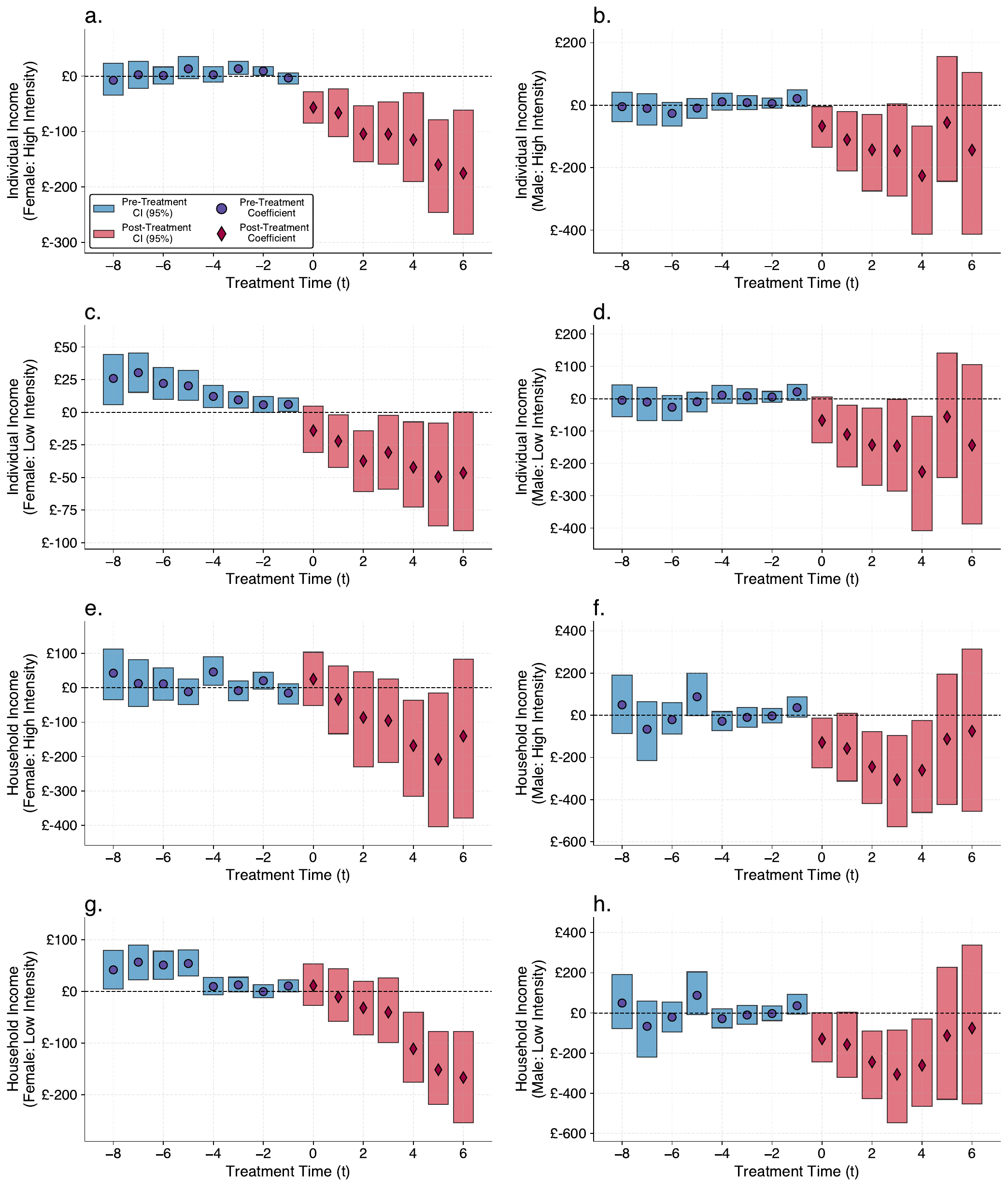}
    \caption{\small{\textbf{Inflation Adjusted Individual and Household Income by sex.} Individual Synthetic Control. Average Treatment Effect on the Treated. The blue shaded areas and blue circles represent the pre-treatment confidence intervals at 95\% and the pre-treatment coefficients, respectively. The red shaded areas and red diamonds denote the post-treatment confidence intervals at 95\% and post-treatment coefficients, respectively. For the full set of individual controls see Table \ref{descr}. Panels a. and e. report the difference between high-intensity informal female carers and their counterfactual; Panels b. and f. report high-intensity informal male carers; Panels c. and g. report low-intensity informal female carers and Panels d. and h. report low-intensity informal male carers. Panels a.-d. report individual income, and e.-h. report household income. Source: UKHLS data (years 2009-2020), authors' calculations.
    }}\label{ci_ii_sex}
\end{figure}

\newpage 
\begin{figure}[H]
    \centering
    \includegraphics[width=0.955\linewidth]{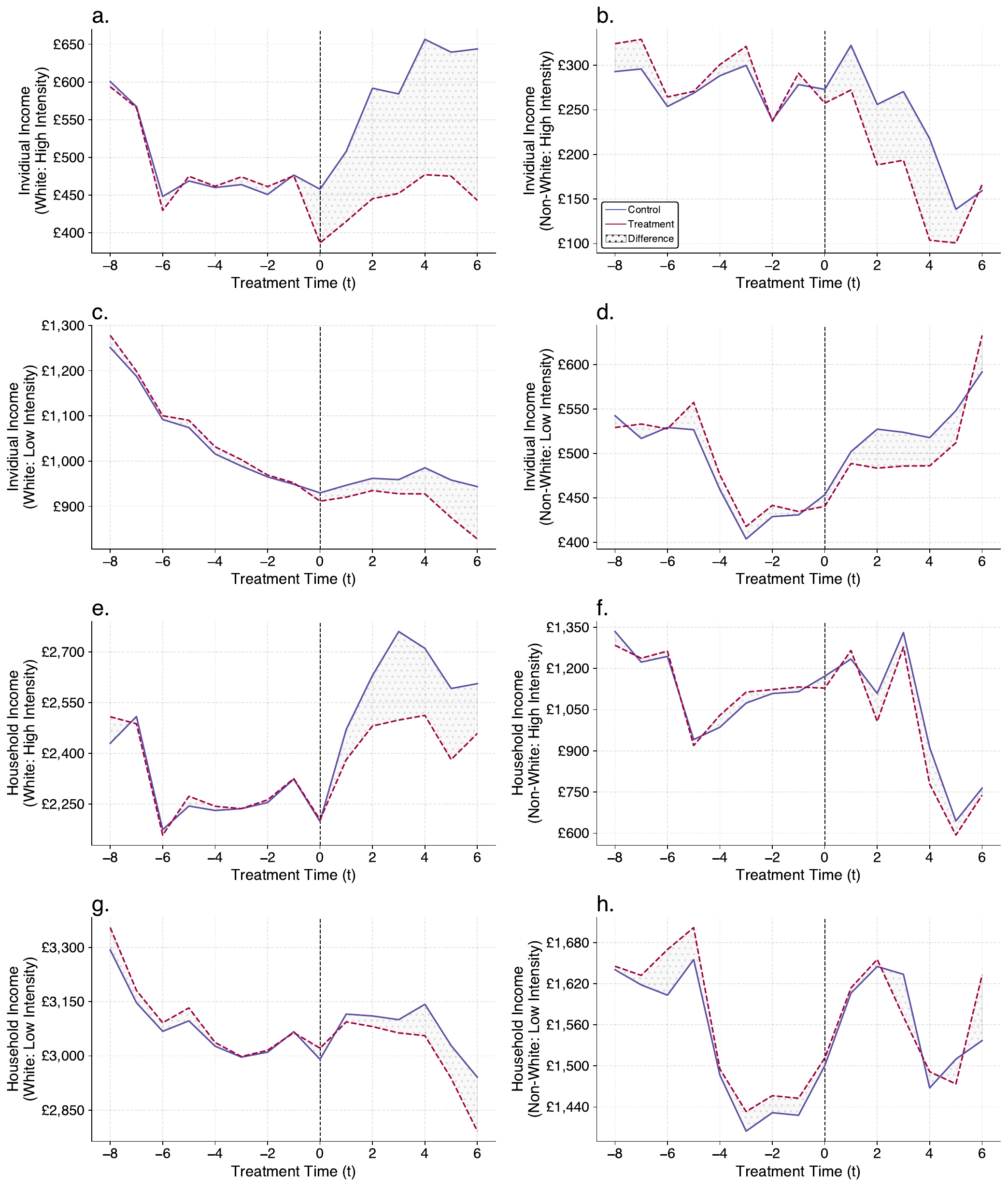}
    \caption{\small{\textbf{Inflation Adjusted Individual and Household Income - Difference between treatment and control groups by ethnicity.} 
    Individual Synthetic Control. The solid violet line depicts non-carers' income trajectories; the burgundy dashed line represents informal carers'. The grey area represents the difference. For the full set of individual controls see Table \ref{descr}. 
    Panels a. and e. report the difference between high-intensity informal `White' carers and their counterfactual; Panels b. and f. report high-intensity informal `non-White' carers; Panels c. and g. report low-intensity informal `White' carers; Panels d. and h. report low-intensity informal `non-White' carers. Panels a.-d. report individual income, while e.-h. report household income. Source: UKHLS data (years 2009-2020), authors' calculations.
    }}\label{ii_eth}
\end{figure}

\newpage
\begin{figure}[H]
    \centering
    \includegraphics[width=0.955\linewidth]{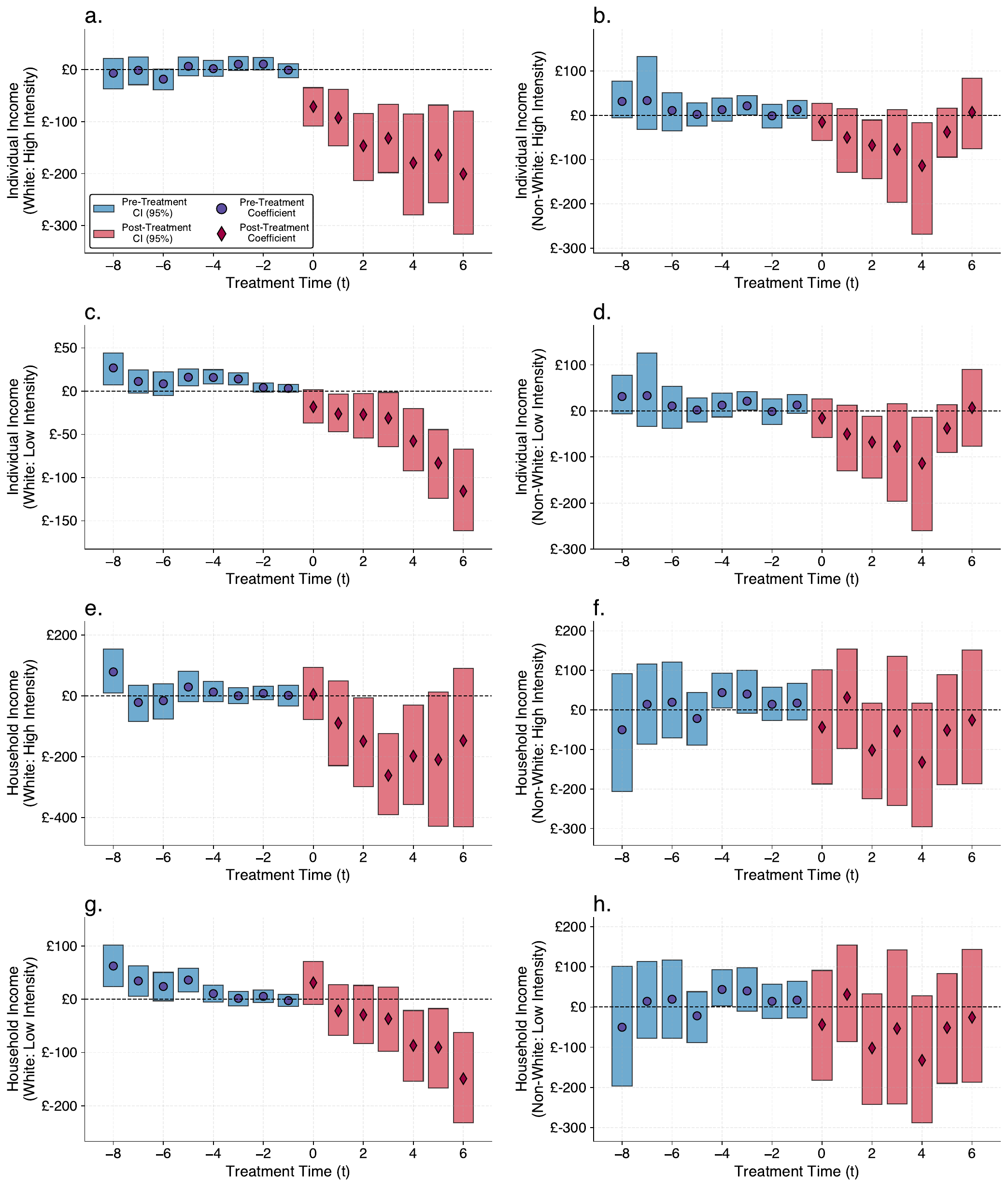}
    \caption{\small{\textbf{Individual and Household Income by ethnicity.} Individual Synthetic Control. Average Treatment Effect on the Treated. The blue shaded areas and blue circles represent the pre-treatment confidence intervals at 95\% and the pre-treatment coefficients, respectively. The red shaded areas and red diamonds denote the post-treatment confidence intervals at 95\% and post-treatment coefficients, respectively. For the full set of individual controls see Table \ref{descr}. 
     Panels a. and e. report the difference between high-intensity informal `White'  carers and their counterfactual; Panel b. and f. report high-intensity informal `non-White' carers; Panel c. and g. report low-intensity informal `White' carers; Panel d. reports low-intensity informal `non-White' carers. Panels a.-d. report individual income, while e.-h. report household income. Source: UKHLS data (years 2009-2020), authors' calculations.
     }}
    \label{ci_ii_eth}
\end{figure}

\newpage
\begin{figure}[H]
    \centering
    \includegraphics[width=0.955\linewidth]{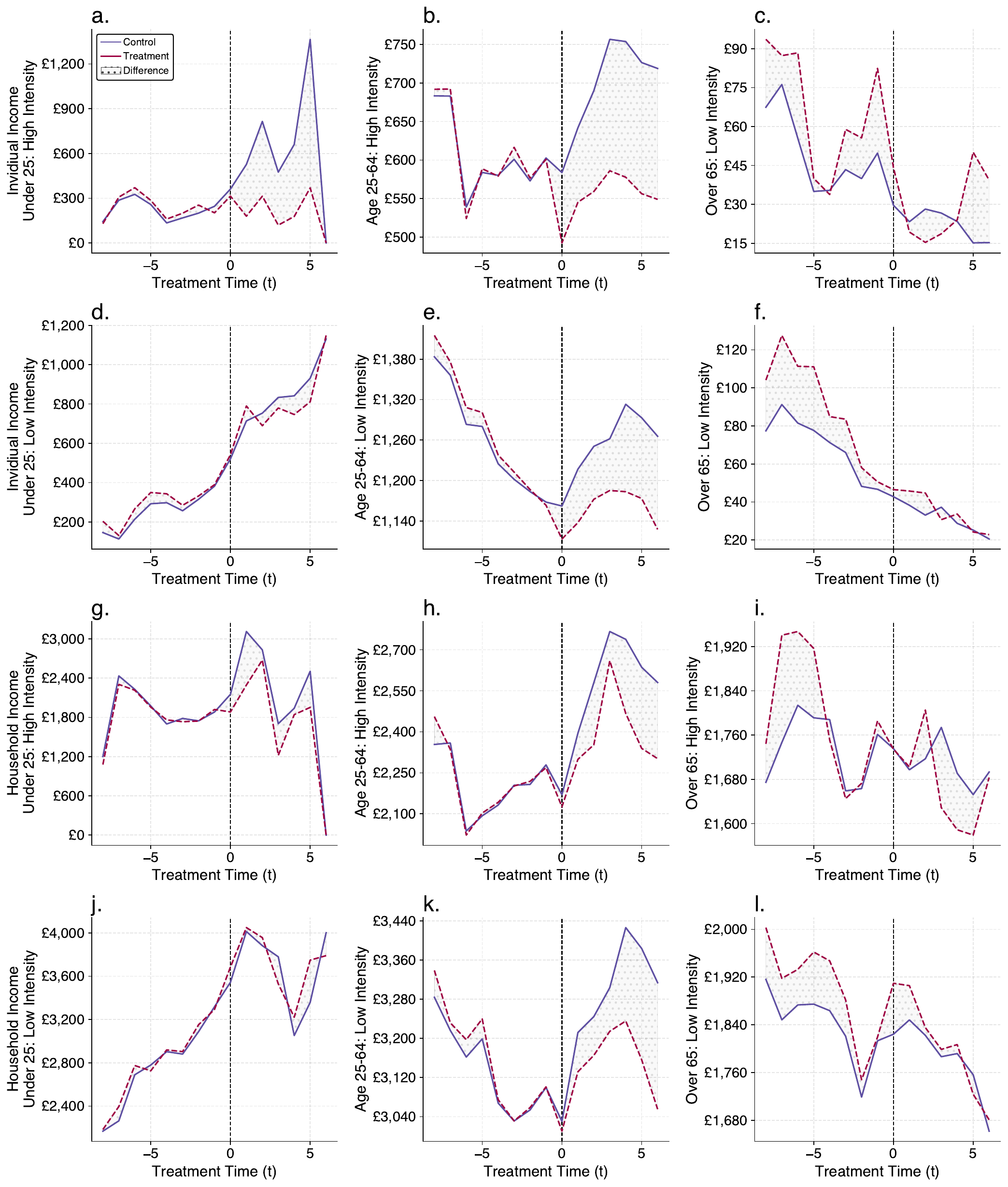}
    \caption{\small{\textbf{Inflation Adjusted Household and Individual Income - Difference between treatment and control groups by age groups.} 
    Individual Synthetic Control. The solid violet line depicts non-carers' income trajectories; the burgundy dashed line represents informal carers'. The grey area represents the difference. For the full set of individual controls see Table \ref{descr}.  Panels a. and g. report the difference between high-intensity informal carers aged below 25 and their counterfactual; Panels b. and h. report high-intensity informal carers aged 25-64; Panels c. and i. report high-intensity informal carers aged 65 and above; Panels d. and j. report low-intensity informal carers aged below 25; Panels e. and k. report low-intensity informal carers aged 25-64; Panels f. and l. report low-intensity informal carers aged 65 amd above Panels a.-f. report individual income, while g.-l. report household income.Source: UKHLS data (years 2009-2020), authors' calculations.
    }}\label{age_ind}
\end{figure}

\newpage
\begin{figure}[H]
    \centering
    \includegraphics[width=0.955\linewidth]{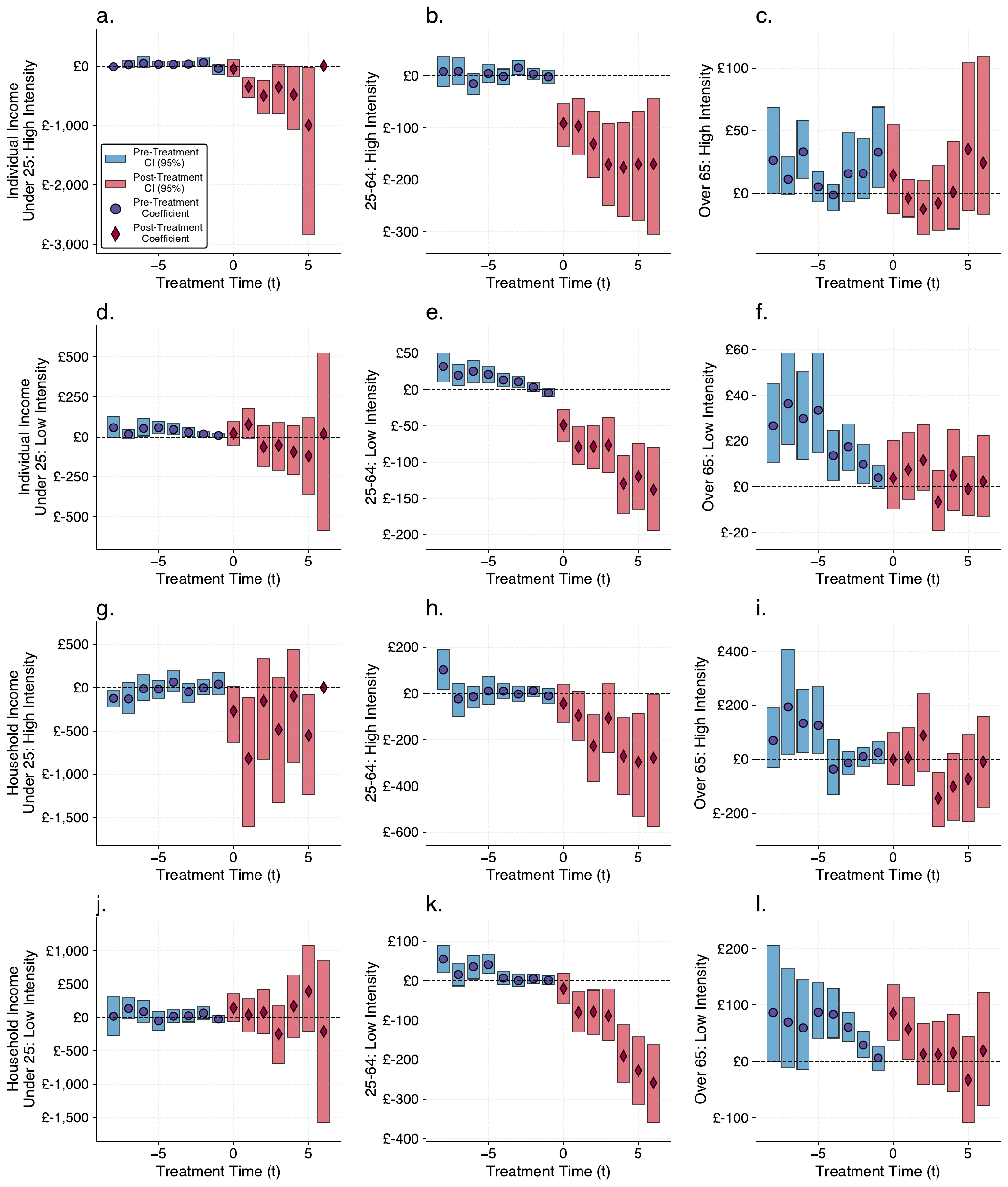}
    \caption{\small{\textbf{Inflation Adjusted Household and Individual Income by age groups.} Individual Synthetic Control estimation. Average Treatment Effect on the Treated. The blue shaded areas and blue circles represent the pre-treatment confidence intervals at 95\% and the pre-treatment coefficients, respectively. The red shaded areas and red diamonds denote the post-treatment confidence intervals at 95\% and post-treatment coefficients, respectively. For the full set of individual controls see Table \ref{descr}. 
     Panels a. and g. report the difference between high-intensity informal carers aged below 25 and their counterfactual; Panels b. and h. report high-intensity informal carers aged 25-64; Panels c. and i. report high-intensity informal carers aged 65 and above; Panels d. and j. report low-intensity informal carers aged below 25; Panels e. and k. report low-intensity informal carers aged 25-64; Panels f. and l. report low-intensity informal carers aged 65 and above Panels a.-f. report individual income, while g.-l. report household income. Source: UKHLS data (years 2009-2020), authors' calculations.
     }}\label{ci_ii_age}
\end{figure}

\newpage
\begin{figure}[H]
    \centering
    \includegraphics[width=0.955\linewidth]{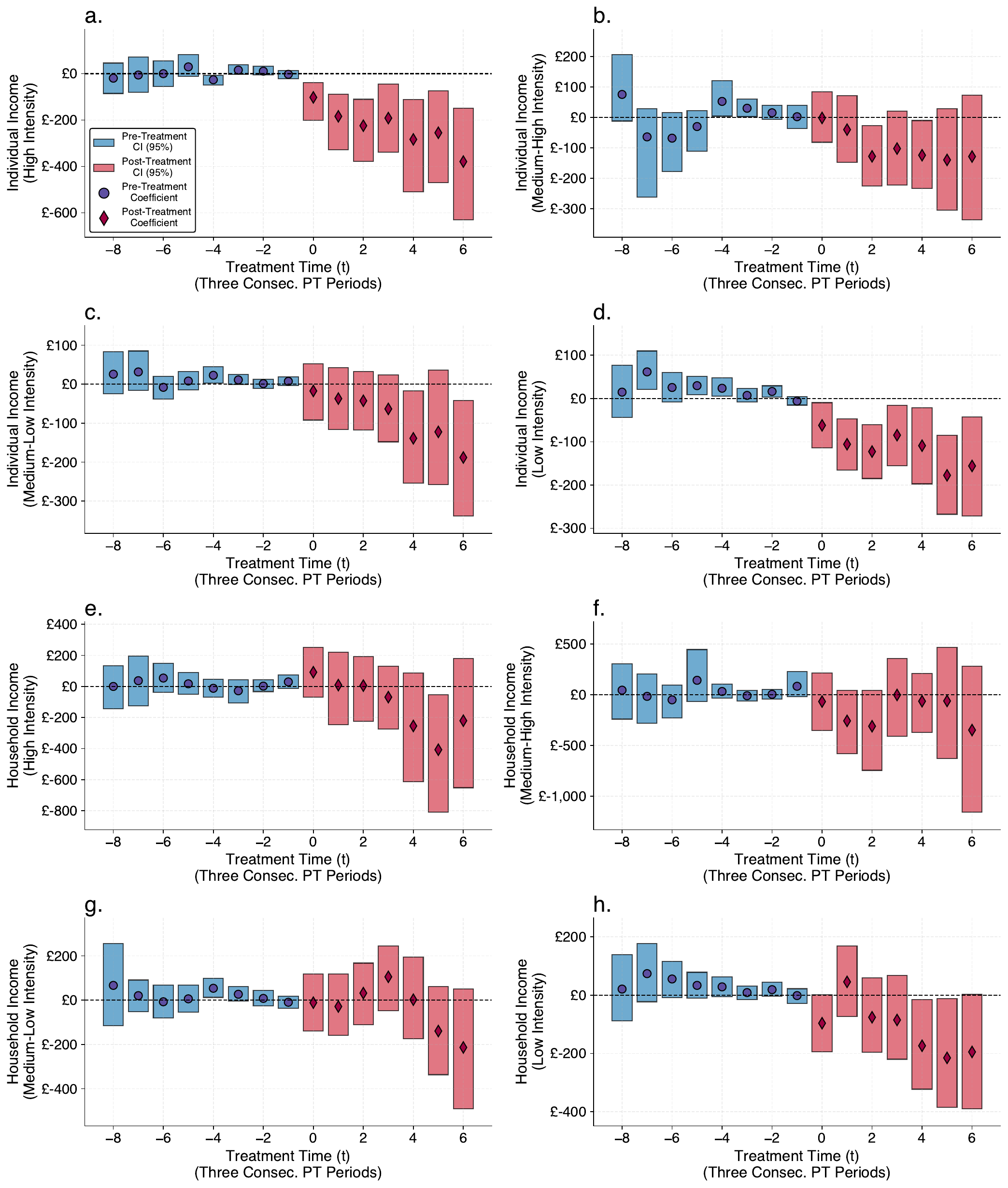}
    \caption{\small{\textbf{Inflation Adjusted Individual and Household Income - Three consecutive pre-treatment periods.} Individual Synthetic Control. Average Treatment Effect on the Treated. The blue shaded areas and blue circles represent the pre-treatment confidence intervals at 95\% and the pre-treatment coefficients, respectively. The red shaded areas and red diamonds denote the post-treatment confidence intervals at 95\% and post-treatment coefficients, respectively. For the full set of individual controls see Table \ref{descr}.
    Panels a. and e. report the difference between high-intensity informal carers and their counterfactual; Panels b. and f. report medium-high-intensity informal carers; Panels c. and g. report medium-low-intensity informal carers; Panel d. and h. report low-intensity informal carers. Panels a.-d. report individual income, while e.-h. report household income. Source: UKHLS data (years 2009-2020), authors' calculations.
    }}\label{hh_and_ii_rc_3}
\end{figure}

\newpage
\begin{figure}[H]
    \centering
    \includegraphics[width=0.955\linewidth]{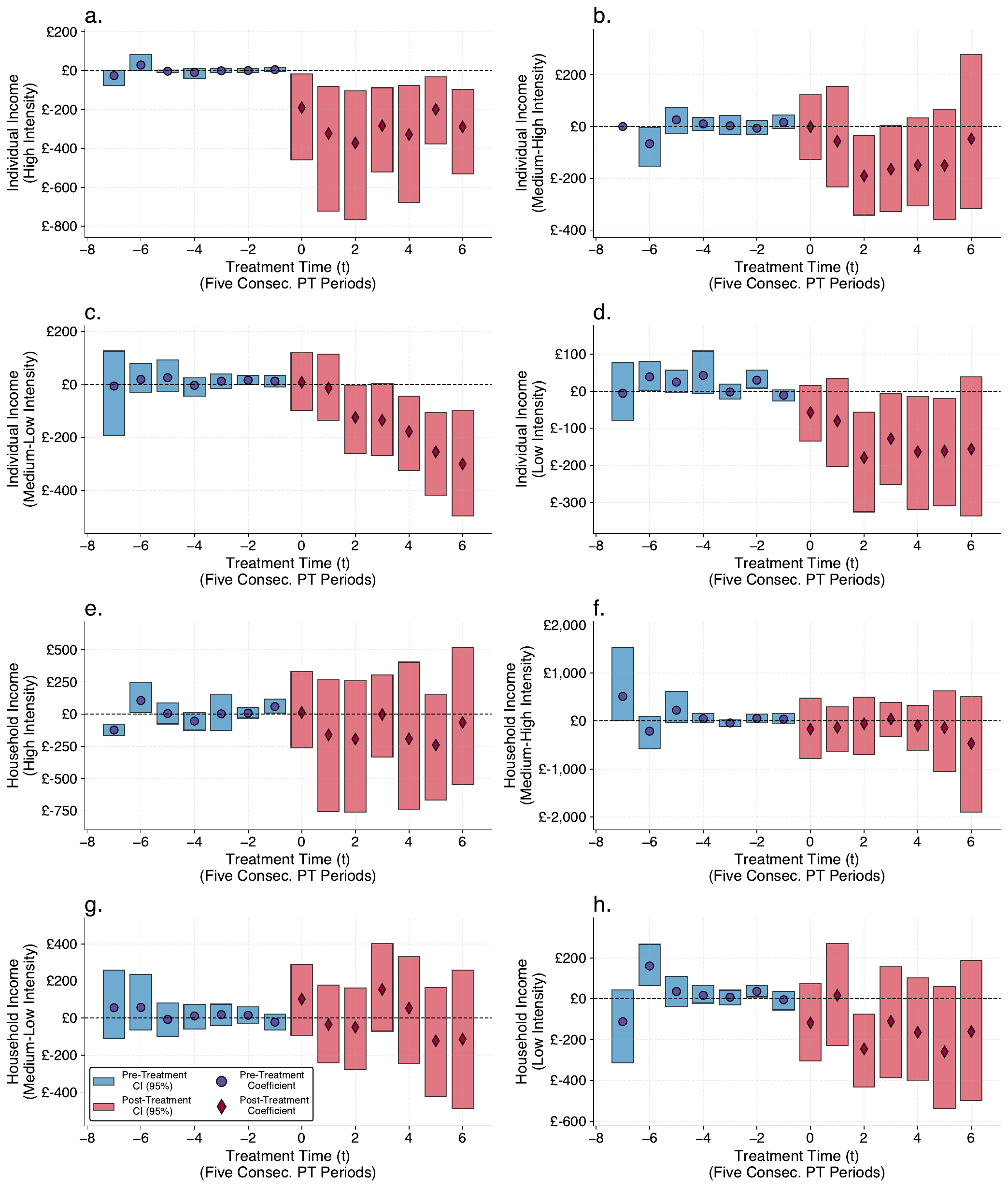}
    \caption{\small{\textbf{Inflation Adjusted Household and Individual Income - Five consecutive pre-treatment periods.} Individual Synthetic Control estimation. The blue shaded areas and blue circles represent the pre-treatment confidence intervals at 95\% and the pre-treatment coefficients, respectively. The red shaded areas and red diamonds denote the post-treatment confidence intervals at 95\% and post-treatment coefficients, respectively. For the full set of individual controls see Table \ref{descr}. Panels a. and e. report the difference between high-intensity informal carers and their counterfactual; Panels b. and f. report medium-high-intensity informal carers; Panels c. and g. report medium-low-intensity informal carers; Panel d. and h. report low-intensity informal carers. Panels a.-d. report individual income, while e.-h. report household income. Source: UKHLS data (years 2009-2020), authors' calculations.
    }}\label{ii_rc_5}
\end{figure}

\newpage
\begin{figure}[H]
    \centering    \includegraphics[width=0.955\linewidth]{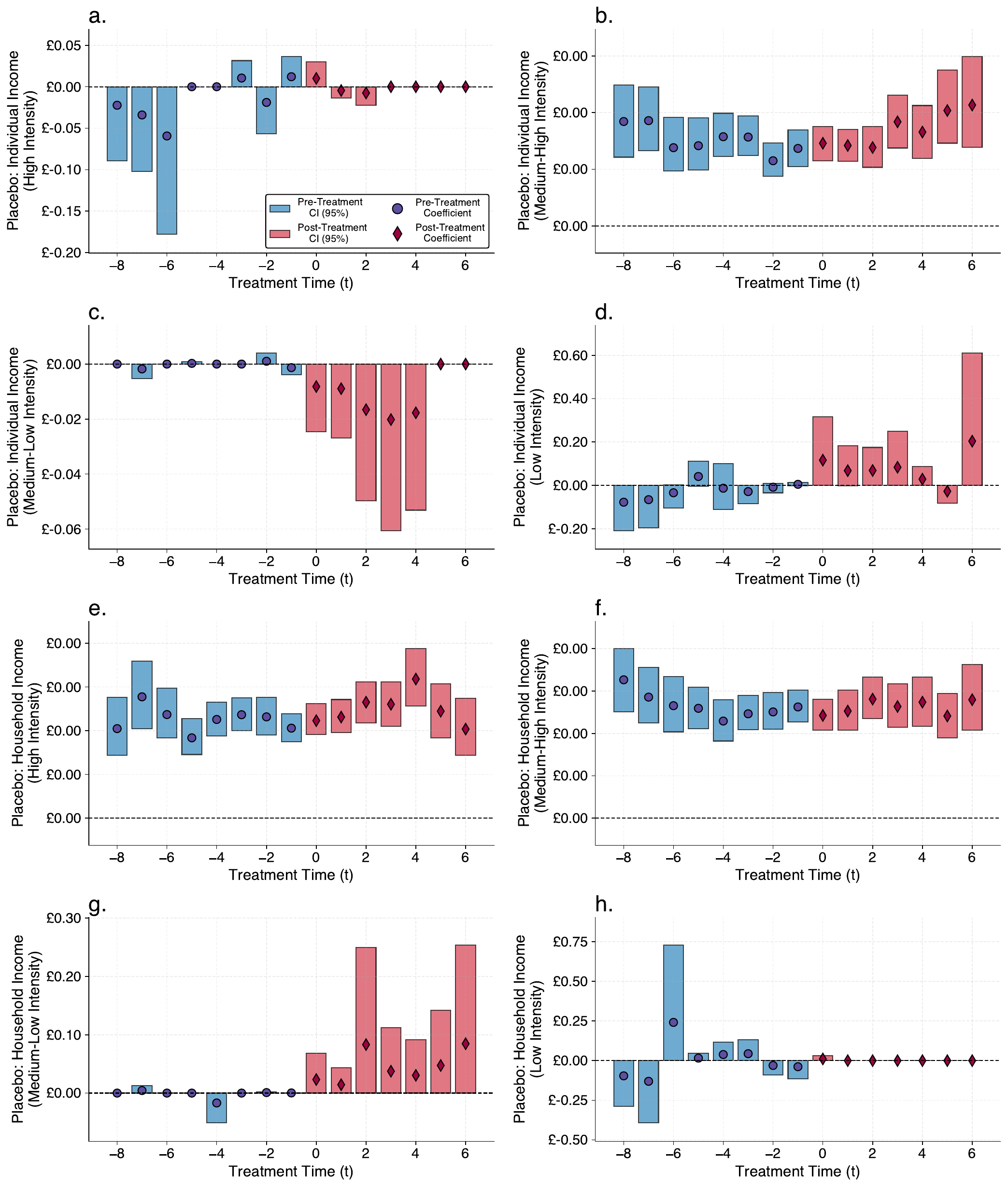}
    \caption{\small{\textbf{Inflation Adjusted Household and Individual Income - Placebo test.} 
     Individual Synthetic Control estimation. The blue shaded areas and blue circles represent the pre-treatment confidence intervals at 95\% and the pre-treatment coefficients, respectively. The red shaded areas and red diamonds denote the post-treatment confidence intervals at 95\% and post-treatment coefficients, respectively. For the full set of individual controls see Table \ref{descr}. Panels a. and e. present the difference between high-intensity informal carers and their counterfactual; Panels b. and f. report medium-high-intensity informal carers; Panels c. and g. report medium-low-intensity informal carers; Panels d. and h. report low-intensity informal carers. Panels a.-d. report individual income, while e.-h. report household income. Source: UKHLS data (years 2009-2020), authors' calculations.}} \label{placebo_ind}
\end{figure}

\newpage 
\begin{figure}[H]
    \centering
    \includegraphics[width=0.955\linewidth]{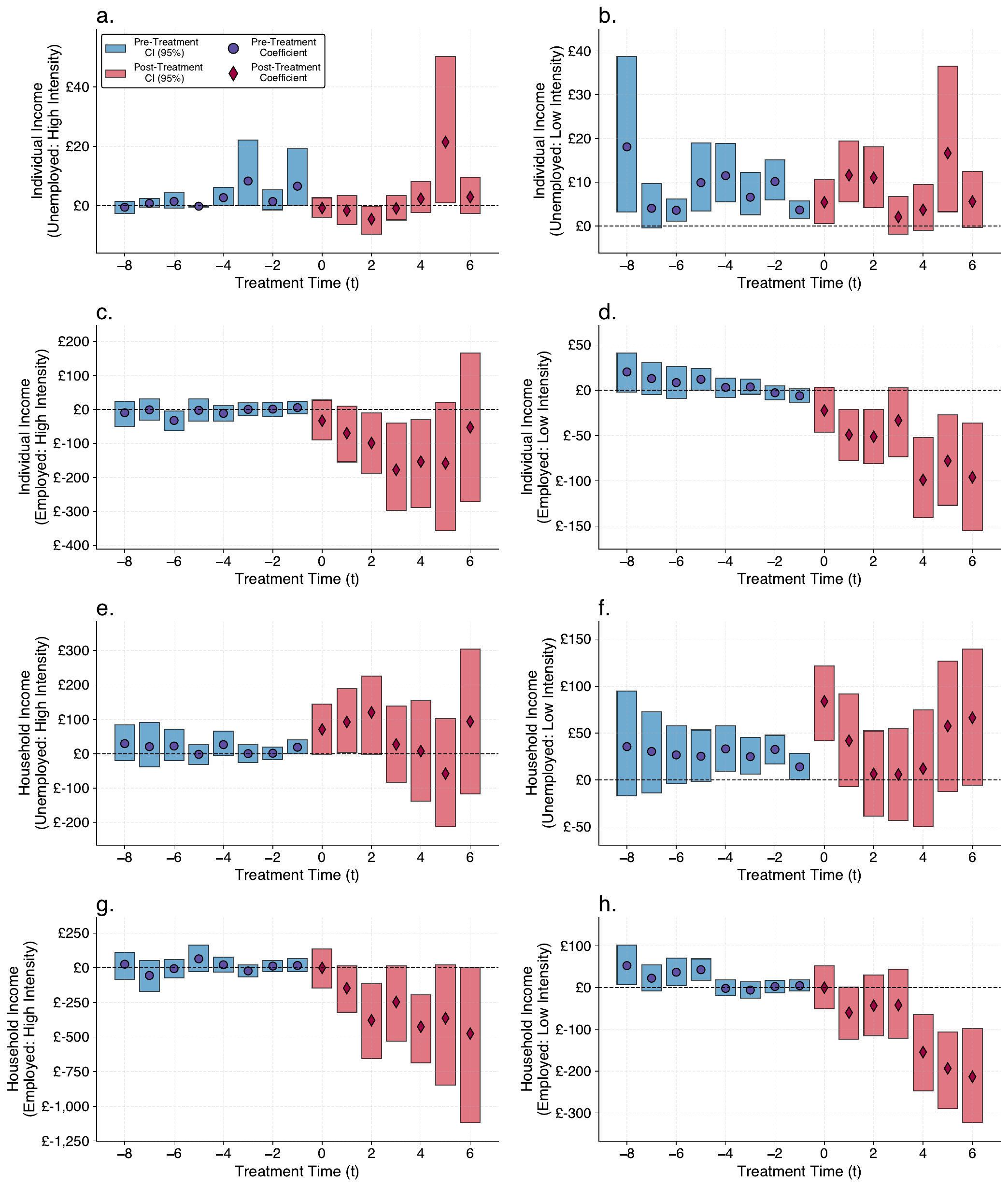}
    \caption{\small{\textbf{Inflation Adjusted Household and Individual Income by employment status.} 
    Individual Synthetic Control. Average Treatment Effect on the Treated. The blue shaded areas and blue circles represent the pre-treatment confidence intervals at 95\% and the pre-treatment coefficients, respectively. The red shaded areas and red diamonds denote the post-treatment confidence intervals at 95\% and post-treatment coefficients, respectively. For the full set of individual controls, see Table \ref{descr}. 
 Panels a. and e. report high-intensity unemployed informal carers; Panels b. and f. report low-intensity unemployed informal carers; Panels c. and g. report high-intensity employed informal carers; Panels d. and h. report low-intensity employed informal carers. Panels a.-d. report individual income, while e.-h. report household income. Source: UKHLS data (years 2009-2020), authors' calculations.
    }}\label{employed_ind}
\end{figure}

\newpage
\begin{figure}[H]
    \centering
    \includegraphics[width=0.955\linewidth]{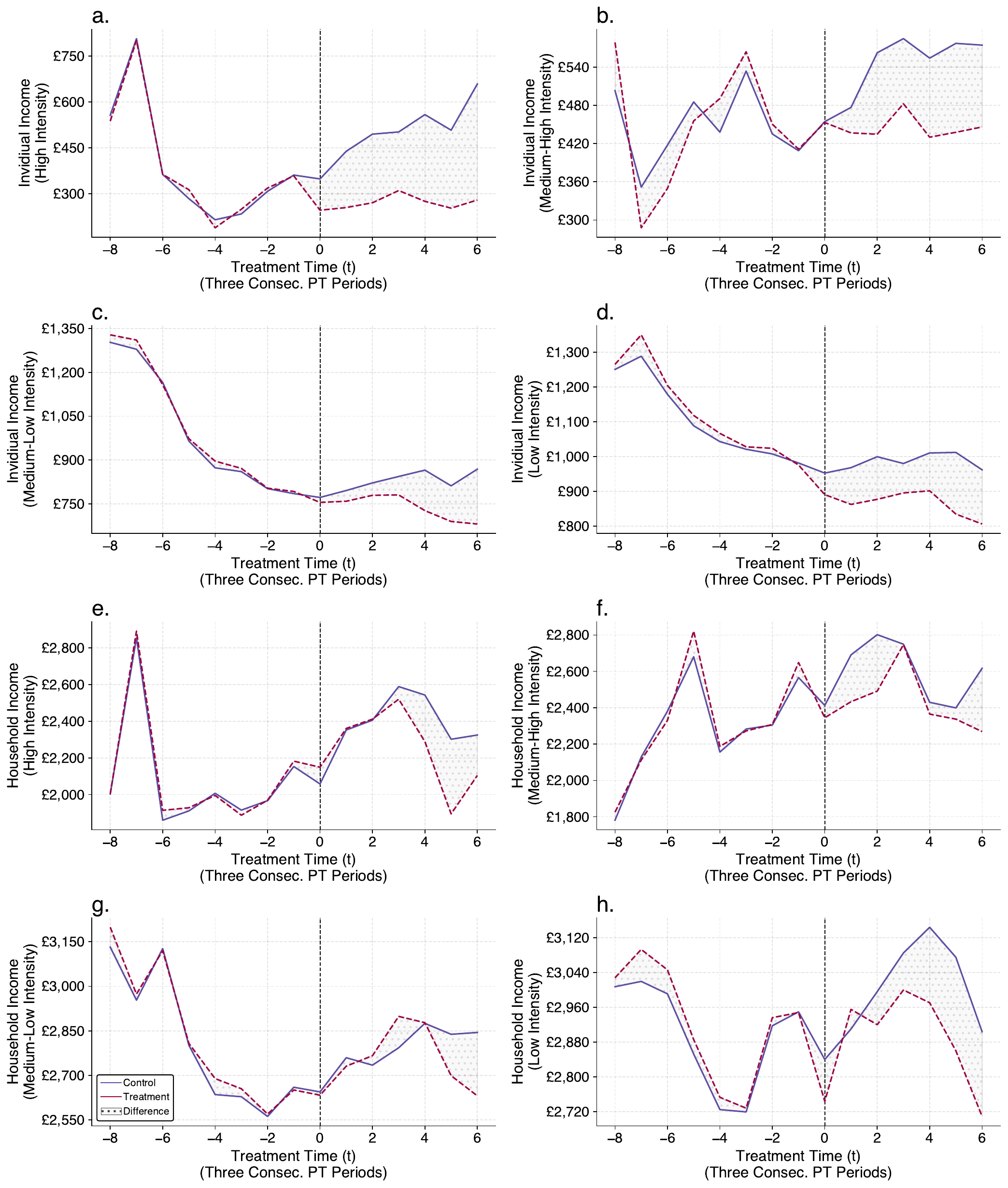}
    \caption{\small{\textbf{Inflation Adjusted Individual and Household Income - Difference between treatment and control groups with a three years continuous treatment period.} 
    Individual Synthetic Control. The solid violet line depicts non-carers' income trajectories; the burgundy dashed line represents informal carers'. The grey area represents the difference. For the full set of individual controls see Table \ref{descr}. Panels a. and e. represent the difference between high-intensity informal carers and their counterfactual; Panels b. and f. report medium-high-intensity informal carers; Panels c. and g. report medium-low-intensity informal carers; Panels d. and h. report low-intensity informal carers. Panels a.-d. represent individual income, while e.-h. represent household income. Source: UKHLS data (years 2009-2020), authors' calculations.  }}\label{ci_hh_and_ind_3consecutive}
\end{figure}

\newpage

\begin{figure}[H]
    \centering
     \includegraphics[width=0.955\linewidth]{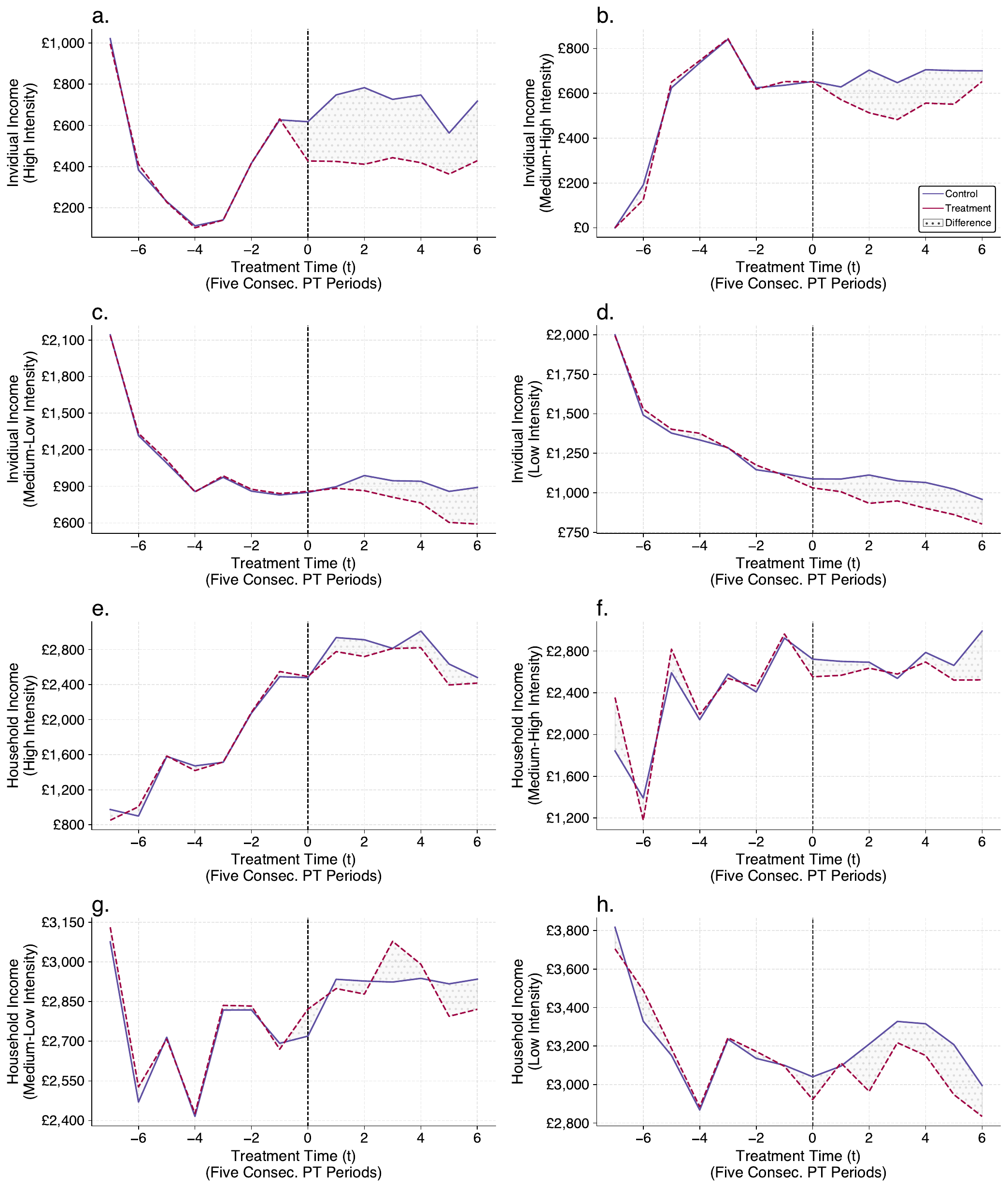}
    \caption{\small{\textbf{Inflation Adjusted Individual and Household Income: Difference between treatment and control groups with a five years continuous treatment period.} 
    Individual Synthetic Control. The solid violet line depicts non-carers' income trajectories; the burgundy dashed line represents informal carers'. The grey area represents the difference. For the full set of individual controls see Table \ref{descr}. Panels a. and e. represent the difference between high-intensity informal carers and their counterfactual; Panel b. and f. report medium-high-intensity informal carers; Panels c. and g. report medium-low-intensity informal carers; Panels d. and h. report low-intensity informal carers. Panels a.-d. report individual income, while e.-h. report household income. Source: UKHLS data (years 2009-2020), authors' calculations.
    }}\label{ci_ind_5consecutive}
\end{figure}
\newpage

\end{document}